\documentclass[longauth]{aa}
\usepackage{xcolor}
\usepackage{graphicx}
\usepackage{longtable}
\usepackage{amsmath}
\usepackage{chemfig}
\usepackage{mwe}
\usepackage{fontawesome}
\usepackage{verbatim}

\definecolor{dgreen}{rgb}{0.0, 0.5, 0.0}
\usepackage{txfonts}
\usepackage{hyperref}
\hypersetup{
    colorlinks=true,
    linkcolor=blue,
    filecolor=blue,
    citecolor=blue,
    urlcolor=blue,
}

\begin{document} 

   \title{JWST Observations of Young protoStars (JOYS+): Detection of icy complex organic molecules and ions}
   \subtitle{I. CH$_4$, SO$_2$, HCOO$^-$, OCN$^-$, H$_2$CO, HCOOH, CH$_3$CH$_2$OH, CH$_3$CHO, CH$_3$OCHO, CH$_3$COOH}

   \author{W. R. M. Rocha\inst{1,2}
   \and E. F. van Dishoeck\inst{2,3}
   \and M. E. Ressler\inst{4}
   \and M. L. van Gelder\inst{2}
   \and K. Slavicinska\inst{1,2}
   \and N. G. C. Brunken\inst{2}
   \and H. Linnartz\inst{1}
   \and T. P. Ray\inst{5}
   \and H. Beuther\inst{6}
   \and A. Caratti o Garatti\inst{7}
   \and V. Geers\inst{8}
   \and P. J. Kavanagh\inst{9}
   \and P. D. Klaassen\inst{8}
   \and K. Justannont\inst{10}
   \and Y. Chen\inst{2}
   \and L. Francis\inst{2}
   \and C. Gieser\inst{3}
   \and G. Perotti\inst{6}
   \and Ł. Tychoniec\inst{11}
   \and M. Barsony\inst{12}
   \and L. Majumdar\inst{13,14}
   \and V. J. M. le Gouellec\inst{15}
   \and L. E. U. Chu\inst{15}
   \and B. W. P. Lew\inst{16}
   \and Th. Henning\inst{6}
   \and G. Wright\inst{8}    
    }

   \institute{Laboratory for Astrophysics, Leiden Observatory, Leiden University, P.O. Box 9513, NL 2300 RA Leiden, The Netherlands.\\
    \email{rocha@strw.leidenuniv.nl}
          \and
             Leiden Observatory, Leiden University, PO Box 9513, NL 2300 RA Leiden, The Netherlands
          \and
             Max Planck Institut f{\"u}r Extraterrestrische Physik (MPE), Giessen-bachstrasse 1, 85748 Garching, Germany
          \and
             Jet Propulsion Laboratory, California Institute of Technology, 4800 Oak Grove Drive, Pasadena, CA 91109, USA
          \and
             Dublin Institute for Advanced Studies, Dublin, Ireland
          \and
             Max Planck Institute for Astronomy, K{\"o}nigstuhl 17, 69117 Heidelberg, Germany
          \and
             INAF-Osservatorio Astronomico di Capodimonte, Salita Moiariello 16, 80131 Napoli, Italy
          \and
             UK Astronomy Technology Centre, Royal Observatory Edinburgh, Blackford Hill, Edinburgh EH9 3HJ, UK
          \and
             Department of Experimental Physics, Maynooth University, Maynooth, Co Kildare, Ireland
          \and
             Department of Space, Earth and Environment, Chalmers University of Technology, Onsala Space Observatory, 439 92 Onsala, Sweden
          \and
             European Southern Observatory, Karl-Schwarzschild-Strasse 2, 85748 Garching bei M{\"n}chen, Germany
          \and
             SETI Institute 189 Bernardo Avenue, 2nd Floor, Mountain View, CA 94043, USA   
          \and
             School of Earth and Planetary Sciences, National Institute of Science Education and Research, Jatni 752050, Odisha, India
          \and
             Homi Bhabha National Institute, Training School Complex, Anushaktinagar, Mumbai 400094, India
          \and
             NASA Postdoctoral Program Fellow, NASA Ames Research Center, Moffett Field, CA, USA
          \and
             Bay Area Environmental Research Institute and NASA Ames Research Center, Moffett Field, CA 94035, USA
            }

   \date{Received xxxx; accepted yyyy}

 
  \abstract
   {Complex organic molecules (COMs) are ubiquitously detected in the gas phase and are thought to be mostly formed on icy grains. Nevertheless, no unambiguous detection of COMs larger than CH$_3$OH in ices has been reported so far, but exploring this matter in more detail has become possible with the unprecedented possibilities offered by the {\it James Webb} Space Telescope (JWST) within the infrared (IR) spectral range with its very high sensitivity and spectral resolution in the critical 5$-$10~$\mu$m range, the fingerprint region of oxygen-bearing COMs.}
   {In the program JWST Observations of Young protoStars (JOYS+), more than 30 protostars are being observed with the Medium Resolution Spectrograph (MRS) of the Mid-IR Instrument (MIRI). The goal of this study is to comprehensively explore the COMs ice signatures in one low- and one high-mass protostar, NGC~1333~IRAS~2A and IRAS~23385+6053, respectively.}
   {We perform global continuum and silicate subtractions of the MIRI-MRS spectra, followed by a local continuum subtraction in optical depth scale in the range around 6.8 and 8.6~$\mu$m, the ice COM fingerprint region. Different choices of local continuum and silicate subtraction were explored. Next, we fit observational data with a large sample of available IR laboratory ice spectra. We use the \texttt{ENIIGMA} fitting tool, a genetic algorithm-based code that not only finds the best fit between the lab data and the observations but also performs statistical analysis of the solutions, such as deriving the confidence intervals and quantifying fit degeneracy.}
   {We report the best fits for the spectral ranges between 6.8 and 8.6~$\mu$m in NGC~1333~IRAS~2A and IRAS~23385+6053, originating from simple molecules and COMs, as well as negative ions. In total, 10 chemical species are needed to reproduce the astronomical data. The strongest feature in this range (7.7~$\mu$m) is dominated by CH$_4$ and has contributions of SO$_2$ and OCN$^-$. Our results indicate that the 7.2 and 7.4~$\mu$m bands are mostly dominated by HCOO$^-$. We also find statistically robust detections of COMs based on multiple bands, most notably CH$_3$CHO, CH$_3$CH$_2$OH, and CH$_3$OCHO. The likely detection of CH$_3$COOH is also reported. Based on the ice column density ratios between CH$_3$CH$_2$OH and CH$_3$CHO of NGC~1333~IRAS~2A and IRAS~23385+6053, we find compelling evidence that these COMs are formed on icy grains. Finally, the derived ice abundances for NGC~1333~IRAS~2A correlate well with those in comet 67P/GC within a factor of 5.}
   {Based on the high-quality JWST (MIRI-MRS) spectra, we conclude that COMs are present in interstellar ices, thus providing additional proof for a solid-state origin of these species in star-forming regions. In addition, the good correlation between the ice abundances in comet 67P and NGC~1333~IRAS~2A is fully in line with the idea that cometary COMs can be significantly inherited from the early protostellar phases.}

\authorrunning{Rocha et al.}
\titlerunning{Frozen complex organic molecules}
\maketitle
%

\section{Introduction}

Complex organic molecules (COMs) are molecules with 6 atoms or more and have at least one atom of Carbon \citep{Herbst2009}. They are intrinsically important to comprehend the chemical complexity developed in star-forming regions since these materials are the feedstock for future exoplanetary systems. Once available in primitive planetary systems, this material can potentially promote the habitability of planets. An important question for delivering organic material to new solar systems is whether the molecules are in the gas phase or in ices as part of icy dust grains. Only in the latter case water and organic molecules are expected to be effectively delivered to terrestrial planets as discussed by \citet{Morbidelli2012, vanDishoeck2014, Morbidelli2018, OBrien2018, vanDishoeck2021} and the lower UV photodestruction cross-section of those molecules in the solid-phase \citep[e.g.,][]{Oberg2016}. Yet, very little information is available about these organic molecules in ices. The {\it James Webb} Space Telescope (JWST) provides a tool to change the situation by observing ice features during the early protostellar phases with higher resolution and sensitivity than before.

Gas-phase observations have been exceptionally successful in probing the chemical complexity towards low-mass and massive young stellar objects (LYSOs and MYSOs, respectively) as shown in the literature \citep[e.g.,][]{Blake1987, Cazaux2003, Bergner2017, Manigand2020, Belloche2020, vanGelder2020, Jorgensen2020, Nazari2021, Gieser2021}, and the observations and abundances of many COMs are summarized in the review by \citet{Jorgensen2020}. The current consensus is that COMs are efficiently formed in the solid phase, and desorb into the gas phase by thermal and non-thermal mechanisms. Nevertheless, methanol (CH$_3$OH) is the only COM securely identified in the solid phase based on the infrared (IR) spectra from the United Kingdom Infrared Telescope \citep[UKIRT;][]{Grim1991, Skinner1992, Dartois1999}, Infrared Space Observatory \citep[{\it ISO}, e.g.,][]{Gibb2004}, Very Large Telescope \citep{Pontoppidan2004, Dartois2003, Thi2006}, Infrared Telescopy Facility \citep{Chu2020}, {\it Spitzer} Space Telescope \citep{Boogert2008, Bottinelli2010}, AKARI \citep{Shimonishi2010, Perotti2021}, and recently with JWST \citep[e.g.,][]{Yang2022, McClure2023}. Larger COMs compared to CH$_3$OH have been tentatively identified based on only a single infrared (IR) vibrational mode or were proposed as possible carriers for yet unidentified features, such as ethanol - CH$_3$CH$_2$OH and acetaldehyde - CH$_3$CHO \citep{Schutte1999_weak, Oberg2011, Scheltinga2018}. These two COMs were also tentatively identified in the JWST high spectral resolution and sensitivity spectrum of IRAS~15398$-$3359 \citep{Yang2022} and towards two background stars \citep{McClure2023}, although at lower spectral resolution. Other upper limits have been determined for 
methyl formate \citep[CH$_3$OCHO;][]{Scheltinga2021}, methylamine \citep[CH$_3$NH$_2$;][]{Rachid2021}, methyl cyanide \citep[CH$_3$CN;][]{Rachid2022}, and formamide \citep[NH$_2$CHO;][]{Schutte1999_weak, slav2023}. \citet{Boudin1998} also estimated upper limits for molecules such as ethane (C$_2$H$_6$), acetylene (C$_2$H$_2$), hydrazine (N$_2$H$_4$), hydrogen peroxide (H$_2$O$_2$) and the hydrozonium ion (N$_2$H$_5^+$).

In parallel, gas phase formation of COMs through ion-molecule reactions has been proposed as well \citep[e.g.,][]{Balucani2015, Skouteris2018, Vazart2022}. The recent detection in a protoplanetary disk of gas phase CH$_3^+$ \citep{Berne2023}, an important intermediate in such reactions, supports the idea that COM formation is not exclusively realized on icy grains. This study focuses on the latter.

In the mid-IR spectral range, icy COMs have multiple absorption features, at for example, $\sim$5.8~$\mu$m, 6.8$-$7.0~$\mu$m, 7.2~$\mu$m, 7.4~$\mu$m, $\sim$8~$\mu$m, and ~$\sim$9.5$-$9.8~$\mu$m. However, due to the spectral overlap with the features of other molecules (including dust grain species), the bands at 7.2 and 7.4~$\mu$m bands have been considered the main fingerprint of COMs \citep[e.g.,][]{Schutte1999_weak, Oberg2011, Scheltinga2018}, which is also emphasized in this work. For example, \citet{Schutte1999_weak} proposed that the 7.2~$\mu$m band can be due to HCOOH (formic acid) ice, notably, the O$-$H bending mode of the carboxylic functional group and the C$-$H bending mode. In particular, \citet{Bisschop2007} finds a good match between the 7.2~$\mu$m band of the high-mass protostar W33A and the tertiary mixture HCOOH:CH$_3$OH:H$_2$O. The assumption that HCOOH would be the only carrier of the 7.2~$\mu$m band would lead to an unrealistic high solid-state abundance which is not in line with gas-phase observations \citep{Bisschop2007}. Nevertheless, this is not an argument for excluding HCOOH as a component of the interstellar ice, because it is also visible at other wavelengths (e.g., 5.8~$\mu$m). Instead, this indicates that HCOOH is not the main carrier of the 7.2~$\mu$m band. As an alternative, \citet{Boudin1998} and \citet{Oberg2011} propose that the deformation mode in the methyl functional group (CH$_3$) of ethanol can be the carrier for the 7.2~$\mu$m band. In the case of the 7.4~$\mu$m absorption feature, \citet{Schutte1999_weak} propose that the formate ion (HCOO$^-$) and acetaldehyde can be the carriers of this band. In all cases, these tentative assignments need a convincing profile fit with laboratory data to confirm these chemical species as carriers of those bands. These fits must also consider a larger spectral range around the 7$-$8~$\mu$m range where other typical vibrational modes of COMs are also detectable. 

In addition to COMs, ions such as OCN$^-$ (cyanate ion), HCOO$^-$ and NH$_4^+$ have been proposed to be present in interstellar ices mostly as part of salts produced by acid-base reactions \citep[e.g.,][]{Geballe1984, Grim1987, Schutte1999_weak, Schutte2003, Galvez2010, Mate2012, Moreno2013, Bergner2016, Kruczkiewicz2021}. The presence of OCN$^-$ in ice mantles can be considered a secure detection based on comprehensive laboratory experiments and extensive analysis towards several protostars \citep[e.g.,][]{Geballe1984, Broekhuizen2004} and towards background stars \citep{McClure2023}, and the good correlation between gas-phase abundances of HNCO and OCN$^-$ ice \citep{Oberg2009ocn}. In the case of HCOO$^-$ and NH$_4^+$, \citet{Boogert2015} argue that a convincing profile fit is still needed to firmly confirm the presence of these ions in ice mantles. If present in interstellar ices, these ions are likely part of refractory salts such as ammonium formate (NH$_4^+$HCOO$^-$) studied in the laboratory and found in comet 67P/G-C \citep[e.g.,][]{Poch2020}. In addition, the presence of OCN$^-$ and HNCO in ices has a strong astrobiological appeal: HNCO participates as a peptide bond between two single amino acids \citep[e.g.,][]{Fedoseev2015, Quenard2018, Colzi2021, Ligterink2022}.

In this work, we address the presence of simple molecules, COMs, and ions in protostellar ices using newly observed JWST spectra of two protostars with the Medium Resolution Spectrograph (MRS) of the Mid-Infrared Instrument (MIRI) under the JOYS+ program\footnote{\url{https://miri.strw.leidenuniv.nl/}} \citep[JWST Observations of Young protoStars;][]{vanDishoeck2023}. The first source is a high-mass \citep[$\sim$220~$M_{\odot}$; d = 4.9~kpc;][]{Molinari1998, Molinari2008} star-forming region called IRAS~23885+6053 (hereafter IRAS~23385). This cluster is highly embedded in its natal molecular cloud and shows maser emissions of H$_2$O \citep{Casoli1986} and CH$_3$OH \citep{Kurtz2004}, characteristic of shocks. {\it Spitzer} observations of this source reveal extended emission of
polycyclic aromatic hydrocarbons (PAHs) that are excited in this region by surrounding sources with spectral types between B1.5 and B5 \citep{Molinari2008}. NOEMA (NOrthern Extended Millimeter Array) observations of this source show a variety of gas-phase molecules, such as OCS, H$_2$CO, HNCO, CH$_3$OH and CH$_3$CN \citep{Cesaroni2019}. No other gas-phase COM have been detected in this region apart from CH$_3$OH and CH$_3$CN \citep{Gieser2021}. Recently, \citet{Beuther2023} present the rich MIRI-MRS spectrum of IRAS~23885, with focus on the outflow tracers H$_2$(0$-$0) S(7), [Fe II] ($^4$F$_{9/2}$ $-$ $^6$D$_{9/2}$) and [Ne II] ($^2$P$_{1/2}$ $-$ $^2$P$_{3/2}$), and an accretion tracer, the Humphreys $\alpha$ HI(7$-$6) emission line detected at a 3$-$4$\sigma$ level. A multiwavelength analysis using MIRI and NOEMA data provides information about the hot, warm and cold molecular components in IRAS~23385 \citep{Gieser2023}. In addition, Francis et al. (submitted) analyse the compact gaseous molecular emission in IRAS~23385.

The second source targeted in this work is the low-mass Class 0 protostar NGC~1333~IRAS~2A (hereafter IRAS~2A) \citep[e.g.,][]{Jorgensen2005, Brinch2009}, a well-studied hot-corino located in the Perseus complex, specifically in the NGC~1333 molecular cloud, at a distance of 299$\pm$3 pc \citep{Ortiz2018, Zucker2018}. This is a protobinary system \citep{Looney2000, Reipurth2002}, hosting two collimated jets \citep{Sandell1994, Tobin2015}. IRAS~2A is also a source where several gas-phase COMs are identified \citep[e.g.,][]{Bottinelli2007}, including glycolaldehyde \citep[HCOCH$_2$OH;][]{Coutens2015, Taquet2015, desimone2017}, an important sugar molecule that participates in the formation of ribose, a component of the ribonucleic acid (RNA). 

Compared to the {\it Spitzer} Space Telescope, which had enough sensitivity to observe the brighter low-mass protostars but low spectral resolution, the JWST (MIRI/MRS) observations are qualitatively superior. The MIRI-MRS resolving power ($R$) around the 7$-$8~$\mu$m region, where COMs signatures are present, is $R$ = 3500$-$4000 \citep{Labiano2021}, whereas the {\it Spitzer} Infrared Spectrograph (IRS) offered a resolving power of only $R$ = 60. \citet{Yang2022} demonstrate for IRAS~15398$-$3359 that the ice absorption features around the 7$-$8~$\mu$m are significantly stronger with much higher S/N profiles due to better resolution and sensitivity in the MIRI/MRS compared with {\it Spitzer}/IRS. 


An important disclaimer to make is that although this paper presents JWST/MIRI spectra of a high- and low-mass protostar, this is not a comparative work of ices in those objects. Instead, we show data from two sources that were observed first in our programs. A suitable comparison of ice features in different sources needs a more extended list of protostars and will be performed in future work. This paper is structured as follows: Section~\ref{obs_sec} introduces the JWST/MIRI observations of IRAS~23385 and IRAS~2A, the method for the data reduction, and the background subtraction. In Section~\ref{method_sec} we describe how the ice bands are isolated, the fitting procedure, the method for quantifying the fitting degeneracy, and which criteria we use for a firm detection. We focus in this paper on the 6.8$-$8.6~$\mu$m range, and therefore molecules with vibrational modes outside this interval are not discussed (e.g., H$_2$O, CO$_2$, NH$_4^+$). The results are shown in Section~\ref{result_sec}, which includes the spectral fits, degeneracy analysis and ice column densities. The discussions of these results are presented in Section~\ref{discuss_sec}, outlining the important insights from the JWST observations, and how the ice COMs abundances correlate between the low- and high-mass protostar, as well as with the comet 67P/G-C. The conclusions are summarized in Section~\ref{conc_sec}.

Below we provide some guidelines for a selective reading of this paper:

\begin{itemize}
    \item JWST data treatment and subtractions performed on the data: Sections~\ref{data_red}, \ref{subsec:background_sub_IRAS23385}, \ref{cont}, \ref{SO2gas}, \ref{local_cont}. Appendices~\ref{silic_comp} and \ref{silic_effect}.\\

    \item Fitting methodology and results: Sections~\ref{global} and \ref{spec_decp}. Appendices~\ref{list_lab} and \ref{fits_increment}.\\

    \item Statistical analysis and degeneracy: Sections~\ref{degen_sec} and \ref{degen}. Appendices~\ref{formateionapd}, \ref{confidence_ap} and \ref{cont_effect}.\\
    
    \item Identification criteria and suggestions for future works: Sections~\ref{sec_crit} and \ref{remarks}. Appendices~\ref{lab_rem} and \ref{spurious_sec}.\\

    \item Ice column densities, abundances, and correlations: Sections~\ref{ice_cd_sec}, \ref{sec_tentative}, \ref{complex_alc}, \ref{GIratio}, and \ref{comet_comp}. Appendix~\ref{Ap_bs} and \ref{water_cd}.\\

    \item Ice chemical complexity discussions: Sections~\ref{spec_decp}, \ref{chem_compl} and \ref{ionice}. Appendix~\ref{hydro}.
\end{itemize}


\section{Observations}
\label{obs_sec}
\begin{figure*}
   \centering
   \includegraphics[width=\hsize]{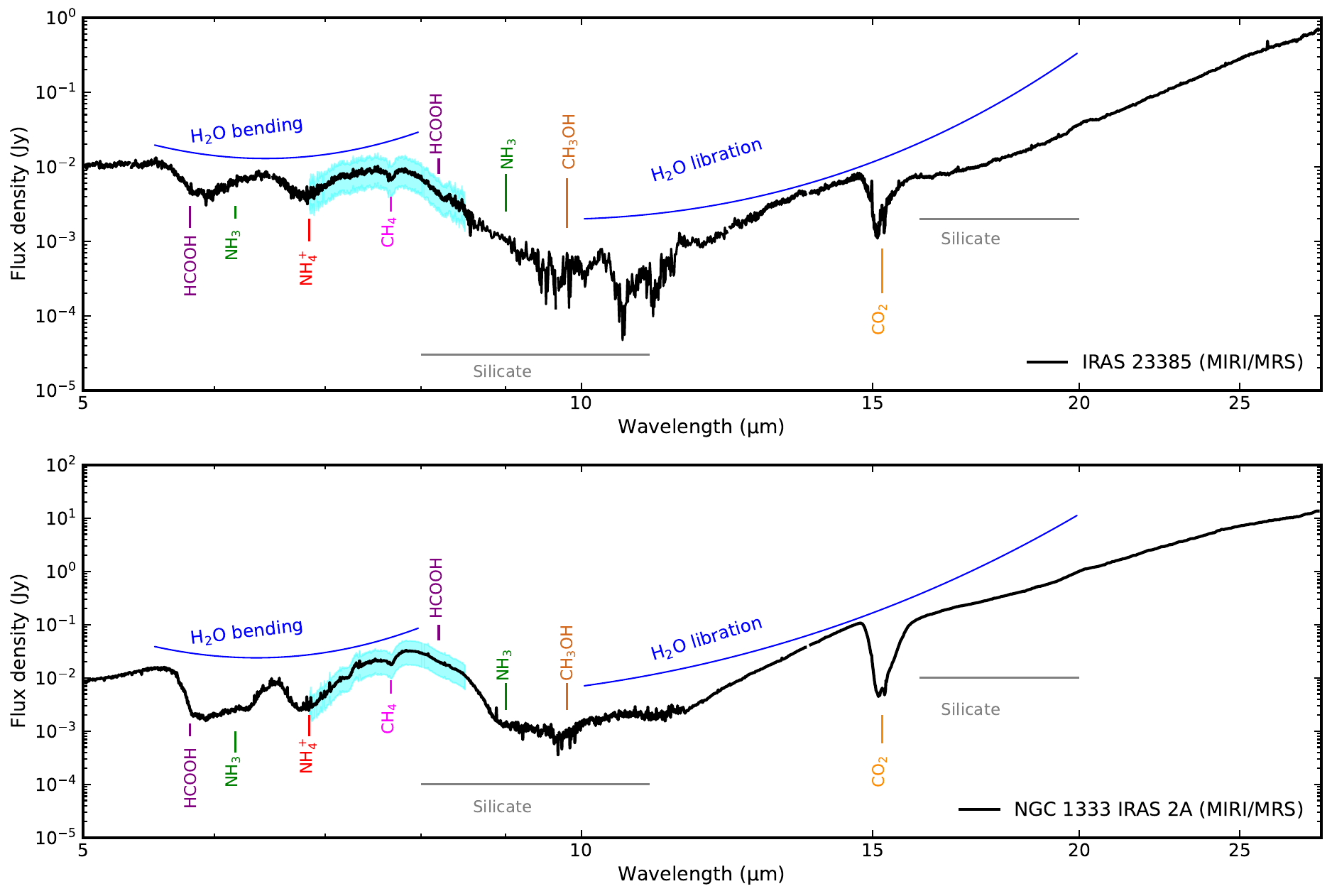}
      \caption{MIRI MRS spectrum of the high-mass protostar IRAS~23385+6053 (top) and the low-mass protostar NGC~1333~IRAS~2A (bottom). Strong gas-phase emission lines are masked in both spectra. The assignments for the absorption bands are given and differentiated by the colours. The shaded cyan area indicates the ``COMs region'' that is studied in this work.}
         \label{mirispec}
   \end{figure*}

\subsection{Data reduction}
\label{data_red}
IRAS~23385+6053 (R.A. 23$^{\rm{h}}$40$^{\rm{m}}$54.5$^{\rm{s}}$, Dec. +61$^{\rm{d}}$10$^{\rm{m}}$28$^{\rm{s}}$) and NGC~1333~IRAS~2A (R.A. 03$^{\rm{h}}$28$^{\rm{m}}$55.57$^{\rm{s}}$, Dec. +31$^{\rm{d}}$14$^{\rm{m}}$36.97$^{\rm{s}}$) were observed with JWST as part of the guaranteed observation time (GTO) 1290 (P.I. E. F. van Dishoeck) and 1236 (P.I. M. Ressler), respectively.

All data presented in this paper were taken with the Mid-InfraRed Instrument \citep[MIRI;][]{Rieke2015, Wright2015, Wright2023} in the Medium Resolution Mode \citep[MRS;][]{Wells2015, Labiano2021, Argyriou2023}. For IRAS~23385, the observation was done in 2-point dither mode in a 4-point mosaic surrounding the central protostellar cluster. For IRAS2A, the observation consists of a single pointing on the protostellar binary in 2-point dither mode. For both targets, also dedicated background observations were performed (no dither for IRAS~23385, 2-point dither for IRAS~2A). For both targets, all three gratings (A, B, C) were used, providing the full wavelength coverage of MIRI (4.9--28~$\mu$m).  All data were taken using the FASTR1 readout mode. The integration time in each grating was 200~s and 111~s for IRAS~23385 (per pointing in the mosaic) and IRAS~2A, respectively.

The data were processed through all three stages of the JWST calibration pipeline \citep{Bushouse2022}, using the reference context {\tt jwst$\_$1017.pmap} (IRAS~23385) and {\tt jwst$\_$0994.pmap} (IRAS2A) of the JWST Calibration Reference Data System \citep[CRDS;][]{Greenfield2016}. The raw {\tt uncal} data were first processed through the {\tt Detector1Pipeline} of the pipeline, followed by the {\tt Spec2Pipeline}. In the latter step, the data are corrected for fringes using the fringe flat for extended sources (Mueller et al. in prep.), followed by applying a residual fringe correction (Kavanagh et al. in prep.). Moreover, in the case of IRAS2A, the telescope background was subtracted in this step using the dedicated background observation. However, the background observation of IRAS~23385 includes significant astronomical background emission across all wavelengths, as well as strong emission of PAHs, and could thus not be used to subtract the telescope background. The background estimation and subtraction procedure for IRAS~23385 is further discussed in Sect.~\ref{subsec:background_sub_IRAS23385}. As mentioned in \citet{Beuther2023}, an astrometric correction was applied for IRAS~23385, i.e., 1.6077$''$ in Right Ascension and 0.3485$''$ in Declination based on identified GAIA-DR3 stars in the parallel images. No such correction was necessary for IRAS2A. The data were further processed with the {\tt Spec3Pipeline} of the pipeline which produces cubes of all 12 subbands. In this step, both the master background and outlier rejection routines were switched off. 

The observation of IRAS~23385+6053 reveals two mid-infrared continuum sources \citep{Beuther2023} that are resolved at shorter wavelengths (channels 1 and 2, $\lambda<12$~$\mu$m), but which start to become marginally resolved at longer wavelengths (channel 3, $12$~$\mu$m$<\lambda<17$~$\mu$m) and are completely unresolved at the longest wavelengths (channel 4, $\lambda>17$~$\mu$m). The spectrum is therefore extracted from a large aperture of 2.5$''$ in diameter (which does not increase with wavelength) centred in between the two sources (R.A. (J2000) 23$^{\rm h}$40$^{\rm m}$54.49$^{\rm s}$, Dec (J2000) 61$^{\rm d}$10$^{\rm m}$27.40$^{\rm s}$) to encompass the flux of both sources across the full wavelength range. The estimated 1$\sigma$ rms increases from about 0.4~mJy below 20~$\mu$m to a few mJy at the longest wavelengths. 

In the observation of IRAS2A, only continuum emission related to the primary component of the binary is detected. The spectrum is therefore extracted from the peak of the continuum emission at 5.5~$\mu$m (R.A. (J2000) 03$^{\rm h}$28$^{\rm m}$55.57$^{\rm s}$, Dec (J2000) 31$^{\rm d}$14$^{\rm m}$36.76$^{\rm s}$). We assume that any contribution of the secondary component to the spectrum is negligible. The diameter of the aperture was set to $4\times1.22\lambda/D$ to capture as much of the source flux as possible without including too much noise. The estimated 1$\sigma$ rms is about 0.4~mJy below 15~$\mu$m and increases to a few mJy at 19~$\mu$m and $>10$~mJy longwards of 22~$\mu$m.

Figure~\ref{mirispec} (top) shows the MIRI/MRS spectra of IRAS~23385+6053 \citep{Beuther2023} and Figure~\ref{mirispec} (bottom) shows IRAS~2A covering the range between 5 and 28~$\mu$m, and with a resolving power of 4000--1500 \citep{Labiano2021}. The spectral absorption features are associated with different ice molecules where the principal molecules are labelled in this figure, and those in the cyan region will be further discussed in Section~\ref{result_sec}. We highlight the broad feature of H$_2$O covering the range between 5.5 and 8~$\mu$m (bending mode) and between 10 and 20~$\mu$m (libration mode). HCOOH shows prominent features at 5.8 and 8.2~$\mu$m which can be distinguished in these sources. Small features attributed to NH$_3$ (ammonia), CH$_4$ (methane) and NH$_4^+$ (ammonium) are also seen in this spectrum. Silicates are the other main solid-state species contributing to the absorption bands around 9.8 and 18~$\mu$m. In addition to the absorption features, the spectra of IRAS~23385 and IRAS~2A have various narrow emission lines which are masked in this work since it is focused on ice absorption features. We also point out that the IRAS~23385 spectrum is binned by a factor of four between 8.6 and 12~$\mu$m because of the saturated silicate band, and IRAS~2A spectrum is binned by a factor of two in the entire MIRI-MRS range.

\subsection{Background subtraction for IRAS~23385+6053}
\label{subsec:background_sub_IRAS23385}
In the case of IRAS~23385+6053, the telescope and other backgrounds could not be subtracted in either the {\tt Spec2Pipeline} or {\tt Spec3Pipeline} since this results in negative fluxes due to significant astronomical emission in the dedicated background observation. It is, however, crucial to remove the telescope background from our observations to derive accurate ice column densities. The background was therefore estimated by extracting a spectrum from the science observation off-source from the main infrared continuum sources at the position within the IFU (Integral Field Unit) where the background flux was the lowest (R.A. (J2000) 23$^{\rm h}$40$^{\rm m}$54.15$^{\rm s}$, Dec (J2000) 61$^{\rm d}$10$^{\rm m}$26.96$^{\rm s}$) using the same aperture size as used for extracting the science data (2.5$''$). The background subtraction also results in the subtraction of the 8.6~$\mu$m and 11.3~$\mu$m PAH features, the emission of which was about equally strong in the background-position as at the source position. However, a possible under- or over-subtraction of PAH emission does not alter the results in this work since the PAH bands are broader than the ice bands at the wavelengths targeted in this paper. The background subtracted spectrum of IRAS~23385+6053 is shown in the top panel of Fig.~\ref{mirispec}. The numerous strong gas-phase emission lines have been masked in this version of the spectrum. The unmasked version is available in \citet{Beuther2023} and \citet{vanGelder2023} for IRAS~23385 and IRAS~2A, respectively.

\section{Methodology}
\label{method_sec}
\begin{figure}
   \centering
   \includegraphics[width=6cm]{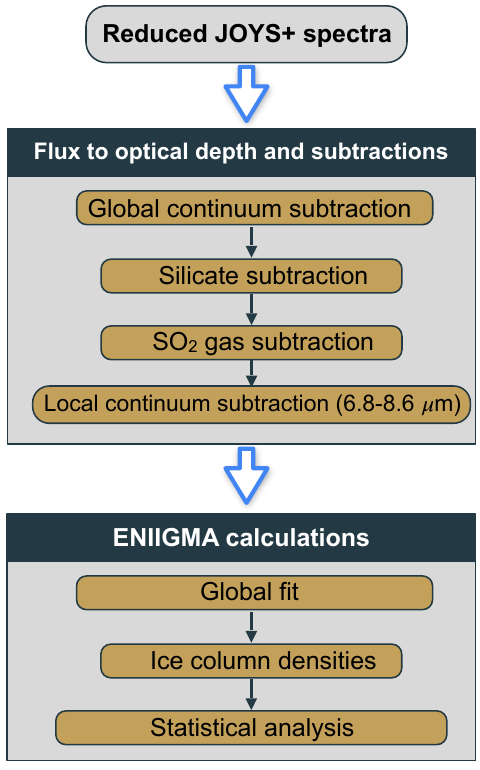}
      \caption{Summary of the methodology used in this paper.}
         \label{flow}
   \end{figure}

In this section, we provide information about the methods used to trace and subtract the dust continuum profile, the procedure to remove the silicate bands, and finally, the technique used to fit and identify the ice absorption bands. The focus of this paper is on the 6.8$-$8.6~$\mu$m range as indicated in Figure~\ref{mirispec}, but the entire MIRI spectrum has to be taken into account for the continuum fitting. For clarity reasons, we summarize the methodology steps in a flowchart shown in Figure~\ref{flow}.

\subsection{Continuum subtraction and silicate removal}
\label{cont}
\begin{figure*}
   \centering
   \includegraphics[width=\hsize]{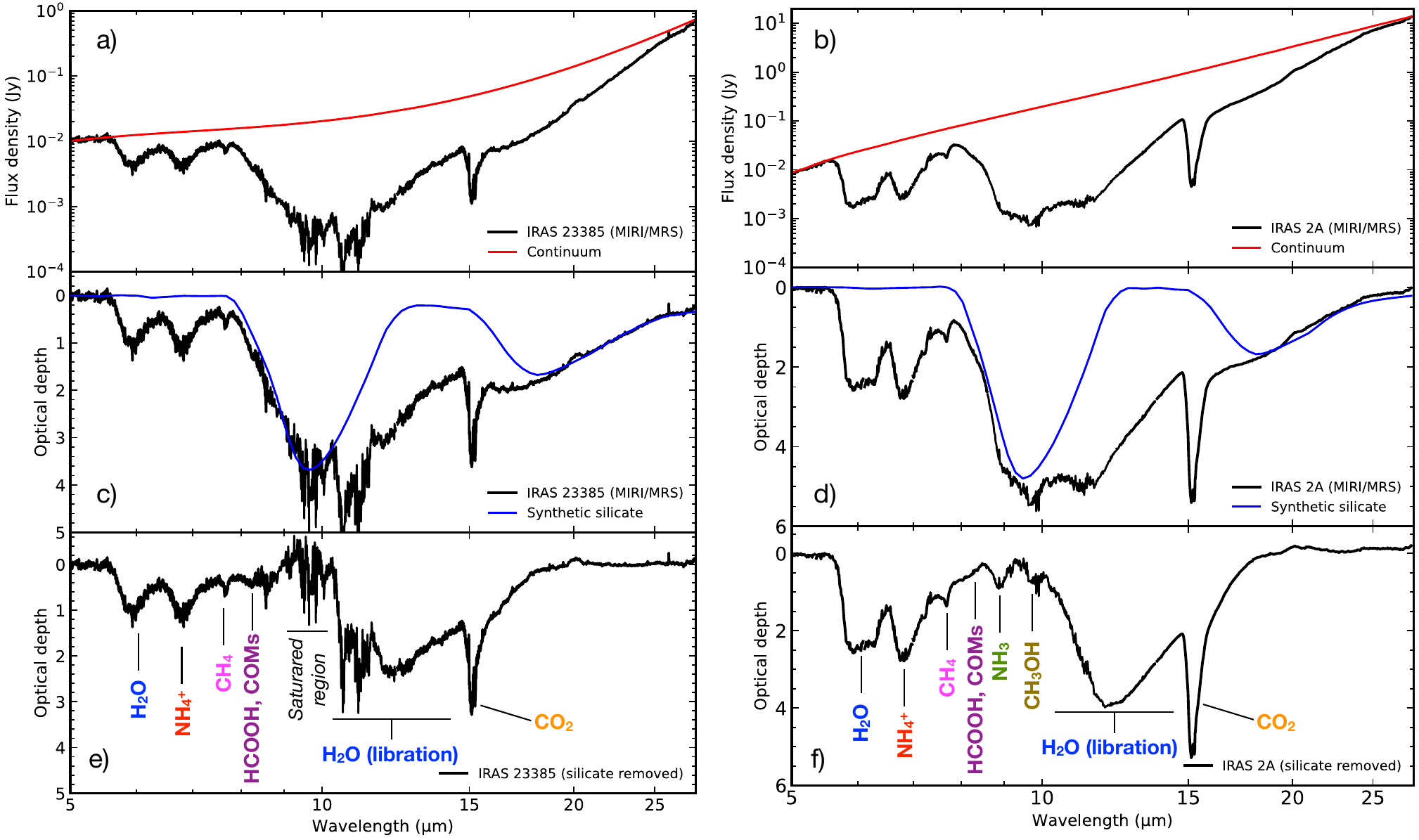}
      \caption{Continuum (a, b) and silicate subtraction (c, d) steps in IRAS~23385 (left) and IRAS~2A (right). The spectrum of IRAS~23385 is binned by a factor of two between 9 and 11~$\mu$m due to the saturated silicate profile. A third-order polynomial function is used to trace the continuum, and the silicate profile is a combination of two laboratory silicate spectra (olivine and pyroxene). Panels e and f show the silicate subtracted spectra of both protostars, with major features labelled.}
         \label{cont_od}
   \end{figure*}

The spectral energy distributions (SEDs) of IRAS~23385+6053 and IRAS~2A show an increasing slope towards long wavelengths which is typical of embedded protostars. Such SED shapes are also observed towards many high-mass protostars \citep[e.g., Orion BN, Orion IRc2;][]{Gibb2004} and low-mass protostars \citep[e.g., CrA~IRAS32, IRAS~03301+311 and L1448~IRS1][]{Boogert2008}. The shorter wavelength SED is composed of contributions by warm dust at a range of temperatures, whereas the increase in flux beyond 20~$\mu$m is due to the coldest envelope material \citep{Adams1987}. Determining the continuum of these protostars in the mid-IR range is not trivial because of the broad absorption bands in this spectral region. Often, a guided polynomial function is used \citep[e.g.,][]{Boogert2008} to trace the continuum SED. In this work, a third-order polynomial function is used to fit selected points between 5.3$-$5.5 and 27$-$28.5~$\mu$m where there is little or no overlap exists with ice absorption bands. An additional point was added at 7.5~$\mu$m slightly above the observational data to avoid unrealistic inflexions in the low-order polynomial that would lead to an unphysical continuum. The reason for using this point is to account for known broad absorption features in this range, for example, the blue wing of the 9.8~$\mu$m silicate feature, the H$_2$O ice broad bending mode, part of the C5 component proposed in \citet{Boogert2008}, and some of the complex molecules targeted in this paper. In this case, the observed flux itself at 7.5~$\mu$m is not suitable to be used as an anchor point. Uncertainties in the position of this guiding point do not affect the conclusions of this work. Figures~\ref{cont_od}a and \ref{cont_od}b, show the polynomial fits used for IRAS~23385 and IRAS~2A, respectively, and the emission lines are masked. Note that there is significant absorption with respect to this continuum over the entire wavelength range in both sources.


Once the continuum SED is determined, we convert the MIRI-MRS spectra of the two protostars to an optical depth scale, as shown in Figures~\ref{cont_od}c and \ref{cont_od}d by using the equation below:
\begin{equation}
    \tau_{\lambda} = -\mathrm{ln} \left( \frac{F_{\lambda}^{\rm{source}}}{F_{\lambda}^{\mathrm{cont}}} \right),
    \label{tau_obs}
\end{equation}
where $F_{\lambda}^{\rm{source}}$ is the source spectrum and $F_{\lambda}^{\mathrm{cont}}$ is the continuum SED.

Among the absorption bands seen in both spectra, silicates significantly contribute to the bands around 9.8~$\mu$m and 18~$\mu$m. Since the profile at 9.8~$\mu$m is positioned on top of a few ice bands, we perform a silicate removal to analyse the absorption features related to icy molecules. Often, the silicate profile observed towards the galactic centre source, GCS~3, is used to remove the silicate profile of other protostars \citep{Boogert2008, Bottinelli2010}. However, this method is not used in this work because the shape of the silicate features towards IRAS~23385+6053 and IRAS~2A are broader than GCS~3 as shown in Figures~\ref{Silic2A} and \ref{Silic23385} (Appendix~\ref{silic_comp}), and would lead to a spurious absorption profile around 8.6~$\mu$m after removing the silicate. As an alternative method, we combine two types of silicate to match the band at 9.8~$\mu$m following the approach described in \citet{Boogert2011, Poteet2015, DoDuy2020} and \citet{McClure2023}. Specifically, we use a mixture of amorphous pyroxene (Mg$_{0.5}$Fe$_{0.5}$SiO$_3$) and olivine (MgFeSiO$_4$) from \citet{Dorschner1995}.

We use the \texttt{optool} code \citep{Dominik2021} to generate optical depth spectra of the two silicate types. Briefly, we assume a mixture of silicate and carbon, typical chemical species of interstellar grains, with volume fractions of 87\% and 13\%, for IRAS~23385 and 82\% and 18\% for IRAS~2A, as typically used in protostar models. The variation in the fraction of carbon in the models allows us to fit the 9.8~$\mu$m and 18~$\mu$m bands simultaneously because a higher volume fraction of amorphous carbon reduces the intensities of the silicate bands. Different carbon fractions have been used in the literature to create dust models, for example, 30\% \citep{Weingartner2001}, and 15\% \citep{Pontoppidan2005, Woitke2016}. For the dust models, we assume a power-law size distribution with an exponent of -3.5 and grain sizes ranging from 0.1 to 1~$\mu$m. We adopt a distribution of hollow spheres \citep[DHS;][]{Min2005} to model the silicate band, as this approach mimics irregular geometries of the dust grains. The generated silicates are combined linearly to match the 18~$\mu$m band without exceeding the absorption at 9.8~$\mu$m and the blue silicate wing around 8.5~$\mu$m as shown in Figure~\ref{cont_od}c and \ref{cont_od}d.

The optical depth ratio between the 9.8 and 18~$\mu$m silicate features is equal to 2.7 for IRAS~23385 and 2.95 for IRAS~2A. These values are higher compared to the silicates found in the diffuse interstellar medium, which ranges from 1.4 to 2.0 \citep{Chiar2006}. It is likely that grain growth plays a role in this case, but a detailed study of this process is beyond the scope of this work. The silicate-removed spectra of IRAS~23385 and IRAS~2A are presented in Figures~\ref{cont_od}e and ~\ref{cont_od}f. By removing the silicate bands, the H$_2$O ice libration band is revealed around 12~$\mu$m. One can also see the ammonia umbrella mode at 9.0~$\mu$m and the methanol C-O stretch mode at 9.8~$\mu$m in IRAS~2A (Fig.~\ref{cont_od}f), but not in IRAS~23385 because of the saturation due to the silicate feature, as well as the higher noise level.

For completeness, we show in Appendix~\ref{silic_effect} the silicate subtraction using different silicates for IRAS~2A, the source with high signal-to-noise ratio (S/N). Those spectra are scaled to an optical depth of $\tau = 4.9$, which is the same as in the synthetic silicate fit (top panel of Figure~\ref{diff_silic}). The subtracted spectra are shown in the bottom panel. It is clear that the synthetic silicate and the MgSiO$_3$ model taken from \citet{Poteet2015} show very similar profiles. On the other hand, the use of the GCS~3 silicate leads to an unrealistic absorption excess. Finally, it is worth mentioning that silica (SiO$_2$) has a blue shoulder at 8.6~$\mu$m and a peak at 9~$\mu$m. However, the presence of silica is associated with other materials such as enstatite and forsterite whose spectral features are not seen in the sources addressed in this paper. For the different silicates considered in this paper, there are no relevant differences in the spectral shapes between 7.8 and 8.5~$\mu$m.

\subsection{SO$_2$ gas emission subtraction}
\label{SO2gas}
\begin{figure}[t]
   \centering
  \includegraphics[width=8cm]{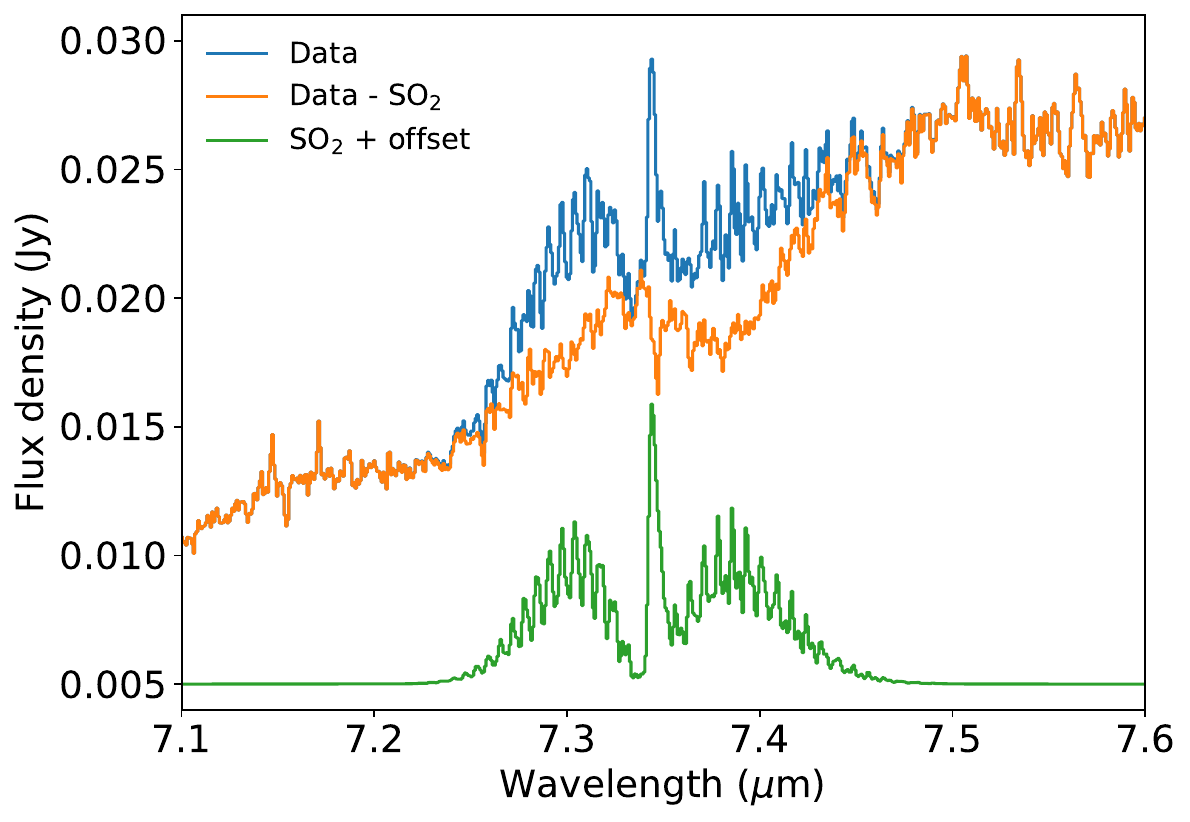}
      \caption{The observed spectrum of IRAS~2A (blue) surrounding the 7.2~$\mu$m and 7.4~$\mu$m ice absorption features with SO$_2$ ($\nu_3$) emission. The best-fit SO$_2$ model of \citet{vanGelder2023} is presented in green and the SO$_2$ subtracted spectrum of IRAS2A is shown in orange. Subtracting the SO$_2$ emission clearly reveals the 7.4~$\mu$m absorption feature.}
         \label{so2_sub}
   \end{figure}

In the spectrum of IRAS~2A, clear molecular emission is superimposed on the ice absorption features between $\sim7.25$~$\mu$m and $\sim7.45$~$\mu$m, see Fig.~\ref{so2_sub}. This emission was recently found to originate from warm gas-phase SO$_2$ \citep{vanGelder2023}. In order to accurately determine the contribution of ices in this wavelength range, the gas-phase emission lines of SO$_2$ ($\nu_3$) have to be subtracted from the spectrum. This was achieved by subtracting the best-fit gas-phase emission line model of SO$_2$, which was recently derived by \citet{vanGelder2023}. This model is very well constrained by the $R$ branch lines at 7.3~$\mu$m.

The SO$_2$ emission subtracted spectrum is also presented in Figure~\ref{so2_sub} as the orange line and clearly reveals an ice absorption feature around 7.4~$\mu$m that was hidden by the SO$_2$ emission. The SO$_2$ emission is slightly over-subtracted around 7.35~$\mu$m (i.e. at the Q-branch) but this does not hamper the analysis of the ice bands since this residual is much more narrow than typical ice absorption bands.

\subsection{Isolating ice features between 6.85 and 8.6~$\mu$m}
\label{local_cont}

Weak absorption features have been measured in the laboratory covering the range between 6.85 and 8.6~$\mu$m \citep[e.g.,][]{Lacy1991, Schutte1999_weak, Oberg2011}, and \citet{Scheltinga2018}, which are compared with observational spectra of massive protostars, by \citet{Boogert2008, Oberg2011}, low-mass protostars \citep[e.g.,][]{Zasowski2009, Yang2022} and background stars \citep{McClure2023}. Most notable are the absorption features around 7.2 and 7.4~$\mu$m. To isolate these bands, previous works used a polynomial fit on the flux scale data to trace a local continuum starting around 7.0$-$7.14~$\mu$m and finishing around 7.8-8.0~$\mu$m \citep[e.g.,][]{Schutte1999_weak}. This approach isolates the 7.2 and 7.4~$\mu$m features but excludes potential absorption features around 7~$\mu$m and at wavelengths long-wards of 7.8~$\mu$m where C$-$H and C$-$O absorption features of many possible molecules contribute.

\begin{figure*}
   \centering
   \includegraphics[width=\hsize]{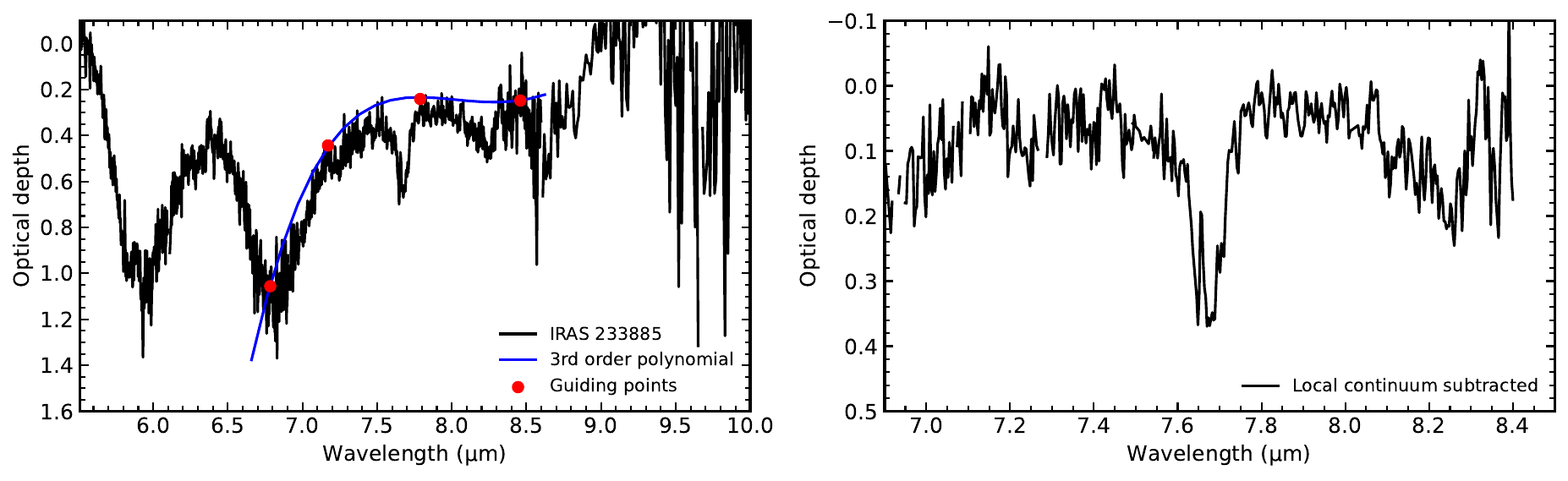}
   \includegraphics[width=\hsize]{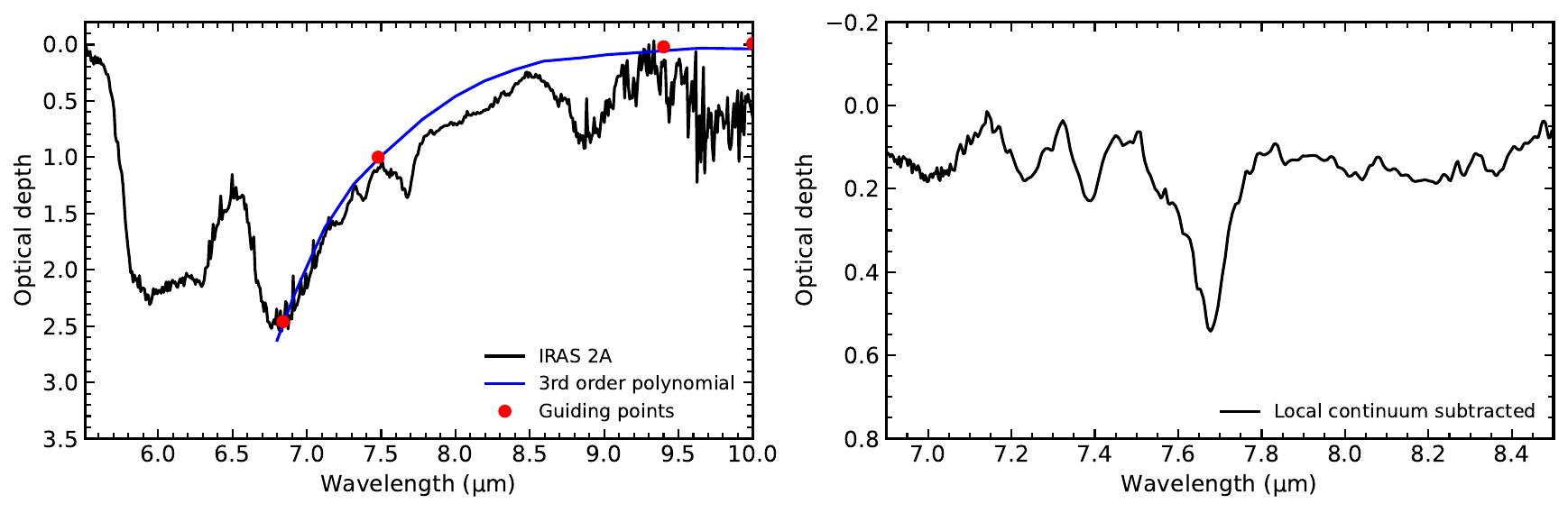}
    \caption{Left: Local continuum in the 6.8$-$8.6 spectral range using a fourth-order polynomial function. Right: Isolated 6.8$-$8.6 optical depth spectrum in both protostars.}
         \label{lc_coms}
   \end{figure*}

In this work, we isolate the 7.2 and 7.4~$\mu$m features using a third-order polynomial function, and following a slightly different approach. First, instead of using the spectrum on the flux scale, we perform the polynomial fit on the optical depth scale after removing the silicate absorption. The strong silicate band makes it difficult to observe small features such as those due to COMs. Second, we use guiding points fitted by a third-order polynomial function, as shown in the left panels of Figure~\ref{lc_coms}. This additional continuum represents blended absorption profiles from the broad H$_2$O ice bending mode, the red wing of the NH$_4^+$ cation, and the C5 component proposed by \citet{Boogert2008}. Another small contribution from the O-H bending mode of CH$_3$OH ice is also considered in this step. The points used for IRAS~23385 are at 6.8, 7.2, 7.7 and 8.5~$\mu$m and at 6.8, 7.3, 7.5, 9.4 and 10~$\mu$m for IRAS~2A. The positions of the guiding points are distinct because of the differences in the absorption profiles of the two sources. In order to account for possible C$-$H absorption bands, the first point at 6.8~$\mu$m is chosen to be marginally above the wing of the strong 6.85~$\mu$m feature. The points long-wards of 7.8~$\mu$m are selected where we expect no or weak ice absorption, to account for the C$-$O features. In the case of IRAS~2A, we use points at 9.4 and 10.0~$\mu$m because of the clear absorption profiles at 9 and 9.8~$\mu$m. 

After subtracting the local continuum fit, we isolate the ice features in the range between 6.8 and 8.5~$\mu$m as shown in the right panels of Figure~\ref{lc_coms}. In these figures, one can see absorption bands at 7, 7.2, 7.4, 7.5$-$7.8 and 8.2~$\mu$m that are fitted and discussed in Sections~\ref{result_sec} and \ref{discuss_sec}, respectively. Notably, the 8.2~$\mu$m band of IRAS~23385 seems broad and asymmetric, whereas in IRAS~2A the same band seems broad, weaker and symmetric. The band around 7.7~$\mu$m has a prominent blue shoulder in both sources, reflecting the interaction with other neighbouring chemical species.

\subsection{Fits and laboratory data}
\label{global}
Since molecules have multiple functional groups, performing simultaneous fits at different wavelengths supports secure detection. Additionally, some molecules are expected to absorb in the wavelength range targeted in this paper, such as CH$_4$, SO$_2$, HCOOH, CH$_3$CHO and CH$_3$CH$_2$OH \citep[e.g.,][]{Schutte1999_weak, Bisschop2007, Oberg2008}, which are used as starting point in the MIRI spectral ice decomposition. We fit the region between 6.8 and 8.6~$\mu$m of IRAS~2A and IRAS~23385 using the \texttt{ENIIGMA} fitting tool \citep{Rocha2021}. This code searches for the global minimum solution that fits the observations by performing a linear combination of laboratory ice data. The genetic algorithm approach benefits from a simple fitness function \citep{baeck2000}, and here we use the root-mean-square error (RMSE) given by the equation below: 
\begin{equation}
    RMSE = \sqrt{\frac{1}{n}\sum_{i=0}^{n-1} \left(\tau_{\nu,i}^{\rm{obs}}  - \sum_{j=0}^{m-1} w_j \tau_{\nu,j}^{\rm{lab}} \right)^2}
    \label{residual_eq}
\end{equation}
where both experimental ($\mathrm{\tau_{\nu,j}^{lab}}$) and observational ($\mathrm{\tau_{\nu,i}^{obs}}$) spectrum are converted to wavenumber space ($\nu$), $w_j$ is the scale factor, and $m$ and $n$ are the $m$th and $n$th data point. The absorbance laboratory data ($Abs$) are converted to an optical depth scale by the equation $\tau_{\nu}^{\rm{lab}} = 2.3 Abs_{\nu}$. In the degeneracy analysis shown in Section~\ref{degen_sec}, the error of the data is taken into account.

In a nutshell, \texttt{ENIIGMA} uses genetic modelling algorithms for searching the optimal coefficients of the linear combination ($w$). Genetic algorithms are robust optimization techniques based on the processes of natural selection that aim to find the global minimum solution for complex problems \citep{Holland1975, Koza1992}. Once the best fit is found, \texttt{ENIIGMA} calculates the ice column density of each component using the following equation:
\begin{equation}
    N_{ice} = \frac{1}{A} \int_{\nu_1}^{\nu_2} \tau_{\nu}^{lab} d\nu,
\end{equation}
where $A$ is the vibrational mode band strength of the molecule, which is listed in Table~\ref{ice_bs}. The band strengths of molecules change depending on the chemical environment. For this reason, we adopt values of corrected band strengths, when available, to derive the ice column densities. The derivation of band strengths is not straightforward because it depends on the ice density. The typical band strength uncertainties are around 15\% and 30\% for pure and mixed ices, respectively \citep{Rachid2022, slav2023}.

\begin{table*}
\caption{\label{ice_bs} List of vibrational transitions and band strengths of molecules considered in this paper.}
\centering 
\begin{tabular}{ccclccccc}
\hline\hline
Structure & Chemical formula & Name & $\lambda \; [\mu \mathrm{m}]$ & $\nu \; \mathrm{[cm^{-1}]}$ & Identification & $\mathcal{A} \; \mathrm{[cm \; molec^{-1}]}$ & References\\
\hline
\raisebox{-.5\height}{\includegraphics[height=0.25in]{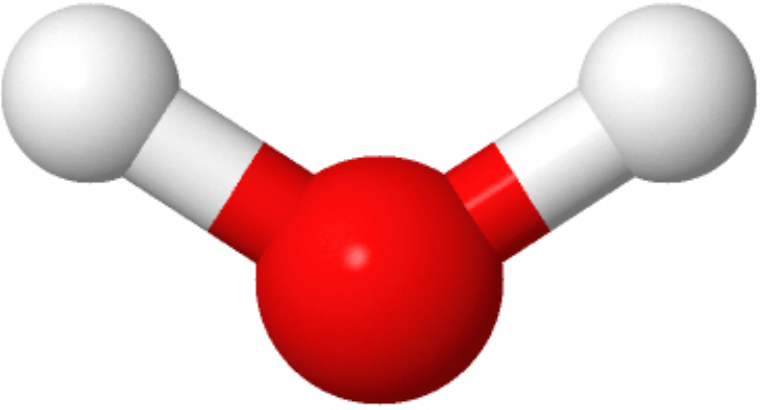}} & H$_2$O & Water & 13.20    & 760 & libration & $\mathrm{3.2 \times 10^{-17}}$ & [1]\\
\raisebox{-.5\height}{\includegraphics[height=0.45in]{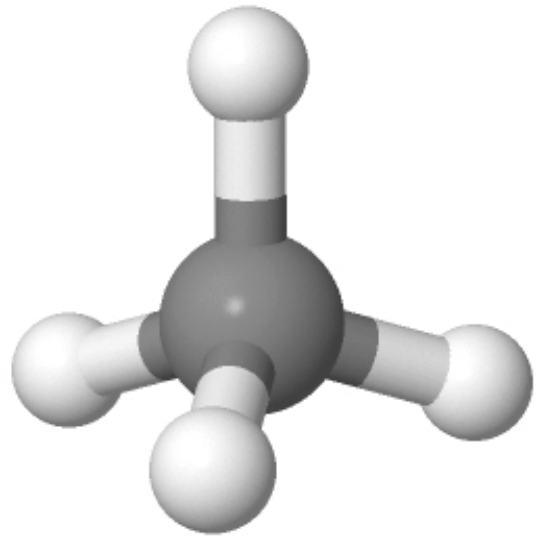}} & CH$_4$    & Methane       & 7.67    & 1303 & CH$_4$ def.   & $\mathrm{8.4 \times 10^{-18}}$ & [1]\\
\raisebox{-.5\height}{\includegraphics[height=0.3in]{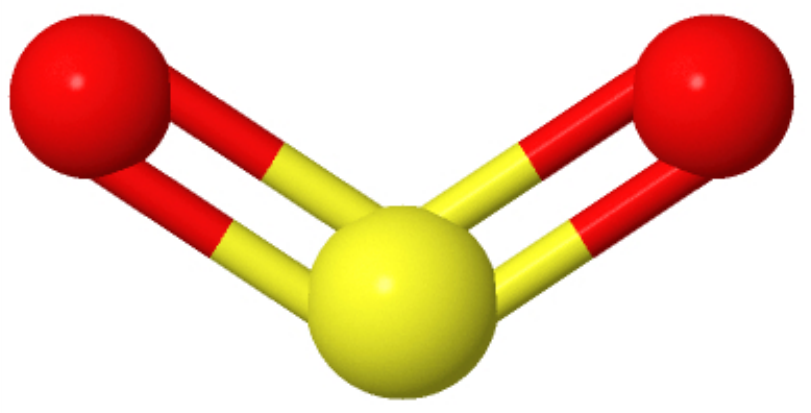}} & SO$_2$   & Sulfur dioxide    & 7.60    & 1320 & SO$_2$ stretch     & $\mathrm{3.4 \times 10^{-17}}$ & [2]\\
\raisebox{-.5\height}{\includegraphics[height=0.5in]{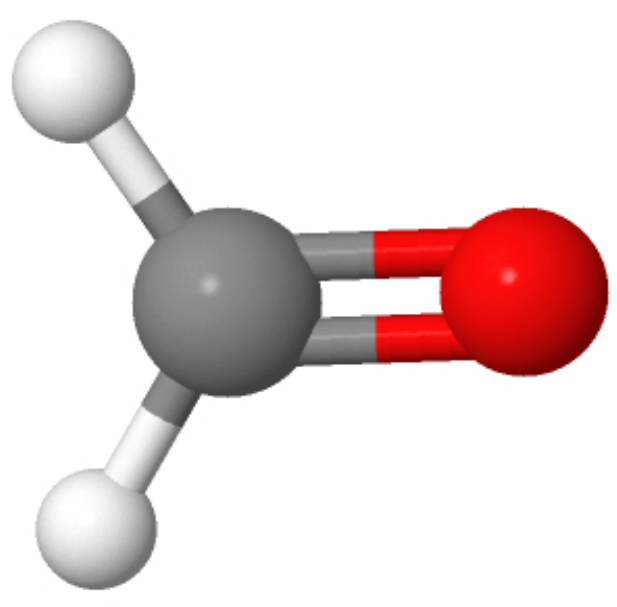}} & H$_2$CO   & Formaldehyde    & 8.04    & 1244 & CH$_2$ rock     & $\mathrm{1.0 \times 10^{-18}}$ & [1]\\
\raisebox{-.5\height}{\includegraphics[height=0.5in]{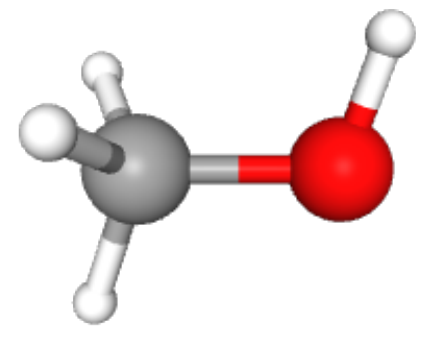}} & CH$_3$OH   & Methanol    & 9.74    & 1026 & C$-$O stretch     & $\mathrm{1.8 \times 10^{-17}}$ & [1]\\
\raisebox{-.5\height}{\includegraphics[height=0.6in]{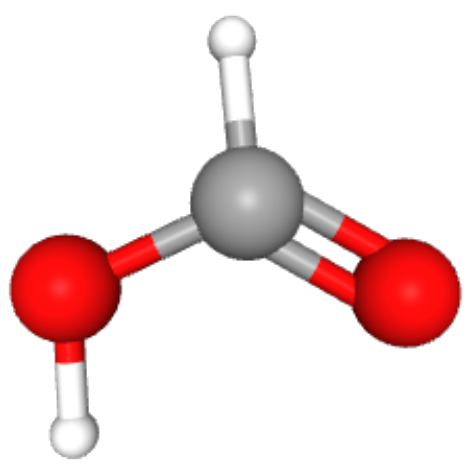}} & HCOOH   & Formic acid     & 8.22    & 1216 & C$-$O stretch     & $\mathrm{2.9 \times 10^{-17}}$ & [1]\\
\raisebox{-.5\height}{\includegraphics[height=0.5in]{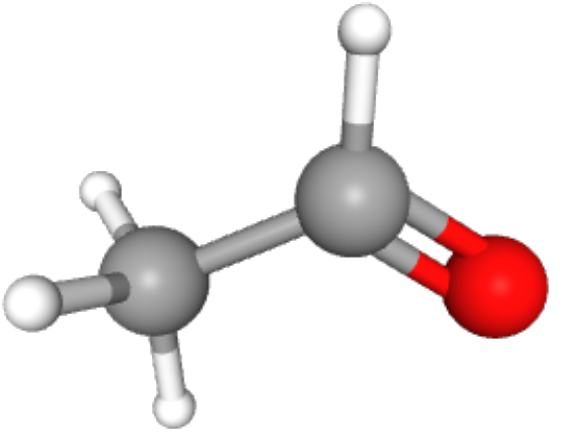}} & CH$_3$CHO   & Acetaldehyde    & 7.41  & 1349 & CH$_3$ s-def./CH wag.     & $\mathrm{4.1 \times 10^{-18,a}}$ & [3]\\
\raisebox{-.5\height}{\includegraphics[height=0.5in]{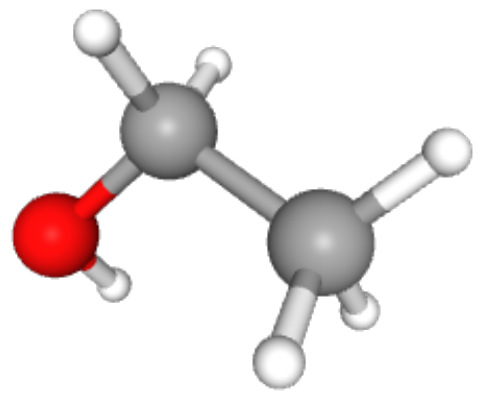}}  & CH$_3$CH$_2$OH      & Ethanol     & 7.23    & 1383 & CH$_3$ s-def.     & $\mathrm{2.4 \times 10^{-18,a}}$ & [4]\\
\raisebox{-.5\height}{\includegraphics[height=0.5in]{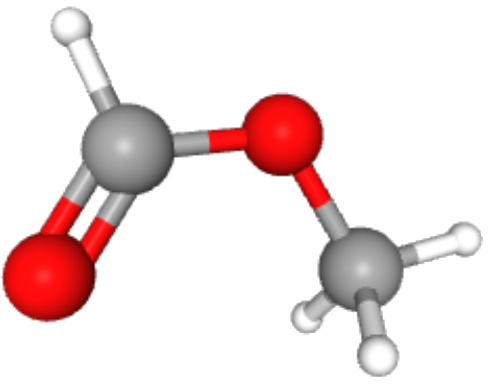}} & CH$_3$OCHO & Methyl formate  & 8.25    & 1211 & C$-$O stretch   & $\mathrm{2.52 \times 10^{-17,a}}$ & [5]\\
 &  &   &     &  &   & $\mathrm{2.28 \times 10^{-17,b}}$ & [5]\\
\raisebox{-.5\height}{\includegraphics[height=0.5in]{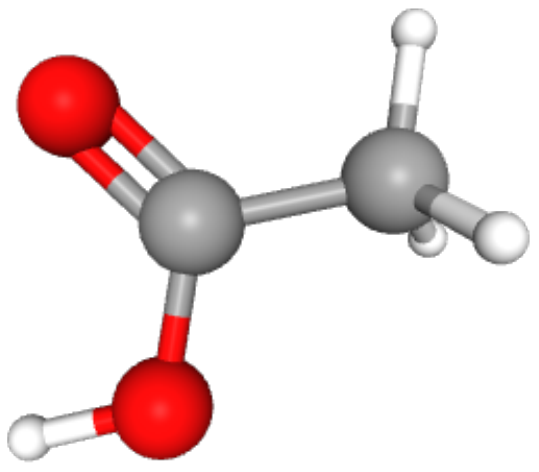}} & CH$_3$COOH & Acetic acid  & 7.82    & 1278 & OH bend      & $\mathrm{4.57 \times 10^{-17}}$ & [6] \\
\raisebox{-.5\height}{\includegraphics[height=0.5in]{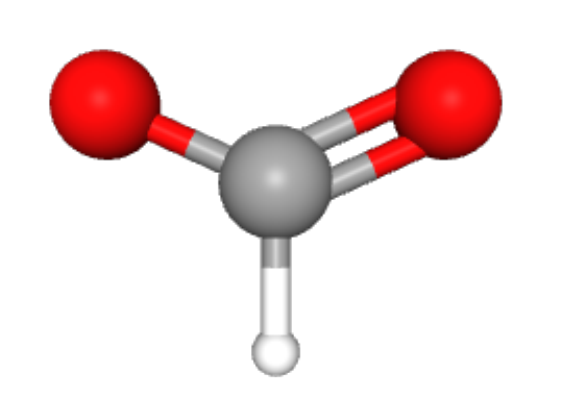}} & HCOO$^-$ (B1) & Formate ion  & 7.23    & 1383 & C$-$O stretch   & $\mathrm{8.0 \times 10^{-18}}$ & [3]\\
\raisebox{-.5\height}{\includegraphics[height=0.5in]{Figures/Formate_pubchem.pdf}} & HCOO$^-$ (B2) & Formate ion   & 7.38    & 1355 & C$-$O stretch  & $\mathrm{1.7 \times 10^{-17}}$ & [3]\\
\raisebox{-.5\height}{\includegraphics[height=0.23in]{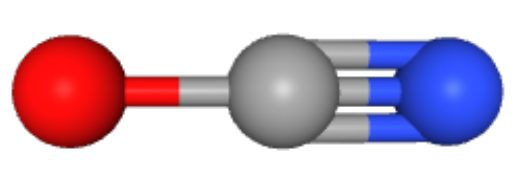}} & OCN$^-$ & Cyanate ion   & 7.62    & 1312 & Comb. (2$\nu_2$)  & $\mathrm{7.45 \times 10^{-18}}$ & [6]\\
\raisebox{-.5\height}{\includegraphics[height=0.40in]{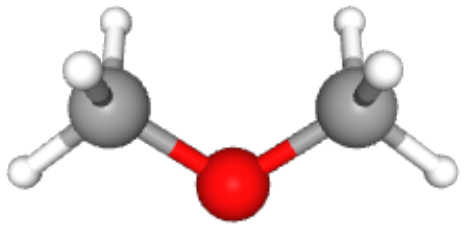}} & CH$_3$OCH$_3$ & Dimethyl ether   & 8.59    & 1163 & COC str. + CH$_3$ rock.  & $\mathrm{4.9 \times 10^{-18,a}}$ & [5]\\

\raisebox{-.5\height}{\includegraphics[height=0.40in]{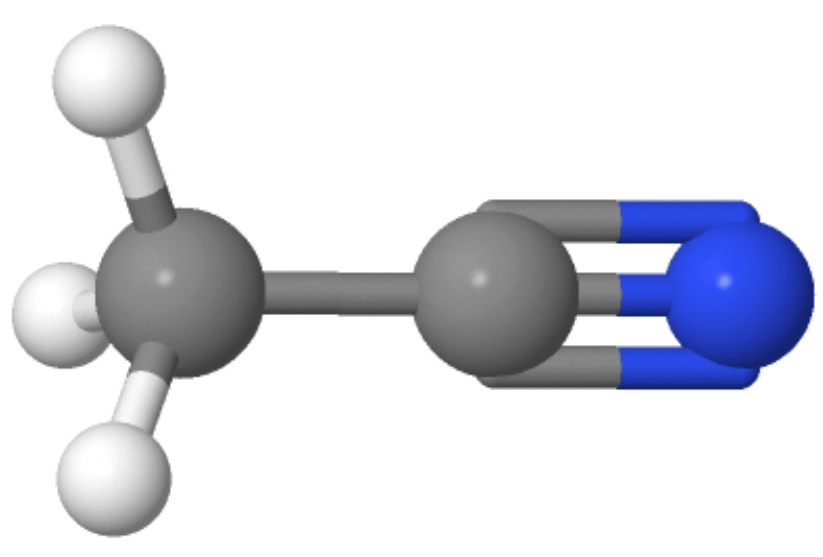}} & CH$_3$CN &Methyl cyanide   & 7.27    & 1374 &  CH$_3$ sym. def.  & $\mathrm{1.2 \times 10^{-18,a}}$ & [5]\\

\raisebox{-.5\height}{\includegraphics[height=0.5in]{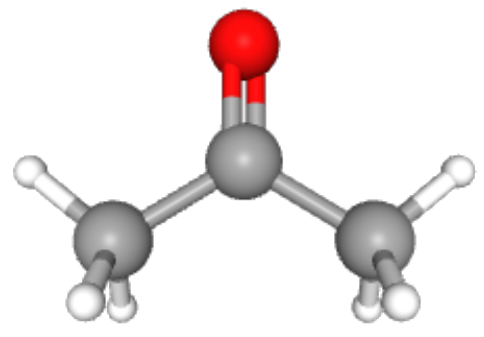}} & CH$_3$COCH$_3$ &Acetone   & 7.33    & 1363 &  CCC asym. str.  & $\mathrm{1.2 \times 10^{-17,a}}$ & [7]\\
\raisebox{-.5\height}{\includegraphics[height=0.45in]{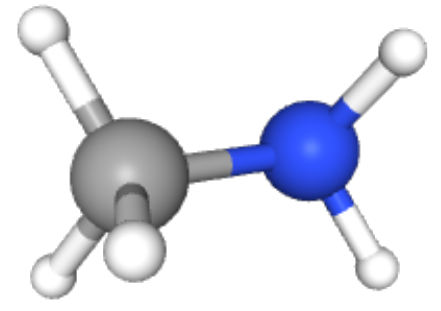}} & CH$_3$NH$_2$ &Methylamine   & 8.5    & 1176 &  CH$_3$ rock  & $\mathrm{1.3 \times 10^{-18,a}}$ & [8]\\
\raisebox{-.5\height}{\includegraphics[height=0.55in]{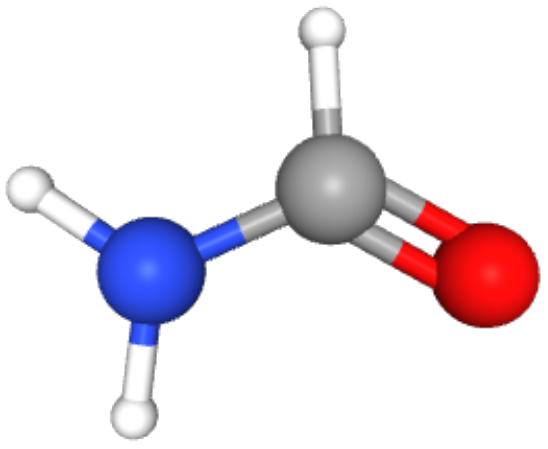}} & NH$_2$CHO &Formamide   & 7.2    & 1388 &  CH bend  & $\mathrm{1.4 \times 10^{-17,a}}$ & [9]\\

\raisebox{-.5\height}{\includegraphics[height=0.55in]{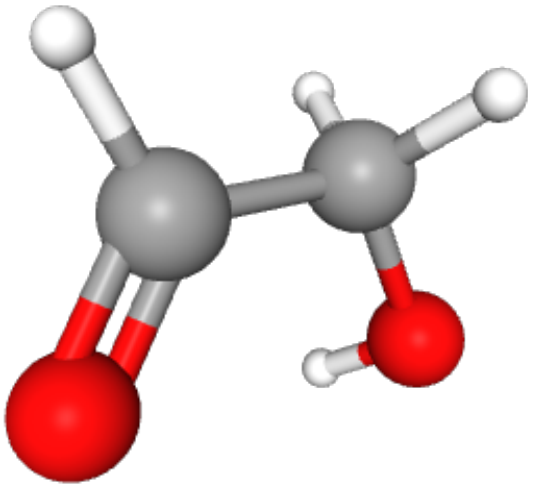}} & HCOCH$_2$OH &Glycolaldehyde   & 7.3    & 1372 &  CH bend  & $\mathrm{7.7 \times 10^{-18}}$ & [10]\\
\hline
\end{tabular}
\tablefoot{\footnotesize
[1] \citet{Bouilloud2015}, [2] \citet{Boogert1997}, [3] \citet{Hudson2020}, [4] \citet{Boudin1998}, [5] \citet{Scheltinga2021}, [6] This work - see Appendix~\ref{Ap_bs}. $^a$Mixture with H$_2$O. $^b$Mixture with CH$_3$OH, [7] \citet{Rachid2020}, [8] \citet{Rachid2021}, [9] \citet{slav2023}, [10] \citet{Hudson2005} 
}
\end{table*}


The laboratory data considered in this paper are listed in Appendix~\ref{list_lab}. These data were taken mainly from the Leiden Ice Database for Astrochemistry\footnote{\url{https://icedb.strw.leidenuniv.nl/}} \citep[LIDA;][]{Rocha2022} and from the Goddard NASA database\footnote{\url{https://science.gsfc.nasa.gov/691/cosmicice/spectra.html}}. The methodology used by \texttt{ENIIGMA} to test the data available is detailed in \citet{Rocha2021}. Here, we provide a brief description of the method. In the first stage \texttt{ENIIGMA} combines IR spectra of pure ice at low temperature with pure ice at high temperature. The best group of solutions is passed to the second step. In the second stage, \texttt{ENIIGMA} combines the previous best solution with pure ice, with all ice mixtures in the \texttt{ENIIGMA} database in a sequential way. At this stage, all data that has an IR feature in the range fitted was tested. Finally, \texttt{ENIIGMA} passes the best-ranked groups of solutions to a final stage, where species from one group of solutions are mixed and combined with species from another group. This allows the code to diversify the number of combinations and increases the possibility for the code to find the global minimum solution not only among the coefficients but also among the laboratory data available. In total, the code tested 3173 different combinations, where the best solution is the best-ranked group of components based on the RMSE value.

We note that for the purpose of this paper, we subtract the absorption profiles of H$_2$O and CH$_3$OH around 6.8~$\mu$m from the ice mixtures using a local subtraction with a polynomial function (see Appendix~\ref{lab_rem}). This enables fitting the observational data after local subtraction. As previously mentioned, the contribution of these two molecules is taken into account in the polynomial fit used to trace the local continuum between 6.8 and 8.6~$\mu$m. Most of the COM laboratory data have a spectral resolving power of $R = 5000$, which is degraded to the nominal spectral resolution of the two sources ($R$ $\sim$ 3500) around 7$-$8~$\mu$m. In the case of HCOO$^-$ and OCN$^-$, we isolated the ice bands of these ions using a local baseline subtraction. These two species are formed from molecules engaged in an acid-base reaction, also known as Bronsted-Lowry acid-base theory \citep{Bronsted1923, Lowry1923}, and also seen in interstellar ices \citep[e.g.,][]{Grim1987, Schutte2003, Broekhuizen2004}. The amount of ions formed depends on the initial abundances of the parent molecules. By isolating these bands, one can mimic at first order, different initial conditions of parent species. It is also worth mentioning that the baselines of HCOOH ice mixture are checked before the analysis. \citet{Oberg2011} comments that one should be careful when deriving the formic acid and formate ion column densities because of baseline issues in some experimental data measured further in the past. In addition, it is important 
to note that HCOO$^-$ and HCOOH share a band at 7.2~$\mu$m. For that reason, we would recommend using the 7.4~$\mu$m to quantify HCOO$^-$.

\subsection{Degeneracy analysis}
\label{degen_sec}

The \texttt{ENIIGMA} fitting tool performs a degeneracy analysis of the coefficients in the linear combination that results in the best fit. Briefly, the code performs a Gaussian variation around each coefficient by using the numpy function \texttt{numpy.random.normal} \citep[][]{Harris2020}, and is given by: 
\begin{equation}
    p(x) = \frac{1}{\sqrt{ 2 \pi \sigma^2 }} \mathrm{exp}\left[ - \frac{ (x - \mu)^2 } {2 \sigma^2} \right]
    \label{prob_dist_eq}
\end{equation}
where $\mu$ values are the optimal coefficients, and $\sigma$ is the standard deviation around $\mu$. This analysis allows us to calculate $\chi^2$ values for each new linear combinati    on, and derive confidence intervals based on a $\Delta\chi^2$ map \citep{Avni1980}, which is formulated as: 
\begin{subequations}
\begin{eqnarray}
\chi^2 &=& \frac{1}{dof}\sum_{i=0}^{n-1} \left(\frac{\tau_{\nu,i}^{\rm{obs}}  - \sum_{j=0}^{m-1} w_j \tau_{\nu,j}^{\rm{lab}}}{\gamma_{\nu,i}^{\rm{obs}}} \right)^2 \\
\Delta \chi^2(\alpha, \epsilon) &=& \chi^2 - \chi_{min}^2
\end{eqnarray}
\label{deltachi}
\end{subequations}
where $dof$ is the number of degrees of freedom, $\gamma$ is the error in the observational optical depth spectrum propagated from the flux error assumed to be 10\%, $\alpha$ and $\epsilon$  are the statistical significance and the number of free parameters, respectively. $\chi_{min}^2$ corresponds to the goodness-of-fit in the global minimum solution.

\texttt{ENIIGMA} also quantifies the statistical significance of a given IR spectrum based on its recurrence, which is defined as:
\begin{equation}
    \mathcal{R} = \frac{f_i}{S}
,\end{equation}
where $f_i$ is the absolute frequency of sample $i$ (i.e. how many times a specific laboratory data participates in the fit) and $S$ is the total number of solutions. If $\mathcal{R} = 100\%$, the chemical species cannot be excluded from the fit. On the other hand, lower percentages mean that the spectrum can be replaced by another one with a similar spectral shape without going outside of the confidence intervals. While the confidence interval analysis evaluates the degeneracy among the components in the best fit, the recurrence analysis quantifies the degeneracy among different spectral data.

\subsection{Criteria for firm COM ice detections}
\label{sec_crit}
\citet{Boogert2015} classify three types of detection in ices:
\begin{enumerate}
    \item Secure: multiple bands or bands of isotopologues are detected in high-quality spectra;

    \item Likely: A single band is detected and the profile matches the laboratory spectra;

    \item Possibly: A single band is detected and there is no exact match between the profile and the laboratory spectra;
\end{enumerate}

COMs bands are naturally weak, and therefore the isotopologues criteria in (1) will be hardly satisfied for solid-phase detections. In addition, to these three criteria, \citet{Jorgensen2020} discuss gauges for the detection of exotic chemical species in the context of gas-phase observations. They mention that a firm identification needs a complete spectral survey with a synthetic spectrum that accounts for all the bands of the identified molecules instead of using independent analytical functions (e.g., Gaussian fits) of individual lines. These criteria can also be applied in the context of COMs in ice, and it is strongly recommended to use IR laboratory data for comparison instead of analytical functions (e.g., Gaussian, Lorentzian). 

We note that different from gas-phase observations, where the emission profiles are narrow and isolated, solid-phase absorption profiles are often blended because of the common functional groups and broader spectral features. In addition, because of the ice matrix in which the molecules are embedded, the shape of the absorption bands change, with mixing ratio and temperature \citep[e.g.,][]{Oberg2007, Bouwman2007}. While ice COMs show peculiar band shapes at high temperatures ($>$ 70~K), they are very similar at lower temperatures (see \citet{Scheltinga2018, Rachid2020, Rachid2021, Scheltinga2021, Rachid2022, slav2023}). In this sense, we add to the criteria presented by \citet{Boogert2015} and \citet{Jorgensen2020} that a degeneracy analysis of molecules sharing similar functional groups is needed in order to claim a firm ice COM detection.

\subsection{Remarks for future works}
\label{remarks}
In this section, we point out some aspects important to guide future works on the analysis of COMs fingerprints in protostellar ices:

\begin{itemize}
    \item The intensities of COM bands in observational data are weaker compared to the major ice components. In this case, special attention must be paid to the baseline correction of COM IR laboratory spectra. Any minor inflexion can mimic a spurious feature and lead to misinterpretation of the observational spectrum. An example is presented in Appendix~\ref{spurious_sec}. The example is given for H$_2$O:CH$_3$CH$_2$OH ice spectrum where a 7th-order polynomial is used to baseline correct the spectrum. We show that if a lower number of data points is considered around 8~$\mu$m, the polynomial function can fluctuate and originate spurious features in the final data. Instead cubic spline functions can be used to mitigate those fluctuations. Another issue, not shown here, is that some weak bands can be removed unintentionally if the user is not familiar with a particular dataset and source details.

    \item Tracing the local continuum on the observed spectrum is as critical as the spectral fitting. In the COMs fingerprint region, where broad bands are present, one should be careful when attributing zero absorption for tracing the local continuum between 6.8-8.6~$\mu$m.

    \item Analysis of observational spectra in the range between 6.8 and 8.6~$\mu$m using laboratory data of H$_2$O- or CH$_3$OH- containing ices must have the features of these two molecules subtracted to allow direct comparison with observational data after local continuum removal. Appendix~\ref{lab_rem} shows an example of how the CH$_3$CH$_2$OH bands in the mixture with H$_2$O and CH$_3$OH were isolated. While a single 4th-order polynomial was used in the case of  H$_2$O:CH$_3$CH$_2$OH, three polynomial functions with different orders were needed in the case of CH$_3$OH:CH$_3$CH$_2$OH, which increases the risk of creating spurious features.
    
\end{itemize}

\section{Results}
\label{result_sec}
In this section, we show the fitting results of the protostars IRAS~23385+6053 and IRAS~2A in the range between 6.8 and 8.4~$\mu$m, as well as the confidence interval analysis. 

\begin{figure*}
   \centering
   \includegraphics[width=18cm]{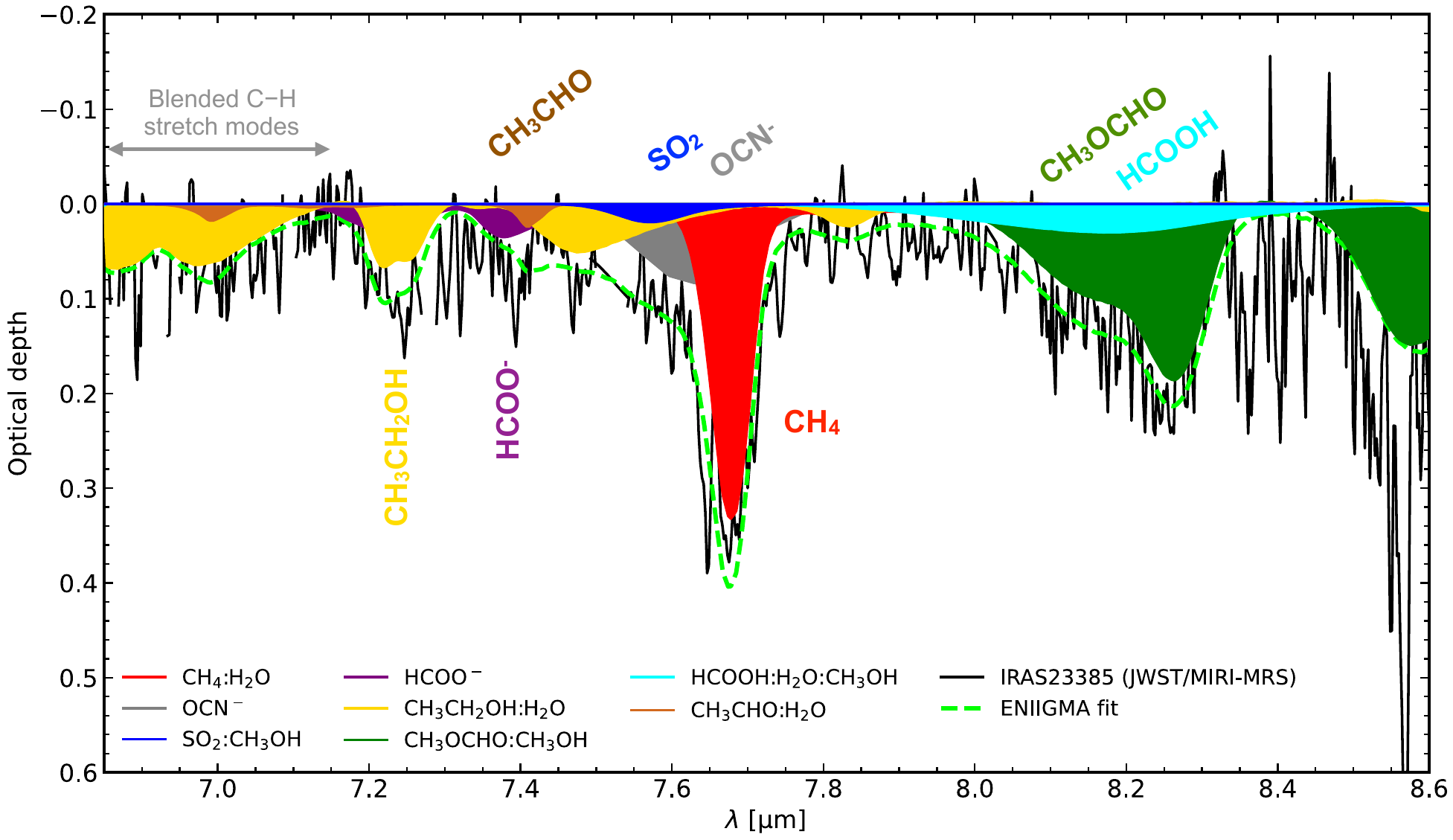}
   \includegraphics[width=18cm]{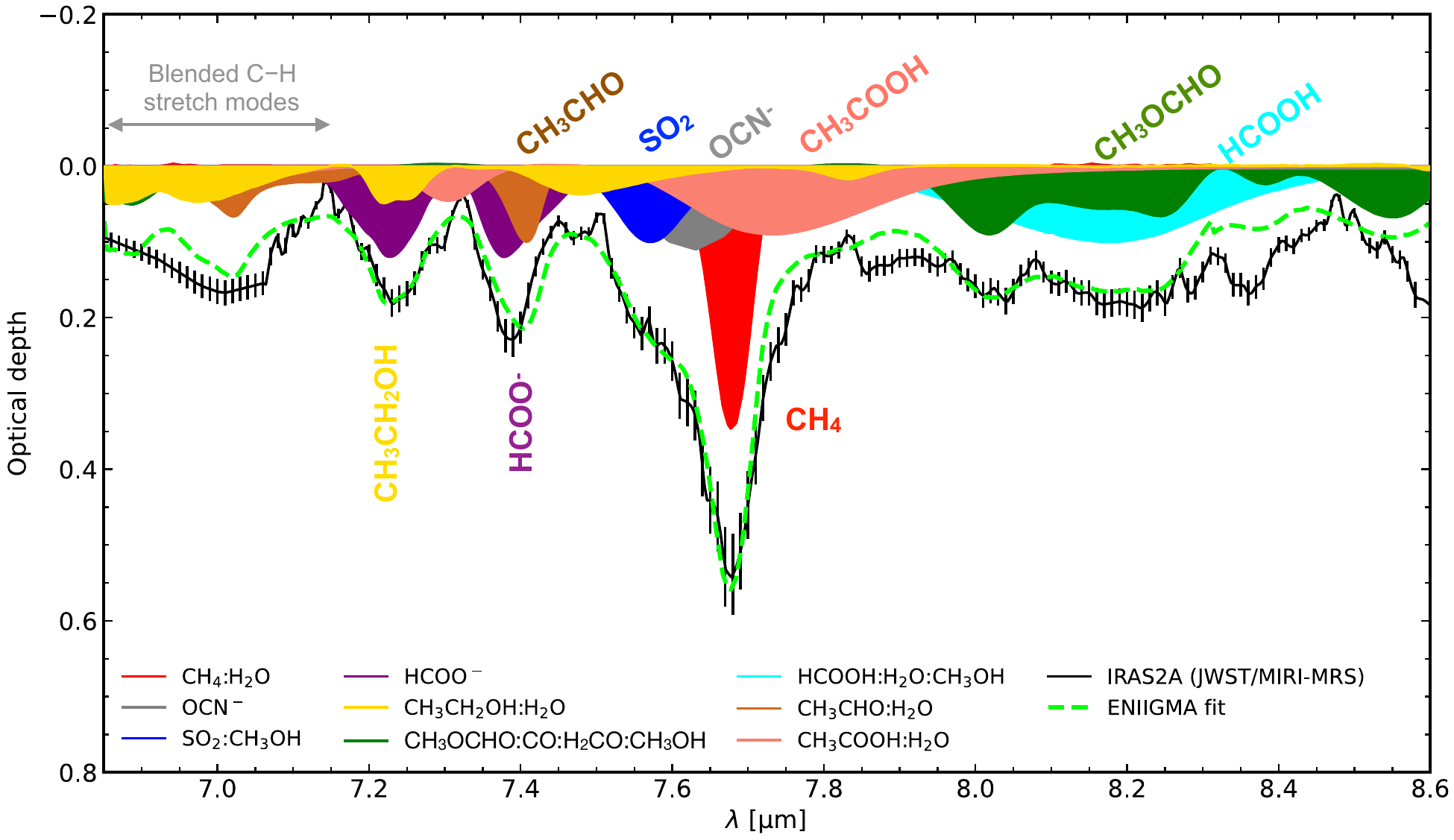}
      \caption{ENIIGMA fits of IRAS~23385 (top) and IRAS~2A (bottom). Gas-phase lines are masked. The figure labels show the ice mixture used in the fits, and a simplified version with the names of the chemical species names is shown close to the bands. The corresponding temperature of these laboratory spectra ranges between 10~K and 15~K.}
         \label{global_fits}
   \end{figure*}

\subsection{Spectral decomposition and feature analysis}
\label{spec_decp}
The fits of the IRAS23385 and IRAS~2A spectra are shown in Figure~\ref{global_fits} top and bottom, respectively (see Appendix~\ref{fits_increment} for an incremental version of these figures, following the individual fitting steps). These two MIRI/MRS spectra are decomposed using nine laboratory spectra which were selected by \texttt{ENIIGMA} and they provide the global minimum solution. Among the COMs fitted in this work are, CH$_3$CHO, CH$_3$CH$_2$OH, CH$_3$OCHO (methyl formate) and CH$_3$COOH (acetic acid), with the first three robustly detected (see \S\ref{sec_crit} and \S\ref{degen}). The simple molecules identified are CH$_4$, SO$_2$, and HCOOH. Additionally, we found a good match of two ions, HCOO$^-$ and OCN$^-$. Because of the contribution of HCOO$^-$ with similar intensities at 7.2 and 7.4~$\mu$m, we stress that these two bands are not only associated with COMs. 

The band around 7.68~$\mu$m was clearly visible in previous ice observations \citep[e.g.,][]{Gibb2004, Oberg2008}, and is now seen in JWST spectra with unprecedented S/N and spectral resolution. CH$_4$ ice is the main carrier of this feature, and \citet{Oberg2008} suggests that SO$_2$ can contribute to the blue wing of this band based on a Gaussian decomposition of the 7.68~$\mu$m feature. In the present work, this band is decomposed into four components, with CH$_4$ mixed with H$_2$O the dominant carrier. This is in line with laboratory experiments suggesting a common formation pathway for CH$_4$ and H$_2$O ices \citep{Qasim2020}. The blue wing has contributions of SO$_2$ mixed with CH$_3$OH, and the negative cyanate ion (OCN$^-$) for both sources. The red wing of this band can be fitted with CH$_3$COOH in the case of IRAS~2A but is not present in IRAS~23385. A good correlation in ice column density between sulfur-bearing molecules and methanol was observed before by \citet{Boogert1997} and \citet{Boogert2022}. OCN$^-$ is one of the ions formed from molecules engaged in acid-base reactions, and it has a band at 7.63~$\mu$m, in addition to its well-known feature at 4.61~$\mu$m, commonly seen towards protostars. In fact, recent observations of IRAS~2A with the JWST/Near-Infrared Spectrometer (NIRSpec) detect the 4.61~$\mu$m band, thus confirming the presence of OCN$^-$ in the ices towards IRAS~2A as part of the JOYS program. The analysis of the NIRSpec data of this source will be presented in a future paper, but we calculate the ice column density of OCN$^-$ at 4.61~$\mu$m to check the consistency of the fit in the mid-IR (see Section~\ref{ice_cd_sec}). As a result, the blue wing of the 7.68~$\mu$m feature is composed both by SO$_2$ and OCN$^-$. We also note that OCN$^-$ in both sources has a similar intensity, whereas SO$_2$ is stronger in IRAS~2A.

Protostar observations in the mid-IR show the 7.2 and 7.4~$\mu$m features with similar strengths across a number of sources, which suggests the dominance of a single species. In fact, we find that the formate ion, initially proposed by \citet{Schutte1999_weak},  matches well the 7.2 and 7.4~$\mu$m bands of IRAS~2A and IRAS~23385. This ion is formed by the acid-base reaction of H$_2$O:NH$_3$:HCOOH (100:2.6:2) ice mixture at 14~K \citep{Galvez2010}. The intensity of the formate ion in IRAS~2A is a factor of three stronger than in IRAS~23385. In Appendix~\ref{formateionapd}, we compare the 7.2 and 7.4~$\mu$m band with the formate ion at other temperatures. The formate ion spectrum at 150~K has a broader profile at 7.4~$\mu$m, whereas laboratory data at 210~K shows a strong feature at 7.3~$\mu$m that is not observed in either source. Thus, the 7.2 and 7.4~$\mu$m features are a signature of ices dominated by ions in cold regions. Our results also show that despite HCOO$^-$ being the main carrier of the 7.2 and 7.4~$\mu$m bands, other components can contribute to these two features separately as discussed below.

The presence of HCOO$^-$ in the ice is supported by the detection of HCOOH at 8.2~$\mu$m, in addition to its band at 5.8~$\mu$m. The best fit is found when formic acid is mixed with CH$_3$OH and H$_2$O as suggested by \citet{Bisschop2007}. The lower intensity of formic acid in IRAS~23385 can be related to the nature of the source. At warmer temperatures ($>$50~K), HCOOH is more efficiently destroyed via an acid-base reaction \citep[e.g.,][]{Schutte1999_weak, Galvez2010}. Another possibility is that HCOOH was not efficiently formed in this high-mass source because of the lower amount of CO ice available in a high-mass star-forming region.

CH$_3$CH$_2$OH is detected in both protostars through the absorption of four vibrational modes at 6.8$-$7.15~$\mu$m (C$-$H stretch), 7.25~$\mu$m (CH$_3$ s-deformation), 7.4-7.7~$\mu$m (OH deformation) and at 7.85~$\mu$m (CH$_2$ torsion). Among the ethanol data tested, the mixture with H$_2$O provides the best fit. Ethanol is stronger and contributes significantly to the 7.2~$\mu$m band in IRAS~23385, whereas it is less prominent in IRAS~2A.

CH$_3$CHO shows weak absorption around 7.03~$\mu$m (C$-$H stretch), and at 7.41~$\mu$m (CH$_3$ s-deformation/CH wagging). These bands are stronger in IRAS~2A than in IRAS~23385. Experimental characterization of the CH$_3$ s-deformation/CH wagging shows a strong dependence on the chemical environment and temperature \citep{Scheltinga2018}. The closer match with both protostars is for acetaldehyde mixed with H$_2$O ice, which has a peak at 7.41~$\mu$m. The observed band, however, has a peak at 7.38~$\mu$m, indicating other carriers for this band (e.g., HCOO$^-$).

CH$_3$OCHO is observed in both sources around 8.2 and 8.6~$\mu$m, but with different spectral shapes and intensities. In IRAS~23385, the methyl formate band is better fitted by a mixture with methanol at 15~K, although the data at 30 and 80~K are also statistically likely (see Section~\ref{temp_degen}). Other methyl formate mixtures do not exhibit such an asymmetric profile, which makes them unlikely to be the carrier of this band (see Section~\ref{polar}). While the shape of the 8.2~$\mu$m is well defined, this is not the case for the 8.6~$\mu$m band, which is affected by saturation due to the strong silicate feature. In IRAS~2A the bands at 8.2 and 8.6~$\mu$m are consistent with a mixture containing CO, H$_2$CO, and CH$_3$OH. This ice mixture of CH$_3$OCHO also fits well the 8.0~$\mu$m band associated with H$_2$CO.

CH$_3$COOH is found to fit well to the IRAS~2A spectrum, but not that of IRAS~23385. In addition, a better match with the observations is found for acetic acid mixed with H$_2$O ice. The two identified bands are located at 7.3 and 7.7~$\mu$m (salmon colour in Figure~\ref{global_fits}, bottom). Specifically, the 7.7~$\mu$m band in IRAS~2A has a broad profile that is more than just CH$_4$ and it is necessary to fit the absorption excess around this region without strong overlap between 7.7 and 7.85~$\mu$m. The 7.3~$\mu$m is slightly over-estimated in the fits, because of the amount of absorption needed to fit the 7.7~$\mu$m band. This can be because of the SO$_2$ emission lines subtraction around 7.3~$\mu$m or the uncertainties involved in the local continuum choice around the 7.7~$\mu$m band.

The robustness of these detections based on the difference in local continuum choice is discussed in Section~\ref{choice_degen} for IRAS~2A, the higher S/N source.

\subsection{Statistical analysis}
\label{degen}

\subsubsection{Confidence intervals}
We derive the confidence intervals (see Section~\ref{degen_sec}) of the fits for IRAS~23385 and IRAS~2A in three different ranges separately: (i) 6.85$-$7.5~$\mu$m, (ii) 7.5$-$7.8~$\mu$m and (iii) 7.8$-$8.6~$\mu$m. The components in these three ranges are relatively isolated, and therefore their contribution in one given interval is kept constant when analysing other ranges. 

\begin{figure}
    \centering
    \includegraphics[width=1\linewidth]{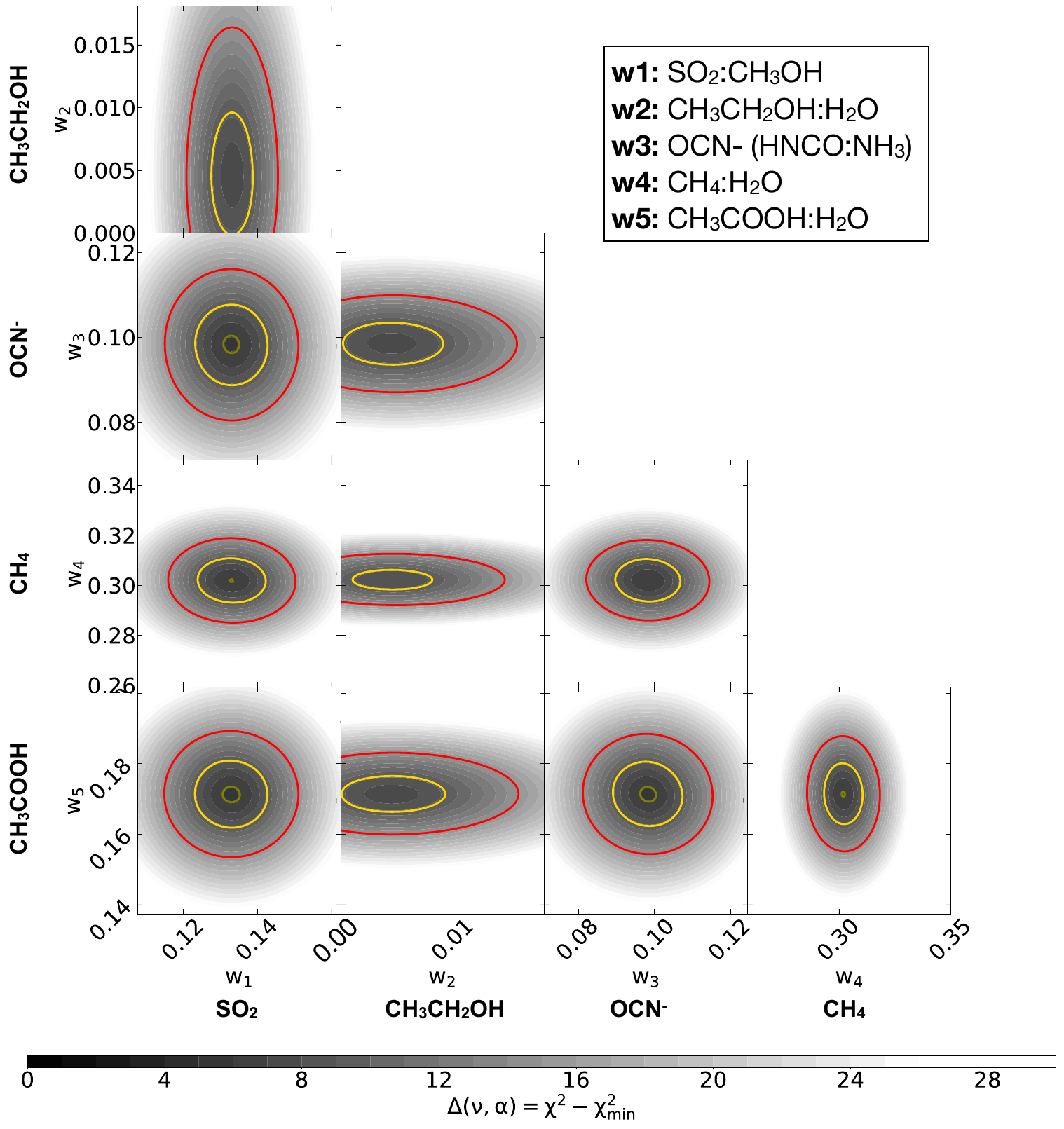}
    \caption{Corner plots showing the IRAS~2A coefficient confidence intervals for the range between 7.5 and 7.8~$\mu$m. The grey scale colour is the $\Delta\chi^2$ map derived from a total of 5000 values. Yellow and red contours represent 2 and 3$\sigma$ significance, respectively.}
    \label{corner_76}
\end{figure}

In Figure~\ref{corner_76}, we show the confidence intervals for the fit of IRAS~2A in the range between 7.5 and 7.8~$\mu$m. The yellow and red contours indicate 2$\sigma$ and 3$\sigma$ confidence intervals. Based on these contours, one can note that all components, but CH$_3$CH$_2$OH, are required to fit IRAS~2A. In this particular spectral range (7.5 and 7.8~$\mu$m), only a small portion of CH$_3$CH$_2$OH spectrum at 7.5~$\mu$m contributes to the absorption. Therefore, the confidence interval analysis shows that CH$_3$CH$_2$OH is not crucial to fit the 7.5 and 7.8~$\mu$m. The contribution of CH$_3$CH$_2$OH in IRAS~2A is better evaluated using the range between 6.86 and 7.5~$\mu$m, which is shown in Appendix~\ref{confidence_ap}. In Figure~\ref{Conf_coeff_ir2a} (top), the CH$_3$CH$_2$OH:H$_2$O is not zero, which reinforces the idea that CH$_3$CH$_2$OH is robustly found in IRAS~2A. Another important result from Figure~\ref{corner_76} is that SO$_2$ (w1) and OCN$^-$ (w3) are both required to fit IRAS~2A since their coefficients cannot be zero. Finally, CH$_3$COOH:H$_2$O cannot be excluded as a solution based on this statistical analysis. Further analysis of the 6.86$-$7.5~$\mu$m range allows us to conclude that CH$_3$CHO, HCOO$^-$, and HCOOH are also robust detections. In addition to these chemical species, Figure~\ref{Conf_coeff_ir2a} (bottom) also shows that CH$_3$COOH cannot be excluded from the fit obtained with one specific continuum subtraction. More details for CH$_3$COOH are presented in Section~\ref{choice_degen}. The analysis of the spectral range between 7.8 and 8.6~$\mu$m, shows that another COM, CH$_3$OCHO is also a robust detection in IRAS~2A spectrum.

A similar analysis is performed for IRAS~23385 (Figures~\ref{Conf_coeff_ir23385_p1} and \ref{Conf_coeff_ir23385_p2}). For the range between 7.5 and 7.8~$\mu$m, CH$_4$ and OCN$^-$ are essential, whereas SO$_2$ can be statistically not required due to the low S/N in IRAS~23385 spectrum. Likewise, the analysis of the 6.86$-$7.5~$\mu$m range shows that CH$_3$CH$_2$OH, CH$_3$CHO and HCOOH are robust detections, whereas the formate ion (HCOO$^-$) absorption may be explained by CH$_3$CH$_2$OH or HCOOH ice features. For the 7.8$-$8.6~$\mu$m interval, both HCOOH and methyl formate (CH$_3$OCHO) are robust detections.

\subsubsection{Recurrence of the ice components}
\begin{figure}
   \centering
   \includegraphics[width=\hsize]{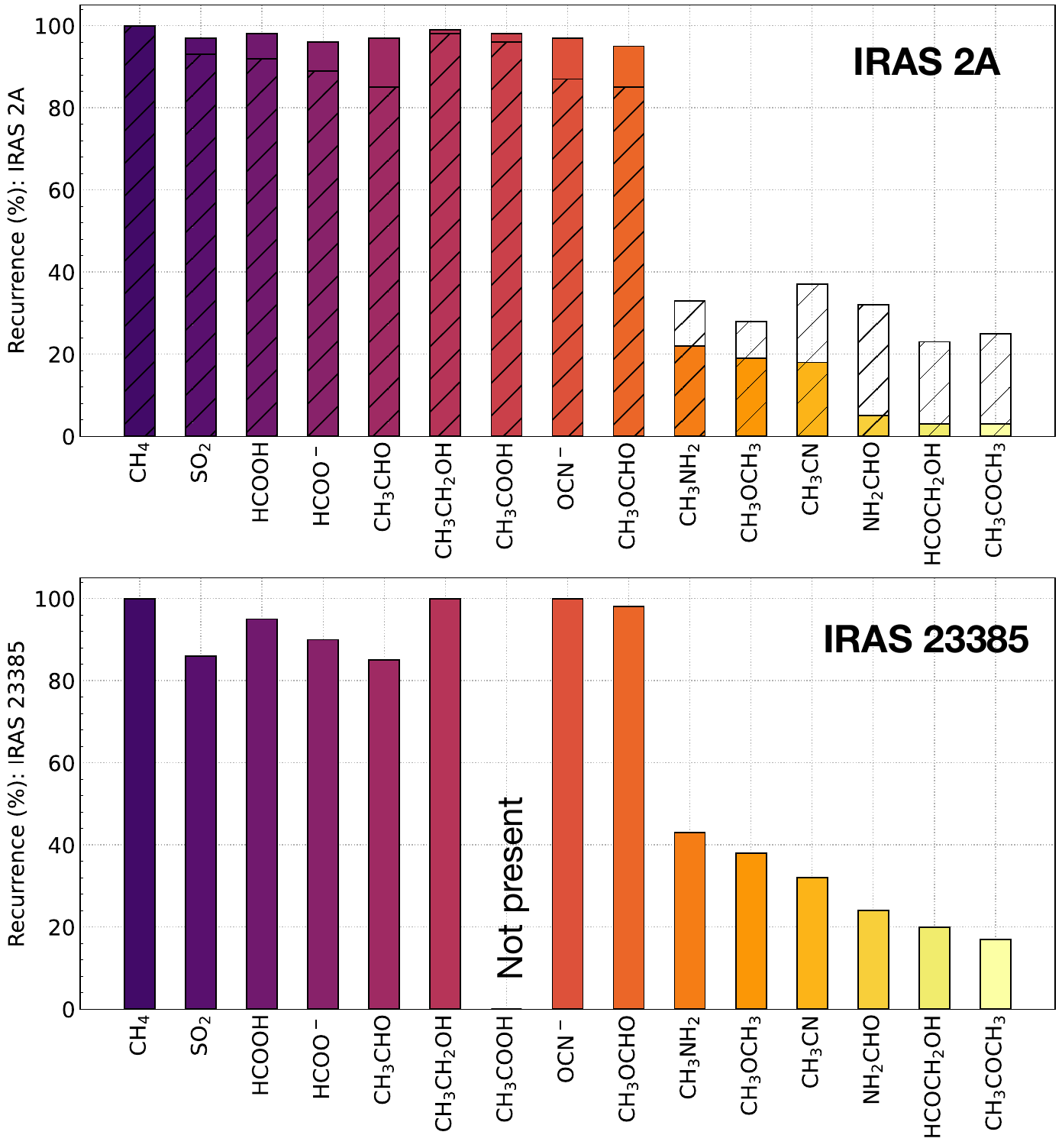}
      \caption{Bar plot showing the recurrence of the difference chemical species to the fit in IRAS~2A (top) and IRAS~23385 (bottom). Recurrences above 50\% are considered essential to the fit and robust detection. CH$_3$COOH in IRAS~23385 does not occur in any of the solutions. In the case of IRAS~2A (top panel) we present hatched bars that show the recurrence of the solutions if the errors in optical depth are increased by a factor of three.}
         \label{piechart}
   \end{figure}

A complementary statistical analysis is performed on the recurrence of all solutions within a given confidence interval. The difference in this method is that we do not vary the coefficient values of each solution, but the laboratory data instead. Figure~\ref{piechart} (top) shows the recurrence of the IRAS~2A fit of 15 chemical species inside 3$\sigma$ confidence interval and considering several solutions. This analysis indicates that the COMs providing the best fits (see Figure~\ref{global_fits}), are the most recurrent in the bar chart (87.5\% $\leq$ R $\leq$ 100\%), and therefore cannot be excluded as a solution. Other chemical species, such as CH$_3$NH$_2$, CH$_3$OCH$_3$, NH$_2$CHO, HCOCH$_2$OH and CH$_3$COCH$_3$, some of which have been suggested previously to contribute in this range, have a recurrence lower than 50\%. This is not sufficient to claim a firm detection and at best upper limits can be derived. 

The reason that formamide and acetone are not part of the global fit is that the formate ion band shape dominates the absorption profile at 7.2 and 7.4~$\mu$m. In addition, acetic acid and ethanol also contribute to these two absorption profiles. In the cases of methylamine and dimethyl ether, the fits indicate that methyl formate contributes more to the 8.5$-$8.6~$\mu$m range.

The same analysis is done for IRAS~23385, shown in Figure~\ref{piechart} (bottom). Since the MIRI data for this source have a lower S/N compared to IRAS~2A, there exist slightly more variations among the recurrence values. The components found in the global fit have a recurrence above 85\%, whereas other COMs are recurrent by $\sim$40\% or less. Interestingly, CH$_3$COOH does not participate in any of the solutions tested because OCN$^-$ and CH$_3$CH$_2$OH account for all the absorption around 7.7~$\mu$m.

\subsubsection{Temperature degeneracy}
\label{temp_degen}
Despite the well-known changes in the band profiles of astrophysical ices with temperature, there are some absorption features that barely vary with temperature. Consequently, the fitting routine is not able to distinguish between these data. \citet{Rocha2021} show that when data that is known to provide a good fit is arbitrarily removed from the database, the \texttt{ENIIGMA} fitting tool uses another data of the same species, but with similar temperature when possible. This slightly increases the fitness function value, but the fit is still good within the confidence intervals. In this section, we perform a statistical analysis to evaluate which temperature ranges are degenerate and provide a good fit, from those temperature ranges that can be excluded as a solution.

\begin{figure}
   \centering
   \includegraphics[width=\hsize]{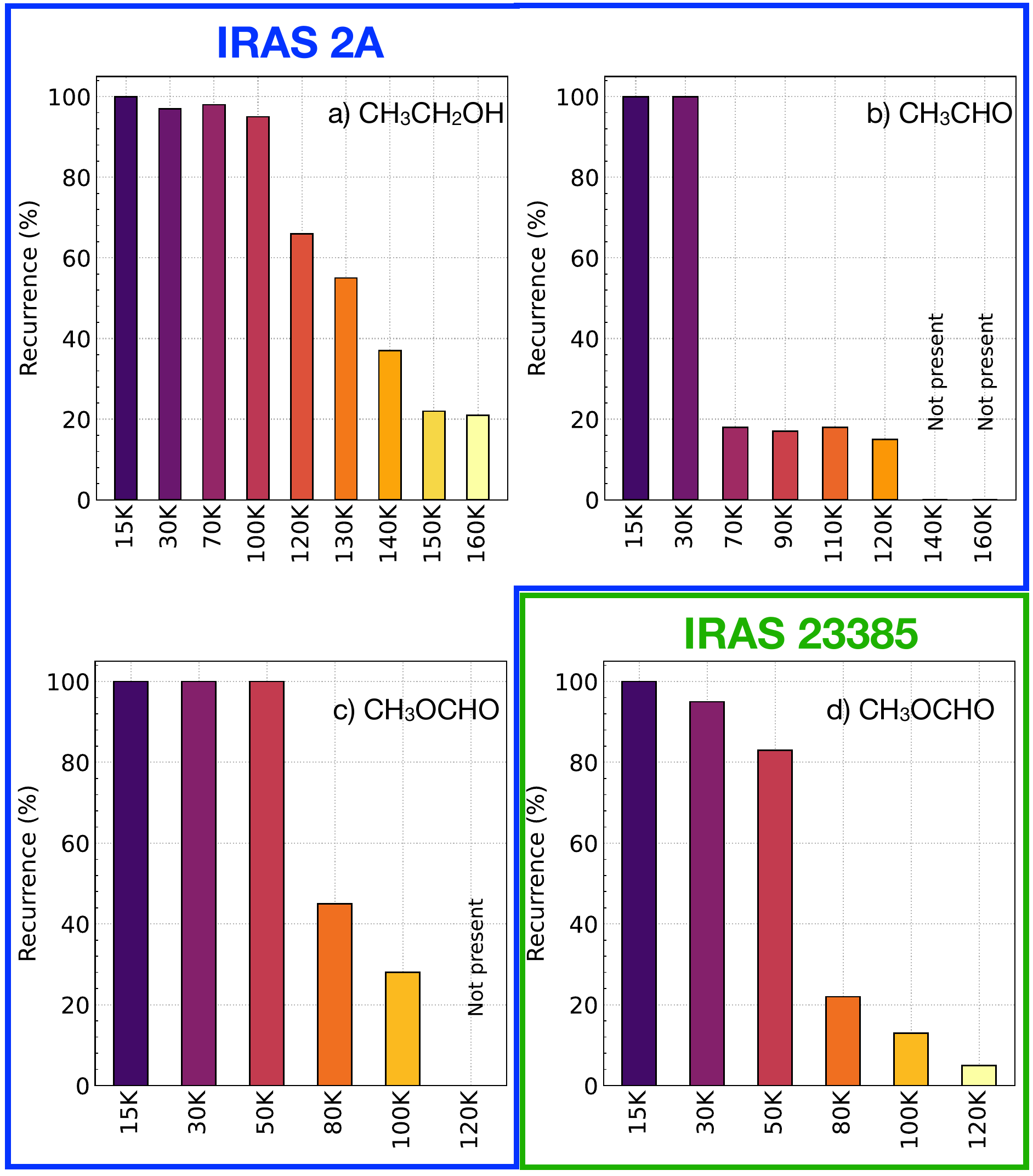}
      \caption{Bar plot showing the recurrence of the same ice mixture, but at different temperatures. Plots inside the blue and green polygons refer to IRAS~2A and IRAS~23385, respectively.}
         \label{bar_comrec}
   \end{figure}

Figure~\ref{bar_comrec} shows the recurrence plots for CH$_3$CH$_2$OH, CH$_3$CHO and CH$_3$OCHO in IRAS~2A and CH$_3$OCHO in IRAS~23385 for different temperatures settings available. The analysis of CH$_3$CH$_2$OH and CH$_3$CHO is not performed for IRAS~23385 because of the low S/N in the spectral range considered for the fits. To calculate the recurrence plot for these COMs, we selected the solutions that are ranked inside a 3$\sigma$ confidence interval. For example, in the case of CH$_3$CH$_2$OH, we found 352 solutions where ethanol is present. The ethanol mixture at 15~K is present in all of these solutions, and therefore it has a recurrence of 100\%. There are solutions that combine the low temperature (15\%) ethanol mixture with other temperatures (e.g., 30, 70, 100~K). Because of the similarity of the ethanol bands in the fitted range, the recurrence of these additional ethanol data is slightly reduced, but still high. For the data at 150 and 160~K, there are around 70 solutions using one of these data, which gives a recurrence of $\sim$20\%. Another way to perform the same analysis is by using the best global solution as the initial guess and running new fits by replacing the IR spectra of specific COMs at different temperatures. This forces \texttt{ENIIGMA} to use only one CH$_3$CH$_2$OH data at a time, and prevents overlaps of data with similar spectral shapes. For ethanol, we obtain a total of 9 solutions if the presence of the other components is fixed. In this case, the temperature range between 15 and 100~K has a recurrence of 100\%, and the higher temperatures, have a recurrence of 0\%. In the case of ethanol, IR spectra with temperatures below 100~K are degenerate, and all CH$_3$CH$_2$OH:H$_2$O mixtures fit IRAS~2A in that temperature range. Above 120~K, the ethanol ice mixture no longer fits the IRAS~2A spectrum well because of band broadening and shift. 

Repeating the same procedure for the other species, we notice a similar behaviour. The fits with acetaldehyde in IRAS~2A are degenerate below 30~K. Above this temperature, CH$_3$CHO:H$_2$O ice mixtures do not offer likely solutions mostly because of band shift. The fits with methyl formate are degenerated below 50~K in IRAS~2A. At higher temperatures, some substructures arise in the CH$_3$OCHO spectrum that deviates from the observational data. Similarly, in IRAS~23385, methyl formate is degenerated below 50~K, whereas solutions at 80~K and above are less recurrent because of changes in the band profile. In conclusion, it is likely that most of the ices towards IRAS~2A and IRAS~23385 are located in regions with temperatures below 50~K.  

\subsubsection{Robustness of detection based on different local continuum choices}
\label{choice_degen}

Other local continuum fits for IRAS~2A (higher S/N) in the 6.8-8.6~$\mu$m region are also investigated and presented in Appendix~\ref{cont_effect}. The first three panels of Figure~\ref{diff_cont} display different continuum profiles, where the top panel is the version adopted for the analysis in this paper that traced a third-order polynomial to the guiding points. The second panel displays the fourth-order polynomial where the red dot is added to the guiding points. In this case, the continuum is slightly elevated at shorter wavelengths to accommodate the fit to the extra point at 8.5~$\mu$m. The third panel presents the continuum when two extra points are added ($\lambda = 7.8$~$\mu$m and 8.5~$\mu$m), and a sixth-order polynomial is used. All subtracted spectra using these three approaches are shown in the bottom panel of Figure~\ref{diff_cont}. The major difference is seen in the last case (orange continuum), which completely removes any absorption excess at 7.8~$\mu$m, thus excluding any band at this wavelength.

Given the variability of the optical depth spectra of IRAS~2A between 6.8$-$8.6~$\mu$m with the choice of the local continuum, it is worth accessing the robustness of the detections reported in the previous sections considering other continuum profiles. This is shown in Figure~\ref{diff_fits} . Figure~\ref{diff_fits} top presents the new fit assuming the red continuum profile (inset panel) from Fig~\ref{diff_cont}. This spectral fit remains good and all components found in the best fit are still present. The only issue is found at 8.5~$\mu$m (added guiding point) where the CH$_3$OCHO band is slightly over-predicted. This situation changes when the orange continuum is used to isolate the absorption features in IRAS~2A. The bottom panel of Figure~\ref{diff_fits} depicts a fit where CH$_3$COOH is no longer needed because of the guiding point at 7.8~$\mu$m to trace the local continuum. In this version of the fit, the HCOO$^-$ and OCN$^-$ components become much stronger than in the other cases, and the COMs bands are reduced by a factor between 1.5$-$2.0. A clear mismatch is seen around 7~$\mu$m because of the reduction of the COM bands. Also, there is more excess around 7.3~$\mu$m that is not fitted with the other COMs tentatively detected in this work (see Section~\ref{sec_tentative}). 

This analysis reinforces that the detections of CH$_3$CHO, CH$_3$CH$_2$OH and CH$_3$OCHO, as well as the ions and the simple molecules, are robust and do not depend on the local continuum choice. Clearly, the only exception is for CH$_3$COOH. However, a valid question is what other chemical species would create an absorption profile similar to the orange continuum. There is no trivial answer to this question. A tentative explanation comes from the five components decomposition from \citet{Boogert2008}, in which only the C5 broad feature contributes to the 7.8~$\mu$m. The nature of the C5 component is not well understood and can be related to a flat profile of high-temperature H$_2$O ice bending mode, the overlap of other negative ions (HCO$_3^-$, NO$_3^-$, NO$_2^-$) or organic refractory residue produced by energetic processing. The flatter bending mode would have little absorption at 7.8~$\mu$m and therefore is less likely, but the other two options are strong candidates. 

\subsection{Ice column densities and abundances}
\label{ice_cd_sec}
The column densities of the molecules fitting the 6.8$-$8.6~$\mu$m range are presented in Table~\ref{ice_cd}. To derive their abundances with respect to H$_2$O ice, we use the libration band around 12~$\mu$m to calculate the water ice column density as shown in Appendix~\ref{water_cd}. These abundances are compared to literature values for low- and high-mass protostars. In Section~\ref{comet_comp} we also show abundances with respect to CH$_3$OH ice, in which the method to derive the column density is demonstrated in Appendix~\ref{water_cd}.

\begin{table*}[h]
\caption{\label{ice_cd} Ice column densities and abundances with respect to H$_2$O ice towards IRAS~2A and IRAS~23385. These values are compared to literature values for other objects.}
\renewcommand{\arraystretch}{1.5}
\centering 
\begin{tabular}{lccccccc}
\hline\hline
Specie &  \multicolumn{2}{c}{$N_{\rm{ice}}$ ($10^{17}$ cm$^{-2}$)} & \multicolumn{2}{c}{$X_{\rm{H_2O}}$ (\%)} & \multicolumn{3}{c}{Literature (\% H$_2$O)}\\
\cline{2-8}
 & IRAS~2A & IRAS~23385 & IRAS~2A & IRAS~23385 & LYSOs & MYSOs & Comet 67P/C-G$^m$\\
\hline
H$_2$O$^{\star}$ & 300$\pm$12 & 158$\pm$36 & 100 & 100 & 100 & 100 & 100\\
CH$_4$$^{\star}$ & 4.9$\pm_{3.2}^{7.5}$ & 5.2$\pm_{4.3}^{6.8}$ & 1.6 & 3.3 & $<$3$^a$ & 1$-$11$^b$ & 0.340$\pm$0.07\\
SO$_2$$^{\star}$ & 0.6$\pm_{0.4}^{1.9}$ & 0.2$\pm_{0.0}^{0.7}$ & 0.2 & 0.1 & 0.08$-$0.76$^a$ & $<0.9-$1.4$^b$ & 0.127$\pm$0.100\\
HCOOH$^{\star}$ & 3.0$\pm_{1.7}^{5.3}$ & 1.8$\pm_{1.3}^{2.7}$ & 1.0 & 1.1 &$<0.5-$4$^c$ & $<0.5-$6$^d$ & 0.013$\pm$0.008\\
CH$_3$OH$^{\star}$ & 15, 23$^{\dagger}$ & ... & 5.0, 7.6 & ... &$<1-$25$^d$ & $<3-$31$^d$ & 0.21$\pm$0.06\\
CH$_3$CHO$^{\star}$ & 2.2$\pm_{1.4}^{2.8}$ & 0.7$\pm_{0.4}^{1.1}$ & 0.7 & 0.4 & ... & $<$2.3$^e$ & 0.047$\pm$0.017\\
CH$_3$CH$_2$OH$^{\star}$ & 3.7$\pm_{0.5}^{4.5}$ & 2.9$\pm_{1.9}^{4.1}$ & 1.2 & 1.8 & ... & $<$1.9$^e$  & 0.039$\pm$0.023\\
CH$_3$OCHO$^{\star}$ & 0.2$\pm_{0.1}^{0.4}$ & 1.1$\pm_{1.0}^{1.3}$ & 0.1 & 0.7 & $<$2.3$^f$ & ... & 0.0034$\pm$0.002\\
CH$_3$COOH$^{\ddagger}$ & 0.9$\pm_{0.6}^{1.3}$ & 0.0 & 0.3 & 0 & ... & ... & 0.0034$\pm$0.002\\
HCOO$^-$$^{\star}$ (7.4~$\mu$m) & 1.4$\pm_{0.4}^{2.4}$ & 0.3$\pm_{0.1}^{0.5}$ & 0.4 & 0.2 & $\sim$0.4$^g$ & $<$0.3$-$2.3$^b$ & ...\\
OCN$^-$$^{\star}$ & 3.7$\pm_{3.3}^{6.6}$ & 0.9$\pm_{0.6}^{1.7}$ & 1.2 & 0.6 & $<0.1-$1.1$^h$ & 0.04$-$4.7$^i$ & ...\\
H$_2$CO$^{\ddagger}$ & 12.4$\pm_{6.6}^{19.7}$ & ... & 4.1 & ... & $\sim$6$^g$ & $\sim$2$-$7$^b$ & 0.32$\pm$0.1\\
\hline
 & \multicolumn{7}{c}{Upper limits}\\
\hline
CH$_3$NH$_2$ & $<$ 4.1 & ... & $<$ 1.4 & ... & $<$ 16$^j$ & $<$ 3.4$^j$ & ...\\
CH$_3$OCH$_3$ & $<$ 2.5 & ... & $<$ 0.8 & ... & ... & ... & 0.039$\pm$0.023\\
CH$_3$COCH$_3$ & $<$ 1.1 & ... & $<$ 0.4 & ... & ... & ... & 0.0047$\pm$0.0024\\
HCOCH$_2$OH & $<$ 0.9 & ... & $<$ 0.3 & ... & ... & ... & 0.0034$\pm$0.002\\
CH$_3$CN & $<$ 5.0 & ... & $<$ 1.6 & ... & $<$ 4.1$^k$ & $<$ 3.4$^k$ & 0.0059$\pm$0.0034\\
NH$_2$CHO & $<$ 1.3 & ... & $<$ 0.4 & ... & $<$ 3.7$^l$ & $<$ 2.1$^l$ & 0.0040$\pm$0.0023\\
Glycine & $<$ 0.1$-$0.6 & ... & $<$ 0.03$-$0.2 & ... & ... & $\sim$0.3$^b$ & ...\\
\hline
\end{tabular}
\tablefoot{$^a$\citet{Oberg2008}, $^b$\citet{Gibb2004}, $^c$\citet{Oberg2011}, $^d$\citet{Schutte1999_weak}, $^e$\citet{Scheltinga2018}, $^f$\citet{Scheltinga2021}, $^g$\citet{Boogert2008}, $^h$\citet{vanBroekhuizen2005}, $^i$\citet{Boogert2022}, $^j$\citet{Rachid2021}, $^k$\citet{Rachid2022}, $^l$\citet{slav2023}, $^m$\citet{Rubin2019}. $^{\dagger}$ The CH$_3$OH ice column density is considered a factor of 2 and 3 higher than the Gaussian fit shown in Figure~\ref{methanol_cd} because of the band saturation. $^{\star}$Chemical species with secure detection. $^{\ddagger}$Tentative detections. CH$_3$COOH depends on the local continuum choice. H$_2$CO is based on a single band of this molecule mixed in the ice with other CO, CH$_3$OH and CH$_3$OCHO.}
\end{table*}

The ice abundances of CH$_4$, SO$_2$, OCN$^-$, HCOO$^-$ and HCOOH are within or close to the range expected for LYSOs and MYSOs. From the analysis in this paper, CH$_4$, SO$_2$, OCN$^-$ compose the band around 7.67~$\mu$m, whereas previous works assigned this band to only CH$_4$, with a possible contribution of SO$_2$ \citep{Oberg2008}. Additionally, instead of using Gaussian profiles to derive the column densities, this paper uses laboratory data of CH$_4$:H$_2$O and SO$_2$:CH$_3$OH. Gaussian profiles provide first-order approach analysis of the profile of the ice bands, but they can under or overestimate the FWHM of real CH$_4$ and SO$_2$ bands. 

In the case of OCN$^-$, we derived ice column densities and abundances from the band at 7.62~$\mu$m for IRAS~2A. For consistence, we also calculated the ice column from the NIRSpec feature at 4.61~$\mu$m, which is 2.5$\times$10$^{17}$ cm$^{-2}$. Both values are consistent within the errors, and they result in OCN$^-$ abundances in agreement with the range estimated in the literature for low-mass protostars \citep{vanBroekhuizen2005}. For IRAS~23385, the OCN$^-$ abundance calculated from the MIRI data is within the range found in MYSOs taken from \citet{Boogert2022}.

Finally, the formic acid (HCOOH) and the formate ion (HCOO$^-$) abundances are closely aligned with the literature values for both LYSOs and MYSOs. In the case of the formate ion in LYSOs, there is only one estimate in the literature for the low-mass protostar HH~46~IRS, which is similar to the abundance calculated for IRAS~2A. The ratios between HCOO$^-$ and HCOOH are 0.46 and 0.16 for IRAS~2A and IRAS23385, respectively.

For the COMs, we derive ice column densities and abundances for CH$_3$CHO, CH$_3$CH$_2$OH, CH$_3$OCHO and CH$_3$COOH. In terms of ice abundance with respect to H$_2$O ice, IRAS~2A is more abundant in CH$_3$CHO by a factor of 1.75 compared to IRAS~23385. CH$_3$COOH is not detected in IRAS~23385, but has an estimated abundance of 0.3\% in IRAS~2A. CH$_3$CH$_2$OH and CH$_3$OCHO are more abundant in IRAS~23385 by a factor of 1.5 and 7, respectively, compared to IRAS~2A. We note that the CH$_3$CH$_2$OH abundance for IRAS~23385 is in line with the upper limit derived from the high-mass protostar W33A \citep{Scheltinga2018}. The other ice abundances are consistent with the upper limits derived in the literature, illustrating that JWST can now probe deeper than previous instruments.

With respect to CH$_3$OH ice, the abundances are between 9.3$-$14.6\% for CH$_3$CHO, 16.1$-$24.6\% for CH$_3$CH$_2$OH, 0.9$-$1.3\% for CH$_3$OCHO and 3.9$-$6.0\% for CH$_3$COOH, depending on the lower and higher CH$_3$OH listed in Table~\ref{ice_cd}. The values are much lower than the upper limits derived in the literature \citep{Scheltinga2018} for CH$_3$CHO (52\%) and for CH$_3$CH$_2$OH (42\%).

\subsection{Tentative detections and upper limits on column densities}
\label{sec_tentative}

\begin{figure}
   \centering
   \includegraphics[width=\hsize]{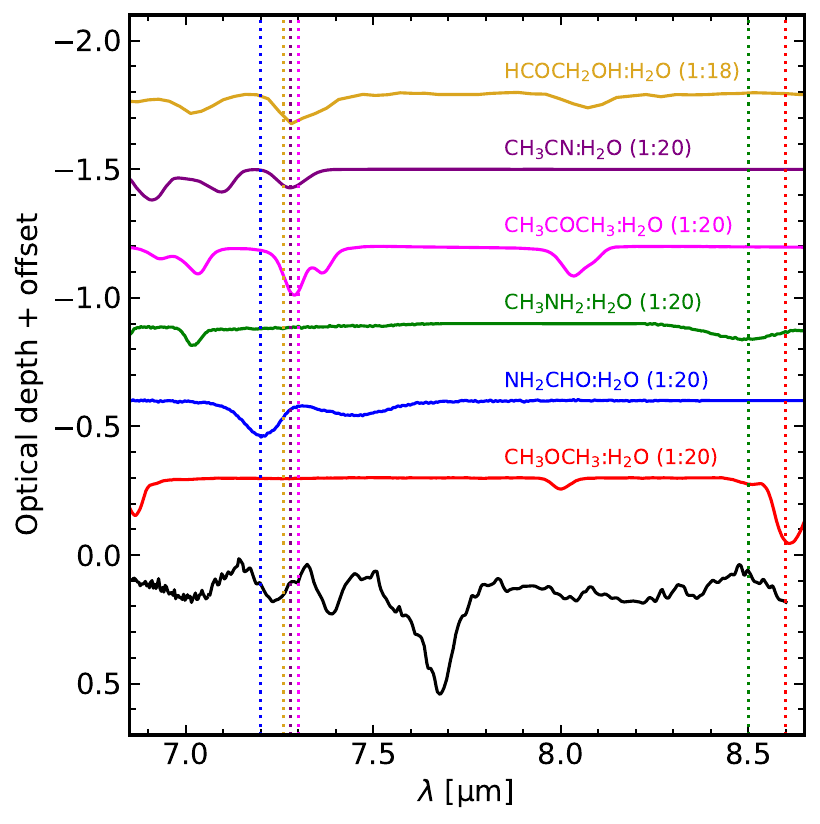}
      \caption{IR spectra of COMs not contributing to the global fit, but with absorption features in the range between 6.8 and 8.6~$\mu$m. These spectra are scaled to the IRAS~2A spectrum (black) to derive upper limit column densities. The vertical dotted lines are colour-coded and indicate the features used to derive the ice column densities. A small horizontal shift in the vertical lines is used to distinguish the bands at 7.3~$\mu$m.}
         \label{COM_ul}
   \end{figure}

A few COMs were not part of the global solution shown in Figure~\ref{global_fits}. In this case, we perform a separate comparison of these data to check for tentative detections and derive upper limit column densities in IRAS~2A. This procedure is not applied to IRAS~23385 due to the low signal-to-noise ratio. For this step, we scale laboratory spectra of COMs at specific wavelengths to the MIRI spectrum. This allows us to take into account both the intensity and width of the IR ice band. We performed separate scaling for COMs that have overlaps of the IR features. The molecules used in this step are CH$_3$OCH$_3$, CH$_3$COCH$_3$, HCOCH$_2$OH, NH$_2$CHO, CH$_3$NH$_2$ and CH$_3$CN mixed with H$_2$O ice. Figure~\ref{COM_ul} shows the COMs spectra superposed to the IRAS~2A MIRI data, with upper limit column densities also listed in Table~\ref{ice_cd}. To scale CH$_3$OCH$_3$, we use the CH$_3$-rock mode at 8.6~$\mu$m as a reference. This band has also contributions of ammonia and methyl formate. CH$_3$COCH$_3$ and CH$_3$CN have a CH$_3$ symmetric deformation mode at $\sim$7.3~$\mu$m, and HCOCH$_2$OH a CH$_2$ deformation mode at the same position. Both CH$_3$COCH$_3$ and CH$_3$OCH$_3$ may contribute to the band at 8.1~$\mu$m due to the CCC asymmetric stretch. In the case of CH$_3$NH$_2$, we use the CH$_3$-rock mode at 8.5~$\mu$m as a reference, whereas the C$-$H bend at 7.2~$\mu$m is considered for NH$_2$CHO.


\begin{figure}
   \centering
   \includegraphics[width=\hsize]{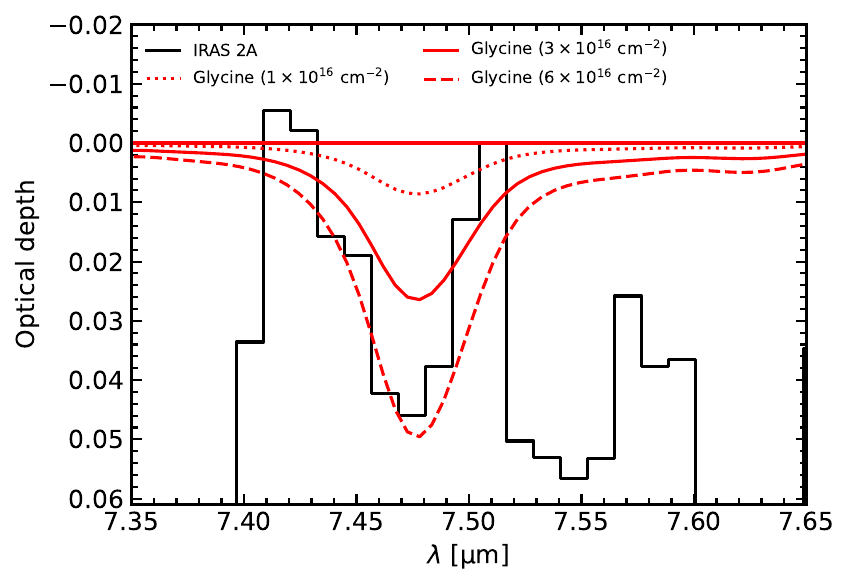}
      \caption{Comparison of the $\omega$CH mode of glycine IR spectrum (red) with local subtracted spectrum of IRAS~2A (black). The intensities of the glycine band are given for three ice column densities.}
         \label{Amino_ul}
   \end{figure}

Finally, we also check the potential presence of solid-phase glycine, the simplest amino acid, in the spectrum of IRAS~2A. In \citet{Ioppolo2021}, the solid-state formation of this simplest amino acid was proven. Glycine has many absorptions features in the IR, and the $\omega$CH$_2$ mode at 7.46~$\mu$m is the strongest band of $\beta$-glycine. Depending on the glycine structure (e.g., neutral - NH$_2$CH$_2$COOH  or zwitterion - NH$_3^+$CH$_2$COO$^-$), the position of this band may shift \citep{Potapov2022} or be narrowed if the molecule is mixed in an argon sample \citep{Ehrenfreund2001}. In this paper, we use the zwitterionic form \citep[NH$_3^+$CH$_2$COO$^-$;][]{Pilling2011}, which is available in the UNIVAP database\footnote{\url{http://www1.univap.br/gaa/nkabs-database/S3.txt}}. In Figure~\ref{Amino_ul}, we compare the local subtracted spectrum around the 7.46~$\mu$m band with the $\beta$-glycine data at three different column densities. We use the band strength from \citet{Holtom2005} of the 7.46~$\mu$m band, calculated as 1.16 $\times$ 10$^{-17}$ cm molecule$^{-1}$. This range of column densities is compatible with \citet{Gibb2004} who estimated an upper limit glycine column density of 3 $\times$ 10$^{16}$ cm$^{-2}$, based on the absorption feature around 5.8~$\mu$m band.  With respect to H$_2$O ice, we find an upper limit of between 0.03$-$0.2\% (See Section~\ref{ice_cd_sec} and Table~\ref{ice_cd}). This is close to the upper limit derived for W33A \citep[$<$0.3;][]{Gibb2004}, and in agreement with the theoretical models \citep[0.03$-$0.7\%;][]{Ioppolo2021}.

\subsection{Testing more complex alcohols and other molecules}
\label{complex_alc}

\begin{figure}
   \centering
   \includegraphics[width=\hsize]{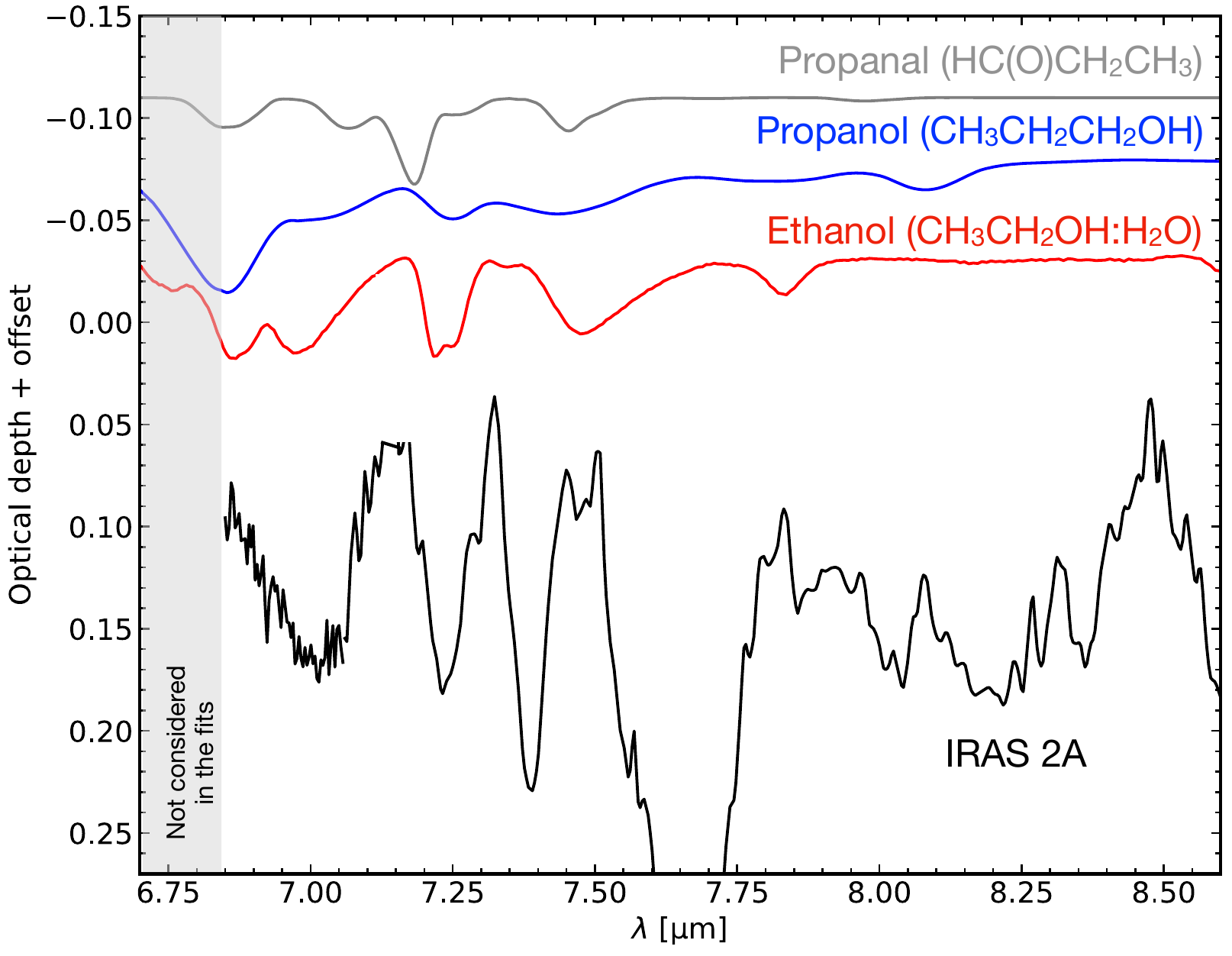}
      \caption{Experimental IR spectra of ethanol, propanal and propanol compared to IRAS~2A spectrum in the range between 6.8$-$8.6~$\mu$m. The grey area is not considered in the fits, but it is shown here to highlight the C$-$H stretching mode of these molecules. For better readability of this figure, the gas-phase emission lines between 6.8$-$7.2~$\mu$m are masked.}
      \label{fig_prop}
   \end{figure}

In this section, we compare the IRAS~2A spectrum with two other alcohols more complex than CH$_3$CH$_2$OH, for instance, propanal (CH(O)CH$_2$CH$_3$) and 1-propanol (CH$_3$CH$_2$CH$_2$OH). The goal is to check for similarities and differences between the functional groups of these alcohols with ethanol since these more complex alcohols are expected to have features at similar locations. It is worth mentioning that propanal and 1-propanol have been synthesized in experiments with ice analogues \citep{Qasim2019} using atom addition reactions, and in the case of propanol, via CH$_3$OH ice UV irradiation \citep{TenelandaOsorio2022}. Both propanal \citep[starless core TMC-1;][]{Agundez2023} and propanol \citep[Galactic Centre;][]{Belloche2022, Jimenez2022} have been securely detected in the gas phase. Figure~\ref{fig_prop} shows that these three alcohols have absorption features around 6.8, 7.2 and 7.5~$\mu$m. However, the relative intensities among these bands for these three molecules are different. For example, ethanol has similar intensities at these three bands, whereas 1-propanol and propanal have different intensities by factors of 3$-$4. In the case of 1-propanol, the fit of the 7.2~$\mu$m band would require twice more absorption at 6.8~$\mu$m than observed in IRAS~2A. On the other hand, propanal could contribute to the blue wing of the 7.2~$\mu$m band. We highlight, however, that in addition to HCOO$^-$, NH$_2$CHO can also contribute to the same absorption feature (see Section~\ref{sec_tentative} and Figure~\ref{COM_ul}). Another caveat is that both 1-propanol and propanal IR spectra correspond to pure molecules, and therefore spectral differences because of the chemical environment are not perceived in this analysis. 

Other than alcohols, hydrocarbons may also contribute to the 7.2 and 7.4~$\mu$m bands. In Appendix~\ref{hydro}, we show a comparison of IRAS~2A spectrum with pure C$_2$H$_2$, C$_2$H$_4$ and C$_2$H$_6$ around the 7~$\mu$m bands and beyond 11~$\mu$m. As a result, it can be seen that C$_2$H$_2$ and C$_2$H$_6$ could contribute to the red wing of the 7.2~$\mu$m band. However, their absorption bands longwards of 11~$\mu$m exceed the absorption profile in IRAS~2A. Based on these comparisons, the statistical analysis, and different choices for the silicate (Figure~\ref{diff_silic}) and local continuum (Figure~\ref{diff_cont}) we can conclude that CH$_3$CH$_2$OH is the alcohol that contributes most to the 7.2~$\mu$m band in addition to HCOO$^-$. The systematic analysis of additional high S/N MIRI data is necessary to obtain robust constraints of larger COMs. 

\section{Discussion}
\label{discuss_sec}
In this section, we discuss the implications of our results from the analysis of the 6.8$-$8.6~$\mu$m region in a low- and high-mass source. This discussion is focused on the presence of simple and neutral species (CH$_4$, SO$_2$, HCOOH, H$_2$CO), simple and ionic species (OCN$^-$, HCOO$^-$), and complex organic molecules (CH$_3$CHO, CH$_3$CH$_2$OH, CH$_3$OCHO, CH$_3$COOH). This section ends with a direct comparison between the ice abundances in the low-mass protostar, IRAS~2A, and the ice bulk abundances in comet 67P/G-C. The goal of this comparison is to evaluate the COM inheritance scenario in solar-type protostellar environments. Because of this, we do not perform the same comparison with IRAS~23385.

\subsection{Chemical complexity of protostellar ices}
\label{chem_compl}
In this work, we expand the COMs ice inventory by reporting the detection of at least two vibrational modes of COMs frozen in protostellar ices. We consider secure detections in the cases of CH$_3$CHO, CH$_3$CH$_2$OH, and CH$_3$OCHO. CH$_3$COOH also has two vibrational modes and a high recurrence in the degeneracy analysis of IRAS~2A. However, its presence depends on silicate subtraction and local continuum determination and therefore it is considered a tentative detection. This series of molecules are chemically related and strongly supports laboratory experiments and computational simulations that suggest COMs formation in the solid phase. We also note that all these COMs are commonly detected in the gas phase in hot core sources and are among the most abundant ones \citep[e.g.,][]{Chen2023}. A detailed comparison between gas and ice abundances is left to a future paper (Chen et al. in prep.). Below we discuss particular aspects of the COMs solid phase detections reported in this work.

\subsubsection{A polar ice environment}
\label{polar}
\begin{figure}
   \centering
   \includegraphics[width=\hsize]{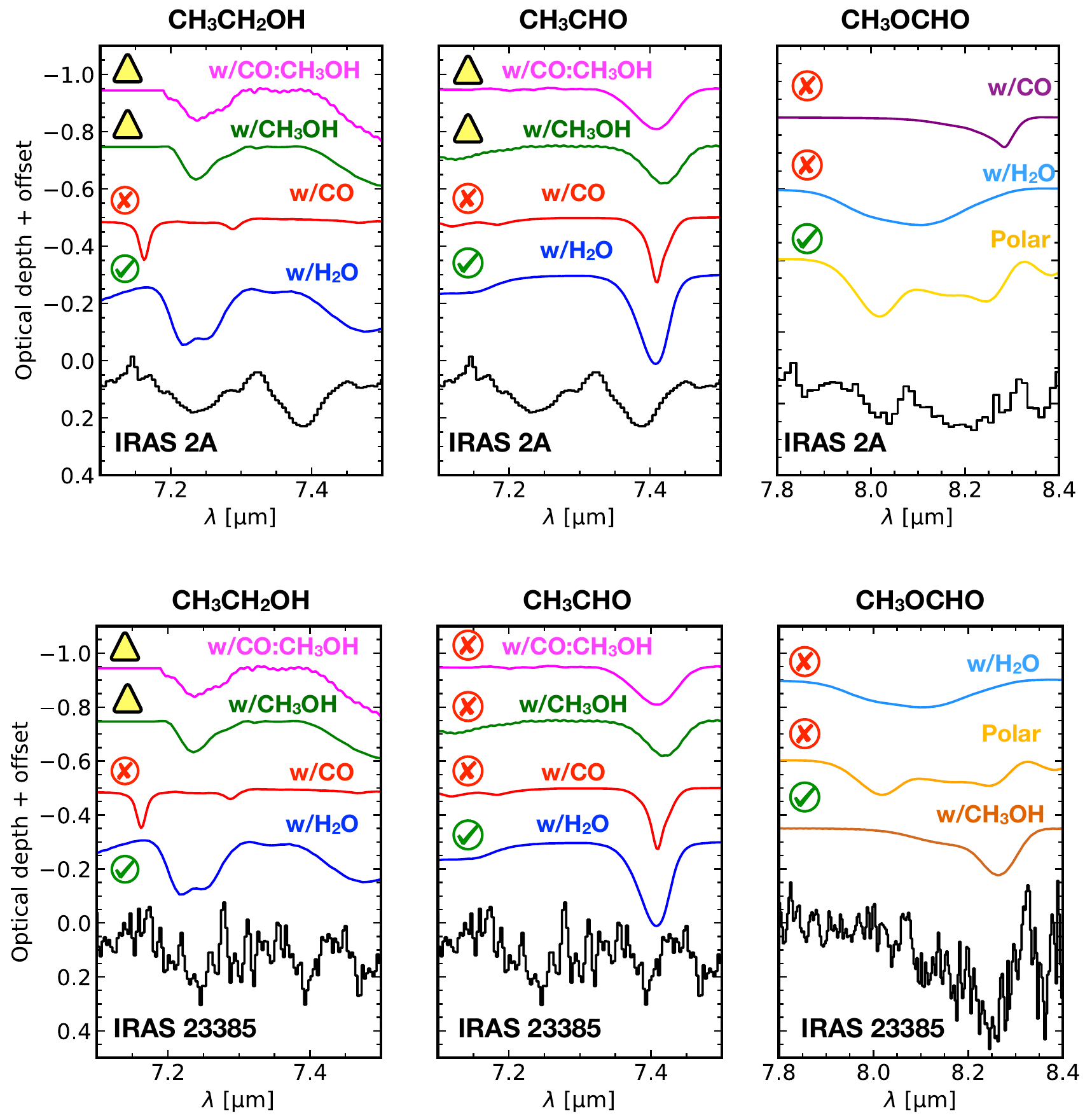}
      \caption{Comparisons between the IR spectra of COMs in different ice matrices with the observed bands of IRAS~2A (top) and IRAS~23385 (bottom). The green checkmarks indicate the data providing the best fit. Excluded data are given by the red cross. Data not part of the best fit and with lower recurrence are indicated by the yellow triangle. The term ``polar'' in the right panels refers to CO:H$_2$CO:CH$_3$OH and it is the original label published in \citet{Scheltinga2021}.}
         \label{mixtures}
   \end{figure}

In both protostars, the CH$_3$CHO, CH$_3$CH$_2$OH, and CH$_3$COOH (continuum-dependent) molecules are diluted in a polar environment (i.e., molecules with high dipole moment), in particular, dominated by H$_2$O ice. In the general case where H$_2$O is the major ice component, this means that COM features are under the strong influence of the water ice polarity. In Figure~\ref{mixtures}, we compare the MIRI observations with the band shape of three COMs in the best fit, the same COMs in an apolar ice matrix, for instance, mixed with CO ice, and finally, in a CH$_3$OH-rich environment. CH$_3$CH$_2$OH mixed with H$_2$O band shape at 7.25~$\mu$m agrees better with the data than the mixture with CO that has narrow and separate bands compared with the observations of IRAS~2A and IRAS~23385. In the case of the IRAS~2A observations, which has a higher signal-to-noise ratio, CH$_3$CH$_2$OH:CO could, potentially, contribute to faint features at 7.16 and 7.29~$\mu$m. However, this particular ice mixture is not part of the possible solutions selected by the \texttt{ENIIGMA} code. If present, it would have an ice column density one order of magnitude lower than ethanol mixed with water ice. The other mixtures with CH$_3$OH are less recurrent because of the narrow 7.2~$\mu$m and the stronger 7.5~$\mu$m that in the global fit makes the 7.2~$\mu$m less prominent. In the case of CH$_3$CHO, the apolar mixture (CH$_3$CHO:CO) makes the CH$_3$ deformation mode narrow by a factor of 2 and red-shifted by 0.02~$\mu$m. In the case of CH$_3$OH-rich mixtures, the peak is red-shifted by 0.03~$\mu$m and the feature is broader.  

This result does not necessarily contradict laboratory experiments that show that COMs are formed via hydrogenation of CO molecules \citep[e.g.,][]{Fuchs2009} and C atoms \citep{Fedoseev2022}, as well as that a fraction of CH$_3$OH ice is mixed with CO \citep{Cuppen2011}. Instead, it points towards a strong effect of H$_2$O ice on the spectral IR bands of COMs. Moreover, results from the JWST-Ice Age program \citep{McClure2023}, suggest that a fraction of CH$_3$OH, the most abundant COM, coexists with H$_2$O in the same ice matrix in cold prestellar clouds. This can be linked to another formation scheme, CH$_4$ + OH, as studied by \citet{Qasim2020}. Regardless of whether CH$_3$CHO and CH$_3$CH$_2$OH reside in a H$_2$O- or CH$_3$OH-rich ice, the important message is that the IR band shapes that resemble better the observations are induced by a polar environment.

Finally, CH$_3$OCHO seems to be primarily mixed with another polar environment that includes CH$_3$OH instead of H$_2$O. The spectral shape of CH$_3$OCHO mixed with H$_2$O has a broader profile centred at 8.1~$\mu$m which does not match well the observations. From the degeneracy analysis of IRAS~23385, it is not completely excluded as a solution, but in the global fits, the presence of HCOOH discards this component as part of the fit. The mixture with CO creates a sharp peak at 8.25~$\mu$m that deviates from the observations. The differences between the chemical environment of CH$_3$CH$_2$OH and CH$_3$CHO compared to CH$_3$OCHO could indicate that methyl formate in IRAS~23385 may have experienced a physical process different from those in IRAS~2A. We discuss this possibility below in Section~\ref{ice_proc}.

\subsubsection{Evidence of ice thermal processing?}
\label{ice_proc}
Among the ice features identified in this work, CH$_3$OCHO has different spectral shapes in IRAS~2A and IRAS~23385, regardless of the local continuum choice. While the fit of the IRAS~2A spectrum contains CH$_3$OCHO mixed with CO:H$_2$CO:CH$_3$OH, in IRAS~23385, the CH$_3$OCHO mixed with CH$_3$OH provides the best fit, and excludes other solutions (see right panels in Figure~\ref{mixtures}). A possible cause for this difference can be due to thermal processing, in particular ice distillation. In this process, the most volatile species desorbs from the ice while other species with higher sublimation temperatures remain. For example, during the CH$_3$OH formation via CO hydrogenation \citep[e.g.,][]{Watanabe2002,Fuchs2009} both CH$_3$O and HCO are formed \citep[e.g.,][]{Chuang2016, Garrod2022, Chen2023}, leading to CH$_3$OCHO. H$_2$CO is also an intermediate step towards the CH$_3$OH formation. In a trivial situation, it is expected that at low temperature (< 30~K) CH$_3$OCHO should be mixed with CO, H$_2$CO and CH$_3$OH, the exact components of the ice mixture used to fit the IRAS~2A spectrum. When this ice is warmed up to 50-100~K, CO ice is fully desorbed and H$_2$CO is partially desorbed, and the ice would be composed mostly of CH$_3$OCHO and CH$_3$OH. Because of the low S/N spectrum of IRAS~23385 we cannot definitely conclude if H$_2$CO is present in the ices toward this source.

\subsubsection{Gas vs. ice: CH$_3$CH$_2$OH/CH$_3$CHO ratio}
\label{GIratio}

CH$_3$CH$_2$OH and CH$_3$CHO are chemically linked since the double hydrogenation of acetaldehyde in the solid state leads to ethanol \citep{Fedoseev2022}. In this regard, the ratio of these two molecules provides a way to understand the hydrogenation efficiency of solid-phase molecules to form larger species in protostars.

\begin{figure}
   \centering
   \includegraphics[width=\hsize]{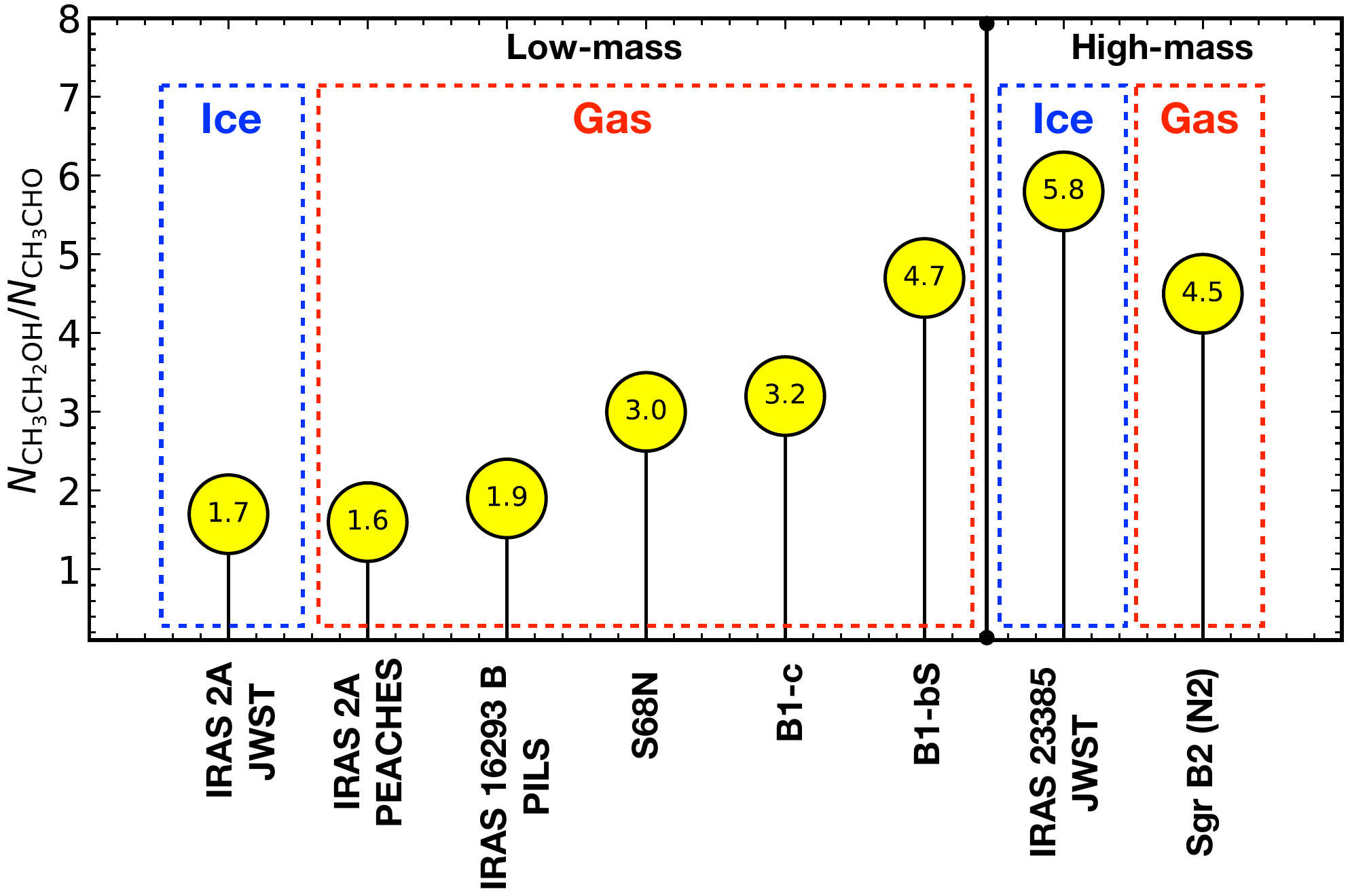}
      \caption{Ethanol/acetaldehyde ratio in the gas and solid phases towards different sources.}
         \label{eth_act}
   \end{figure}

Figure~\ref{eth_act} compares the inferred CH$_3$CH$_2$OH/CH$_3$CHO ratios in the solid phase with those found in the gas phase of many protostars. A persistent ratio of CH$_3$CH$_2$OH and CH$_3$CHO above the unity has been observed in the gas phase with sensitive observations towards protostars in both low- and high-mass star-forming regions \citep[e.g.,][]{Yang2021, Jorgensen2020, vanGelder2020, Chen2023}. For example, the ratios summarized by \citet{Jorgensen2020} for IRAS~16293B and Sgr(B2)~N2 are 1.9 and 4.5, respectively. For low-mass protostars, the CH$_3$CH$_2$OH/CH$_3$CHO ratio taken from \citet{vanGelder2020} are 3.2 (B1-c), 3.0 (S68N), and $<$4.7 (B1-bS). For the first time, a comprehensive analysis of the 6.8-8.6~$\mu$m JWST data allows us to observe the same trend in the solid phase. The CH$_3$CH$_2$OH/CH$_3$CHO ratios derived in this work are 1.7 and 5.8 for IRAS~2A and IRAS~23385, respectively.

This overall trend in protostars suggests that the ethanol/acetaldehyde ratio in the gas phase reflects those in the solid phase. This is a strong hint that ices are the birthplace of COMs detected in the gas phase. Additionally, these COMs are inherited from molecular cloud stages where the ice hydrogenation is efficient to form saturated molecules \citep[e.g.,][]{Fedoseev2022}. We also mention that a full three-phase chemical modelling of these sources, assuming reactions within the bulk of the ice, on the ice surface, and in the gas phase can help us constrain the ice chemistry from the observed ice column densities, and provide a more complete explanation for the ethanol/acetaldehyde ratio in both gas and solid-phase. In fact, recent three-phase chemical modelling work by \citet{Garrod2022} assuming general physical conditions of molecular clouds during collapse and warm-up as well as with non-diffusive ice chemistry confirms that the solid-and gas-phase ratios of ethanol/acetaldehyde are consistent with the results presented in this paper.


\subsection{Ions in icy grain mantles}
\label{ionice}
The presence of ions in interstellar ices has long been proposed in the literature as a result of the energetic processing of ice mantles \citep[e.g.,][]{Grim1987, Strazzulla1988, Allamandola1988, Martinez2014, Pilling2010} and acid-base reactions \citep[e.g.,][]{Novozamsky2001}. Among these ions, OCN$^-$ has been detected in several protostellar envelopes \citep[e.g.,][]{vanBroekhuizen2005}, and recently with JWST in the Chameleon I molecular cloud towards background stars \citep{McClure2023}. Other ions, such as HCOO$^-$, were proposed as one of the carriers of the 7.2 and 7.4~$\mu$m by \citet{Schutte1999_weak} based on the IR peak position.

In the analysis presented in this work, we find that both HCOO$^-$ and OCN$^-$ are important contributors to signals in the range between 6.8$-$8.6~$\mu$m. In particular, an important aspect to highlight regarding HCOO$^-$ is the same relative intensities of the 7.2 and 7.4~$\mu$m bands at low temperature (see Appendix~\ref{formateionapd}), which is considered the main carrier of the absorption profiles in the two protostars presented in this work. The fact that higher temperature HCOO$^-$ profiles do not match well the observations indicates that the acid-base reaction occurs in cold regions of the protostellar envelope without further thermal processing. One could argue that the HCOO$^-$ profile at 14~K is degenerate with the data at 150~K as seen in Appendix~\ref{formateionapd}. However, the full IR spectrum of HCOO$^-$ has an intense peak at 6.3~$\mu$m that is around 8 times stronger than the 7.2 and 7.4~$\mu$m bands, and therefore can not be the main carrier of those bands. This strongly suggests that acid-base reactions occur in IRAS~2A and IRAS~23385 and that ions cannot be neglected when interpreting the IR observations.

The formation of HCOO$^-$ can occur via the acid-base reaction involving H$_2$O:HCOOH:NH$_3$. This is explained by theoretical calculations of HCOOH$\cdot$(H$_2$O)$_n$ clusters \citep{Park2006}. In particular, formic acid partially ionizes when mixed with H$_2$O, forming HCOO$^-$ and H$_3$O$^+$ via proton exchange \citep{Theule2011}. 
Another reaction to form HCOO$^-$ was studied by \citet{Bergner2016}, in which only HCOOH and NH$_3$ are present in the ice, without any H$_2$O.




Similar to HCOO$^-$, OCN$^-$ is efficiently formed via acid-base reaction, which has been studied in laboratory experiments starting with HNCO and NH$_3$ \citep{Schutte2003, Raunier2003, Broekhuizen2004}. A large fraction of OCN$^-$ is produced in these experiments, as well as the counter ion, NH$_4^+$ to maintain the electric neutrality. A possible difficulty in this approach is the high abundance of HNCO in the ice needed to synthesize OCN$^-$, associated with the non-detection of this molecule in ices yet. However, gas-phase observations \citep[e.g.,][]{Hernandez2019} show high abundances of HNCO and laboratory experiments in the ice by \citet{Fedoseev2015} and \citet{Noble2015} show that HNCO can be rapidly formed via an exothermic solid-phase reaction between NH and CO. Additionally, the non-detection of HNCO in ices \citep[e.g.,][]{McClure2023} could be related to the low abundance after being converted into OCN$^-$. The presence of OCN$^-$ and HNCO in ices has a strong astrobiological appeal. HNCO participates as a peptide bond between two single amino acids as shown by \citet{Fedoseev2015}. Moreover, irradiation experiments of ice samples containing OCN$^-$/HNCO would lead to the formation of amino acids and their anions, as well. Finally, we highlight that although NH$_4^+$ is one of the byproducts of this acid-base reaction, it does not contribute to the absorption bands covered in this work, and therefore, it is not discussed here.

Another mechanism for the OCN$^-$ formation is via UV irradiation \citep{vanBroekhuizen2005}, which is discussed as being less dominant in low-mass protostars. This route requires high UV fluxes and an abundance of around 30\% of NH$_3$ in the ice. The former condition is satisfied for high-mass protostars, but not otherwise. In fact, a recent paper by \citet{Onaka2022} shows a clear correlation in the high-mass-source AFGL~2006 between the OCN$^-$ ice column density with the flux intensity of the \texttt{HI} Br$\alpha$ line. This hydrogen recombination line is a tracer of strong UV radiation and supports an OCN$^-$ formation induced by UV photons. Other mechanisms, such as UV-induced flux by cosmic rays are not enough to produce significant amounts of OCN$^-$, and low-mass stellar UV would not reach the regions where ice is located. The latter condition exceeds the abundances estimated in the literature for both low- and high-mass protostars, which is between 2\% and 15\% \citep{Bottinelli2010}.

\subsection{Similarities and differences between IRAS~2A, IRAS~23385 and the Comet 67P/G-C}
\label{comet_comp}
Gas-phase comparisons between the abundances of CHO-bearing COMs with respect to CH$_3$OH in low- and high-mass protostars, and with the bulk composition of the comet 67P/G-C were made by \citet{Drozdovskaya2019} and \citet{Jorgensen2020}. The main conclusion is that there is a good correlation between the abundances of gas-phase COMs in high- and low-mass star-forming regions. On the other hand, there are also differences between the low-mass protostar (IRAS~16293B) and the comet 67P/C-G. In particular, the COMs abundances onto the comet 67P/C-G are enhanced by a factor of up to 10. This difference is interpreted as ice inheritance followed by chemical alteration towards later protostellar phases. 

\begin{figure}
   \centering
   \includegraphics[width=7cm]{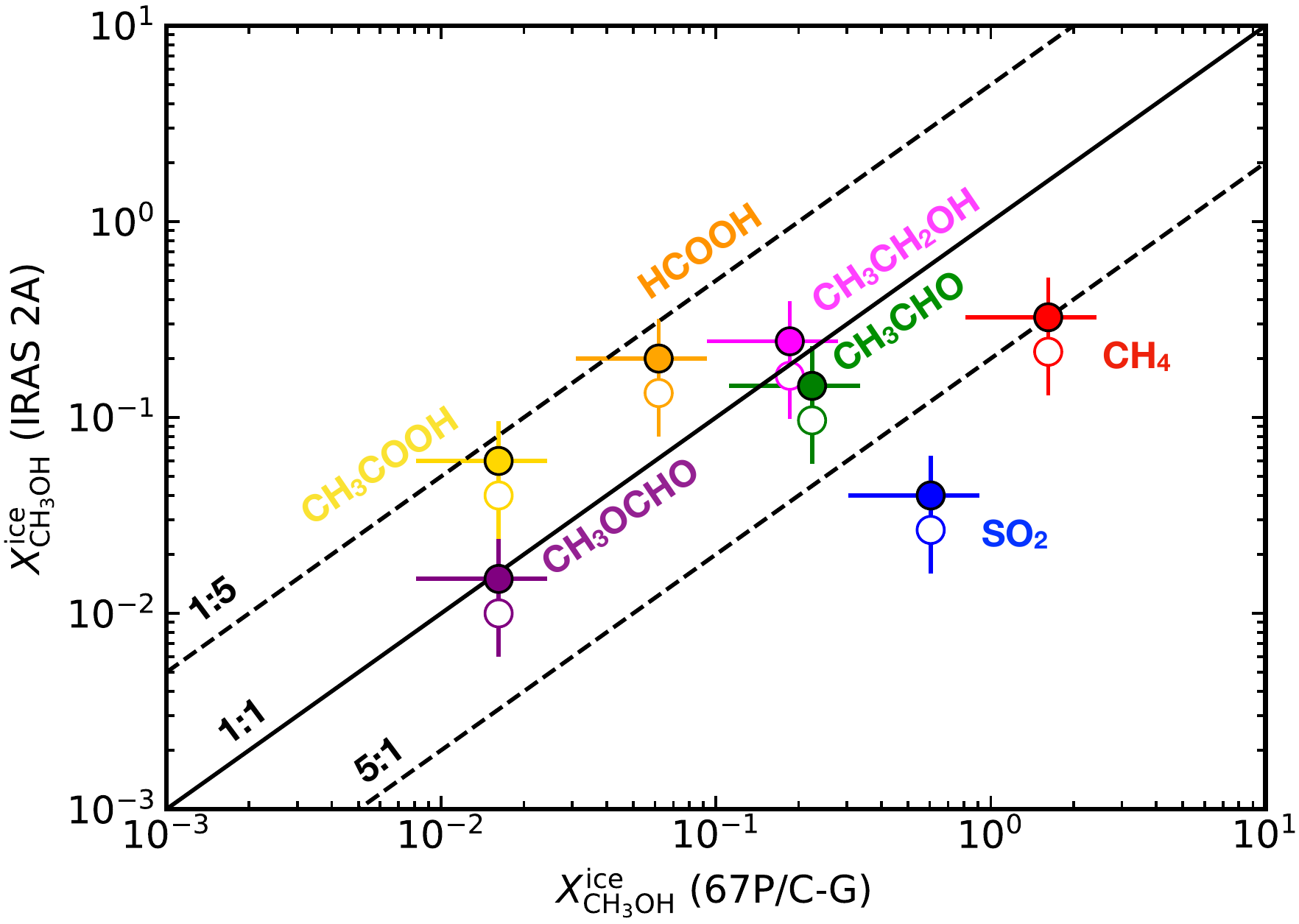}
      \caption{Comparison between ice abundances with respect to CH$_3$OH ice in IRAS~2A and the comet 67P/G-C \citep{Rubin2019}. Full and hollow circles show the abundances considering $N_{\rm{CH_3OH}}^{\rm{ice}} =$ 1.5 $\times$ 10$^{18}$ cm$^{-2}$ and 1.3 $\times$ 10$^{18}$ cm$^{-2}$, respectively. The solid line indicates the 1:1 abundance relation, and the dashed lines indicate a cometary abundance lower and higher by a factor of 5.}
         \label{comp_met}
   \end{figure}

In Figure~\ref{comp_met}, we show a comparison between the ice CHO-bearing COMs abundances in IRAS~2A and in the comet 67P/G-C, both with respect to methanol ice. We stress that the peak of the methanol band in IRAS~2A is saturated, and the comparisons are made by assuming that the real methanol ice column density is higher by a factor of 2$-$3 based on the wings of the C$-$O band at 9.74~$\mu$m (see Appendix~\ref{water_cd}). Under this assumption, we see that COMs and the volatile CH$_4$ correlate well with the cometary abundances within a factor of 5. This agrees with previous conclusions that these COMs are inherited by comets from early protostellar stages. The scatter, however, can be attributed to further chemical alteration at later stages, or slightly different initial composition in the parental molecular cloud. It is also interesting that the volatiles, CH$_4$ and SO$_2$ are enriched in the comet 67P/G-C. This could indicate that these molecules are also formed in the gas phase and condensed at later stages in the comet. However, more analysis of other JWST observations is needed to draw strong conclusions from this correlation. 

\begin{figure}
   \centering
   \includegraphics[width=\hsize]{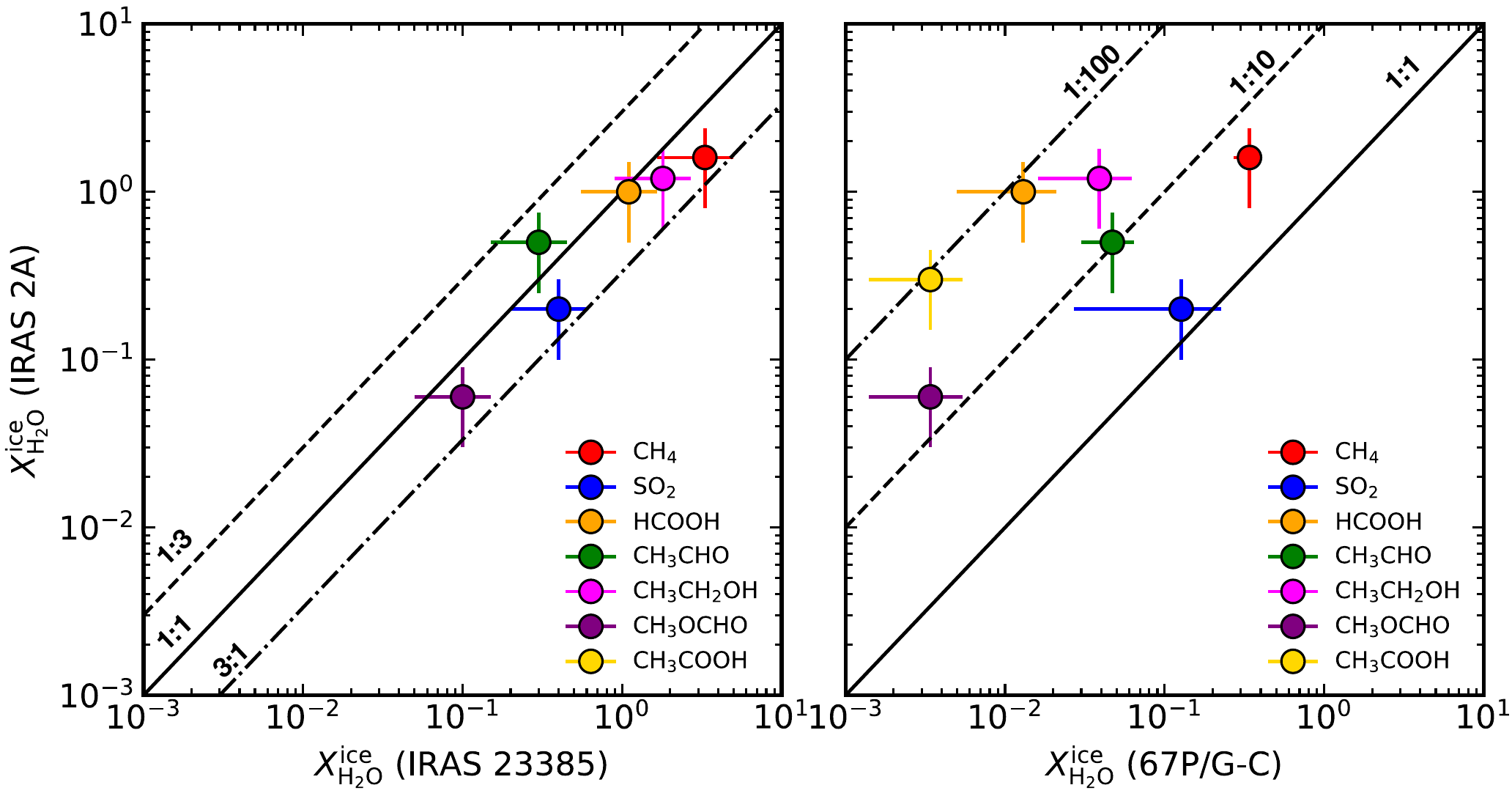}
      \caption{Comparison between ice abundances with respect to H$_2$O ice. The left panel shows the ice abundances in IRAS~2A and IRAS~23385. The solid line indicates the 1:1 abundance relation, whereas the dashed and dot-dashed lines indicate the enhancement and diminution of IRAS~23385 abundance by a factor of 3. The right panel compares the ice abundances between IRAS~2A and the comet 67P/G-C from \citet{Rubin2019}. The solid line indicates the same as in the left panel. The dot-dashed and dashed lines indicate a cometary abundance lower by a factor of 10 and 100, respectively.}
         \label{comp_h2o}
   \end{figure}

In Figure~\ref{comp_h2o}, we show a comparison between the COMs abundances with respect to H$_2$O ice in IRAS~2A, IRAS~23385 and the comet 67P/G-C. The abundances in the low- and high-mass protostar are very close to a linear correlation, which agrees with the results from \citet{Jorgensen2020} between IRAS~16293B and Sgr~B2(N2). On the other hand, the COMs abundances compared to H$_2$O ice are lower in the comet 67P/G-C than in IRAS~2A. Such a high abundance of molecules in low-mass protostars with respect to H$_2$O ice was noticed before for other molecules, such as NH$_3$ \citep{Kawakita2011}, and CO, CH$_4$ and CH$_3$OH \citep{Oberg2011}. The reason is unclear, but it can indicate selective ice destruction of these species compared to H$_2$O ice in the protosolar nebula or that those COMs were formed in a carbon-poor protostellar envelope \citep{Oberg2011}.

\section{Conclusions}
\label{conc_sec}
This paper presents a comprehensive analysis in the range between 6.8$-$8.6~$\mu$m of JWST spectra observed towards the IRAS~2A and IRAS~23385 protostars. We perform spectral fits exploring a vast number of IR laboratory spectra that cover, simple molecules, ions and COMs. The main conclusions are summarised below:

\begin{itemize}
    \item The 7.7~$\mu$m band is fully resolved in the MIRI spectrum of IRAS~2A and IRAS~23385. As found in previous studies, CH$_4$ ice is the main carrier of this band. In addition to CH$_4$, our analysis shows that OCN$^-$ and SO$_2$ contribute to the blue wing of the 7.7~$\mu$m band. SO$_2$ was tentatively detected based on {\it ISO} and {\it Spitzer} observations, and based on a statistical analysis we confirm that SO$_2$ is present in ices towards IRAS~2A. For IRAS~23385, SO$_2$ is classified as likely detection because of a lower S/N and statistical arguments. In the case of OCN$^-$, this is the first time that this ion is attributed to the 7.7~$\mu$m band. Besides the statistical analysis confirming this detection, OCN$^-$ is also seen at 4.59~$\mu$m in IRAS~2A with NIRSpec observation, thus confirming the feature at 7.7~$\mu$m.
    
    \item The 7.2 and 7.4~$\mu$m bands observed in IRAS~2A and IRAS~23385 are mainly due to the formate ion (HCOO$^-$). Since this ion is efficiently formed via acid-base reactions, and given the detection of OCN$^-$ at 7.65~$\mu$m, these results demonstrate that this type of chemical reaction may be rather common in interstellar ices.

    \item The earlier suggestions that CH$_3$CH$_2$OH (ethanol) and CH$_3$CHO (acetaldehyde) contribute to the absorption profiles at 7.2 and 7.4~$\mu$m, respectively, are confirmed in this work. In addition, we find that CH$_3$CH$_2$OH contributes to other bands in the range considered in this work, most notably, at 6.8$-$7.05~$\mu$m and 7.4$-$7.6~$\mu$m. Other possibilities, such as hydrocarbons (C$_2$H$_2$, C$_2$H$_4$, C$_2$H$_6$) and more complex alcohols (1-propanol, propanal) are less likely based on their absorption profiles. In the case of CH$_3$CHO, it also contributes to the range between 6.9$-$7.2~$\mu$m. 

    \item We find robust evidence that CH$_3$OCHO is present in the ices towards IRAS~2A and IRAS~23385. At least two strong bands of this molecule are found between 8.1$-$8.35~$\mu$m and 8.45$-$8.6~$\mu$m. The statistical analysis shows that this spectral component cannot be excluded as a solution for the fit. Based on the criteria for a firm identification, CH$_3$OCHO is classified as a secure detection. We also find that another COM, CH$_3$COOH (acetic acid), is present in the fits of IRAS~2A and cannot be excluded based on the confidence interval analysis for two out of three choices of continuum. However, since it is not found in IRAS~23385, and the local continuum strongly affects the shape of the bands around 7.8~$\mu$m, we classify acetic acid only as a likely detection. More comparisons with other JWST data will elucidate at which level if acetic acid is present in interstellar ices.

    \item The COMs found in this work are likely mixed in a polar environment. For example, CH$_3$CH$_2$OH and CH$_3$CHO fit better the observations when mixed with H$_2$O. These molecules mixed with CO have spectral profiles that deviate from the observed protostellar. In the case of CH$_3$COOH, the mixture with H$_2$O ice also provides a good fit to the IRAS~2A spectrum. In the case of CH$_3$OCHO, the two protostars show different spectral shapes. This differentiation could be related to ice distillation in IRAS~23385, the high-mass protostar.

    \item Ice-gas ratios between CH$_3$CH$_2$OH and CH$_3$CHO show values above unity in both gas and ice phases. This suggests solid-phase reactions for the formation of these molecules. In addition, the high abundance of CH$_3$CH$_2$OH shows that saturated molecules efficiently form in ices because of the high amount of hydrogen available.

    \item For the first time, we compare ice COMs abundances relative to CH$_3$OH ice in a protostar and the comet 67P/G-C. Our results indicate that the COMs ice abundances in the comet 67P/G-C correlate well with those in the protostar within a factor of 5, which strongly suggests that COMs in comets are significantly inherited from earlier protostellar phases. In the case of CH$_4$ and SO$_2$ with respect to CH$_3$OH ice, we find an enhancement in the comet 67P/G-C. On the other hand, the abundance comparison with respect to H$_2$O ice shows that the COMs, SO$_2$ and CH$_4$ are depleted in the comet.
\end{itemize}

The results presented in this paper illustrate how JWST, aided by laboratory experiments, is fully capable of probing the chemical complexity in interstellar ices. Future work using more JWST data with high S/N \citep[e.g., IRAS~15398;][]{Yang2022} will enable us to verify the recurrence of the COMs found in this paper in other sources, and consequently, assess the robustness of these detections. Another outlook would be the determination of ice-gas ratios of COMs more complex than CH$_3$OH. Ultimately, an analysis based on a large sample will help us to further answer the question: to what extent chemical complexity can be reached in interstellar ices?

\begin{acknowledgements}
The following National and International Funding Agencies funded and supported the MIRI development: NASA; ESA; Belgian Science Policy Office (BELSPO); Centre Nationale d’Études Spatiales (CNES); Danish National Space Centre; Deutsches Zentrum fur Luftund Raumfahrt (DLR); Enterprise Ireland; Ministerio De Economiá y Competividad; The Netherlands
Research School for Astronomy (NOVA); The Netherlands Organisation for
Scientific Research (NWO); Science and Technology Facilities Council; Swiss
Space Office; Swedish National Space Agency; and UK Space Agency. We thank the anonymous reviewer for the careful reading of our manuscript and the comments and suggestions that improved the clarity of this work. WRMR, EvD, K.S, N.B, McG, L.F and HL acknowledge the funding from the 
European Research Council (ERC) under the European Union’s Horizon 2020 research and innovation programme (grant agreement No. 291141 MOLDISK). WRMR thanks Niels Ligterink and Maria Drozdovskaya for useful discussions about the 67P/G-C comet abundances. We are grateful for continuing support through NOVA, the Netherlands Research School for Astronomy, the NWO through its Dutch Astrochemistry Program (DANII). The present work is closely connected to ongoing research within INTERCAT, the Center for Interstellar Catalysis located in Aarhus, Denmark. The work of MER was carried out at the Jet Propulsion Laboratory, California
Institute of Technology, under a contract with the National Aeronautics and Space Administration (80NM0018D0004). L.M. acknowledges the financial support of DAE and DST-SERB research grants (SRG/2021/002116 and MTR/2021/000864) from the Government of India. T.R. acknowledges support from ERC grant no. 743029 EASY. T.H. acknowledges support from the ERC Advanced grant no. Origins 83 24 28. H.B. acknowledges support from the
Deutsche Forschungsgemeinschaft in the Collaborative Research Center (SFB
881) “The Milky Way System” (subproject B1). P.J.K. acknowledges financial support from the
Science Foundation Ireland/Irish Research Council Pathway programme under
Grant Number 21/PATH-S/9360. A.C.G. has been supported by PRIN-INAF
MAIN-STREAM 2017 “Protoplanetary disks seen through the eyes of new generation instruments” and from PRIN-INAF 2019 “Spectroscopically tracing
the disk dispersal evolution (STRADE)”. K.J. acknowledges the support from
the Swedish National Space Agency (SNSA).    
\end{acknowledgements}

\bibliographystyle{aa}
\bibliography{References}


\appendix
\section{9.8~$\mu$m band compared to ISO and Spitzer sources}
\label{silic_comp}
\begin{figure*}
   \centering
   \includegraphics[width=17cm]{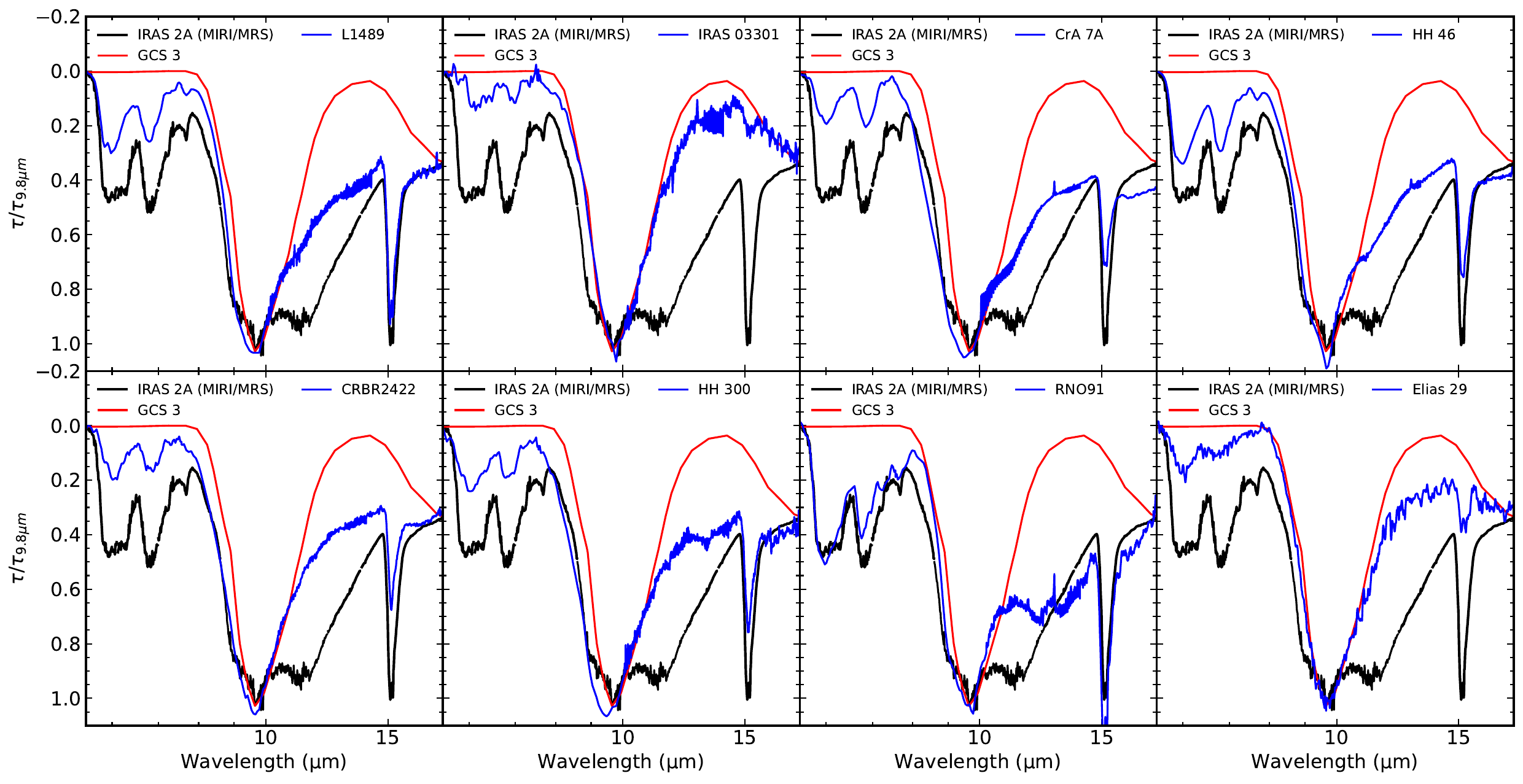}
      \caption{Comparison among MIRI/JWST spectrum of IRAS~2A, the silicate profile of GCS~3 and {\it Spitzer}/IRS spectrum of different low-mass protostars. These spectra are normalized by the optical depth at 9.8~$\mu$m.}
         \label{Silic2A}
   \end{figure*}

\begin{figure*}
   \centering
   \includegraphics[width=17cm]{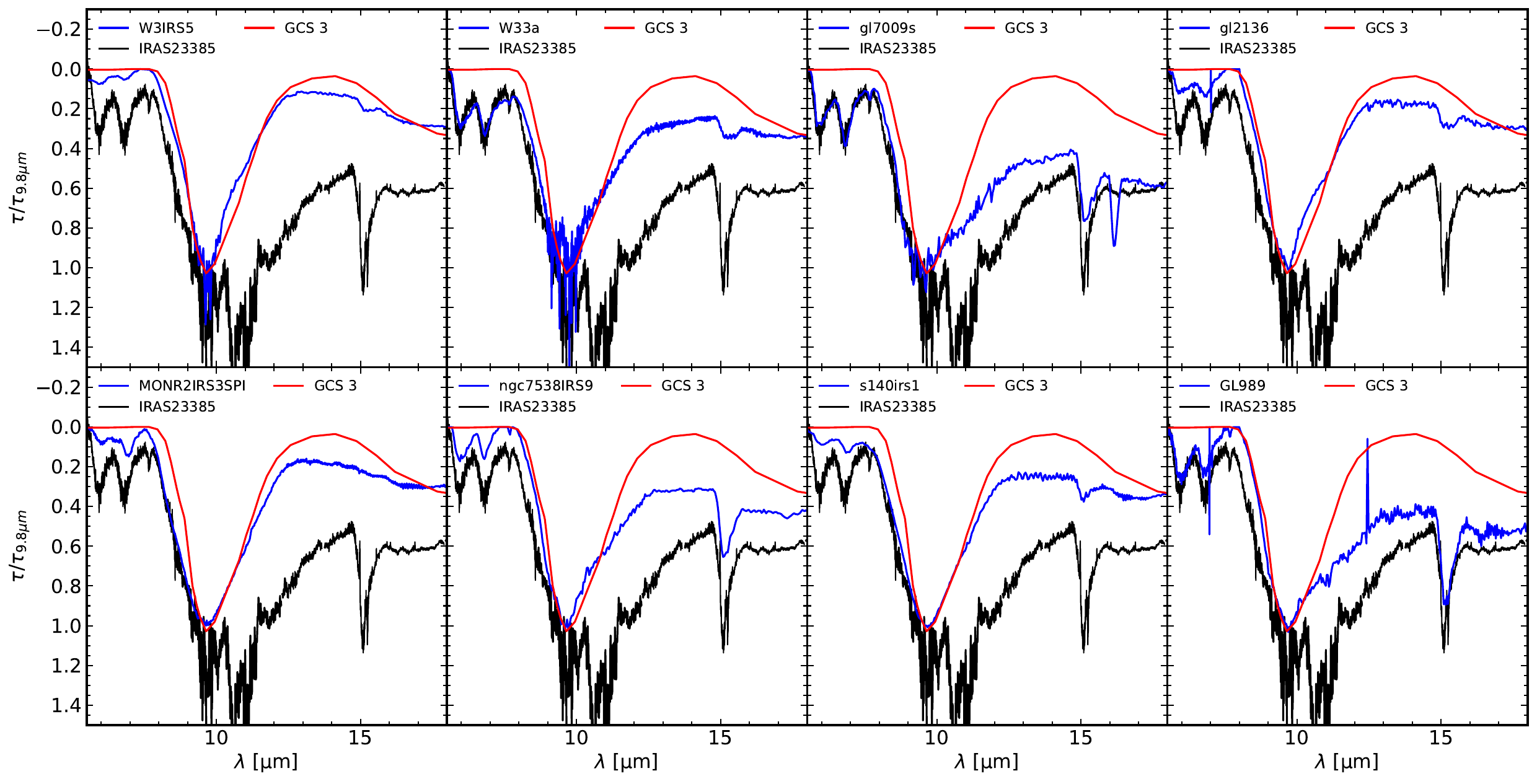}
      \caption{Comparison among MIRI/JWST spectrum of IRAS~23385, the silicate profile of GCS~3 and {\it Spitzer}/IRS spectrum of different low-mass protostars. These spectra are normalized by the optical depth at 9.8~$\mu$m.}
         \label{Silic23385}
   \end{figure*}

Figures~\ref{Silic2A} and \ref{Silic23385} show comparisons with the silicate feature towards GCS~3 and with other low- and high-mass protostars observed with {\it Spitzer} and {\it ISO}. The first note is that both IRAS~2A and IRAS~23385 have a broader 9.8~$\mu$m silicate profile compared to GCS~3. This highlights the need of considering different grain compositions when subtracting the silicate absorption band. Secondly, IRAS~2A and IRAS~23385 show a similar blue side of the spectrum with other sources. On the other hand, the red profile has more differences, which are related to the amount of H$_2$O ice towards the source. In fact, icy-grain models by \citet{Ossenkopf1994} show that coagulated icy grains have a prominent spectral bump around 12~$\mu$m because of the H$_2$O ice libration band.

\section{Effect of using different silicates for subtraction in IRAS~2A}
\label{silic_effect}

\begin{figure}
   \centering
   \includegraphics[width=\hsize]{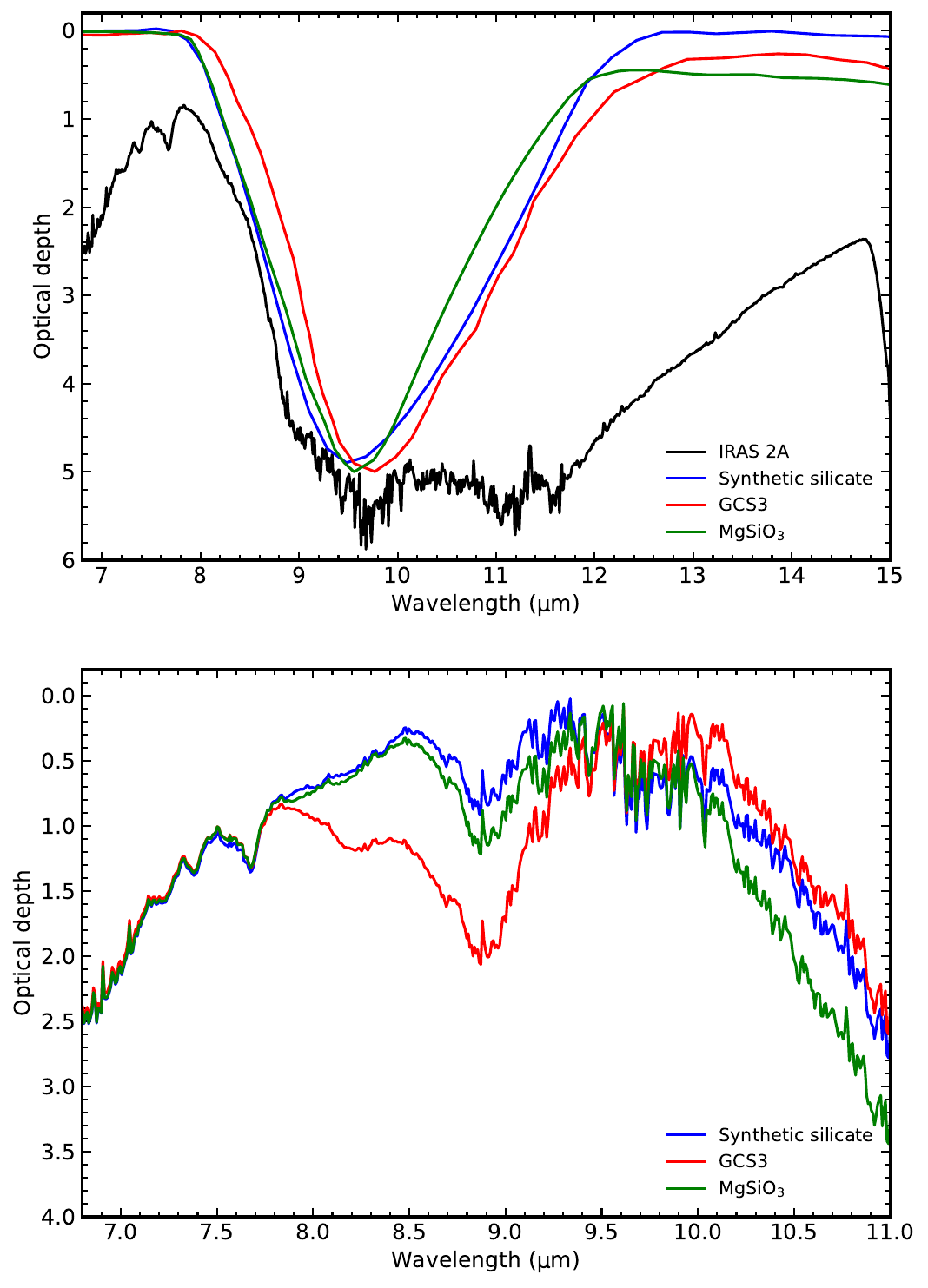}
      \caption{Comparing silicate profiles in IRAS~2A. The top panel shows the synthetic silicate providing the best fit in IRAS~2A (blue) compared to GCS~3 (red) and enstatite (green) silicate model from \citet{Poteet2015}. The bottom panel shows the IRAS~2A spectrum after removing the three silicate models.}
         \label{diff_silic}
   \end{figure}

In Section~\ref{cont}, we describe how the silicate feature is removed from IRAS~23385 and IRAS~2A MIRI-MRS spectra using a synthetic silicate profile combining olivine and pyroxene. In this section, we compare subtractions using different profiles to check how this step affects the shape of the 6.8$-$8.6~$\mu$m region. Figure~\ref{diff_silic} (top) shows three silicate profiles, scaled to IRAS~2A spectrum, i) the synthetic silicate used in Section~\ref{cont}, ii) the silicate profile of GCS~3, and iii) the MgSiO$_3$ profile taken from \citet{Poteet2015}. The peak optical depths are set to match the synthetic silicate profile. We point out that this comparison is focused on the 6.8$-$8.6~$\mu$m region, and therefore mismatches of these silicate spectra at longer wavelengths are not relevant for this specific analysis. We briefly mention SiO (silica), as another possible candidate for the blue wing of the 9.8~$\mu$m band. However, silica has not been found in absorption in protostars so far and its relatively narrow profile at 18~$\mu$m is not seen in both sources investigated in this paper. Nevertheless, if present, SiO would not affect the COMs bands investigated in this paper because of its broadband at around 8.3~$\mu$m. At most, it could slightly reduce the HCOOH ice column density.

Figure~\ref{diff_silic} (bottom) shows the silicate subtracted optical depth spectrum of IRAS~2A. Both synthetic and enstatite silicates result in similar spectral profiles between 6.8$-$8.6~$\mu$m. On the other hand, the silicate subtraction using GCS~3 creates an unrealistic absorption excess (also observed in \citet{Boogert2008}) that deviates from the other two profiles.

\section{List of laboratory data}
\label{list_lab}
A comprehensive list of molecules was used in this paper to search for the best fit of the region between 6.8 and 8.6~$\mu$m. This list includes COMs in different mixtures as well as simple molecules. The full list is shown in Table~\ref{tab_list}.

\begin{table*}[h]
\caption{\label{tab_list} Laboratory data tested in the global fit performed with \texttt{ENIIGMA}.}
\renewcommand{\arraystretch}{1.0}
\centering 
\scalebox{0.92}{
\begin{tabular}{lcccc}
\hline\hline
Label & Temperature (K) & Resolution (cm$^{-1}$) & Database$^a$ & Reference\\
\hline
\multicolumn{5}{c}{\bf{Simple molecules (less than 6 atoms) and hydrocarbons}}\\
\hline
H$_2$O & 15$-$160 & 2.0 & LIDA & [1]\\
CH$_4$ & 10$-$30 & 1.0  & OCdb    & [2]\\
HCOOH  & 15$-$165 &  1.0 & LIDA  &[3]\\
SO$_2$:CH$_3$OH (1:1)  & 15 &  1.0 & LIDA  &[4]\\
H$_2$O:CH$_4$ (10:1)  & 15 &  1.0 & UNIVAP  &[5]\\
C$_2$H$_2$  & 15 &  1.0 & NASA  &[6]\\
C$_2$H$_4$  & 15 &  1.0 & NASA  &[7]\\
C$_2$H$_6$  & 15 &  1.0 & NASA  &[7]\\
\hline
\multicolumn{5}{c}{\bf{Ions}}\\
\hline
OCN$^-$: HNCO:NH$_3$ (1:1) & 15 & 1.0 & LIDA &[8] \\
HCOO$^-$: H$_2$O:NH$_3$:HCOOH (100:2.6:2) & 14$-$210 & 1.0 & LIDA &[9] \\
HCOO$^-$: NH$_3$:HCOOH (1.3:1) & 14$-$210 & 1.0 & LIDA &[9] \\
\hline
\multicolumn{5}{c}{\bf{COMs (more than 6 atoms)}}\\
\hline
CH$_3$OH  & 10$-$120 & 1.0 & OCdb    &[2]\\
CH$_3$CHO  & 15$-$120 &  1.0  &  LIDA    &[10]\\
CH$_3$CN  & 15$-$150 &  1.0  &  LIDA    &[11]\\
CH$_3$OCH$_3$  & 15$-$100 &  1.0 & LIDA      &[10]\\
CH$_3$COCH$_3$  & 15$-$100 &  1.0 & LIDA      &[11]\\
CH$_3$CH$_2$OH  &  15$-$150 &  1.0 & LIDA     &[10]\\
CH$_3$OCHO  &  15$-$120 &  1.0 & LIDA     &[12]\\
CH$_3$COOH  &  10 &  1.0 & NASA     &[13]\\
CH$_3$NH$_2$  &  10 &  1.0 & LIDA     &[14]\\
CH$_3$CH$_2$CH$_2$OH  &  13 &  1.0 & NASA     &[15]\\
HC(O)CH$_2$CH$_3$  &  10 &  1.0 & NASA     &[16]\\
CH$_3$CH$_2$OH:H$_2$O (1:20) & 15$-$160 & 1.0 & LIDA &[10] \\
CH$_3$CH$_2$OH:CO (1:20) & 15, 30 & 1.0 & LIDA  &[10] \\
CH$_3$CH$_2$OH:CH$_3$OH (1:20) & 15$-$150 & 1.0 & LIDA  &[10] \\
CH$_3$CH$_2$OH:CO:CH$_3$OH (1:20:20) & 15$-$150 & 1.0 & LIDA  &[10] \\
CH$_3$CHO:H$_2$O (1:20) & 15$-$120 & 1.0 & LIDA &[10] \\
CH$_3$CHO:CO (1:20) & 15, 30 & 1.0 & LIDA  &[10] \\
CH$_3$CHO:CH$_3$OH (1:20) & 15$-$140 & 1.0 & LIDA  &[10] \\
CH$_3$CHO:CO:CH$_3$OH (1:20:20) & 15$-$120 & 1.0 & LIDA  &[10]\\
CH$_3$OCH$_3$:H$_2$O (1:20) & 15$-$160 & 1.0 & LIDA &[10] \\
CH$_3$OCH$_3$:CO (1:20) & 15, 30 & 1.0 & LIDA  &[10] \\
CH$_3$OCH$_3$:CH$_3$OH (1:20) & 15$-$120 & 1.0 & LIDA  &[10] \\
CH$_3$OCH$_3$:CO:CH$_3$OH (1:20:20) & 15$-$100 & 1.0 & LIDA &[10]\\
CH$_3$COCH$_3$:H$_2$O (1:20) & 15$-$160 & 1.0 & LIDA &[11] \\
CH$_3$COCH$_3$:CO (1:20) & 15, 30 & 1.0 & LIDA &[11] \\
CH$_3$COCH$_3$:CO$_2$ (1:20) & 15$-$100 & 1.0 & LIDA &[11] \\
CH$_3$COCH$_3$:CH$_3$OH (1:20) & 15$-$140 & 1.0 & LIDA &[11]\\
CH$_3$COCH$_3$:H$_2$O:CO$_2$ (1:2.5:2.5) & 15$-$160 & 1.0 & LIDA &[11]\\
CH$_3$COCH$_3$:CO:CH$_3$OH (1:2.5:2.5) & 15$-$140 & 1.0 & LIDA &[11]\\
CH$_3$OCHO:H$_2$O (1:20) & 15$-$120 & 1.0 & LIDA &[12]\\
CH$_3$OCHO:CO (1:20) & 15$-$120 & 1.0 & LIDA &[12]\\
CH$_3$OCHO:H$_2$CO (1:20) & 15$-$120 & 1.0 & LIDA &[12]\\
CH$_3$OCHO:CO:H$_2$CO:CH$_3$OH (1:20:20:20) & 15$-$120 & 1.0 & LIDA &[12]\\
CH$_3$COOH:H$_2$O (1:20) & 10 & 1.0 & NASA &[13]\\
CH$_3$NH$_2$:H$_2$O (1:20) & 15$-$150 & 1.0 & LIDA &[14]\\
CH$_3$NH$_2$:NH$_3$ (1:20) & 15$-$150 & 1.0 & LIDA &[14]\\
CH$_3$NH$_2$:CH$_4$ (1:20) & 15$-$150 & 1.0 & LIDA &[14]\\
CH$_3$CN:H$_2$O (1:20) & 15$-$150 & 1.0 & LIDA &[11]\\
CH$_3$CN:CO (1:20) & 15$-$100 & 1.0 & LIDA &[11]\\
NH$_2$CHO:H$_2$O (7:100) & 15$-$160 & 1.0 & LIDA &[17]\\
NH$_2$CHO:CO (4:100) & 15$-$34 & 1.0 & LIDA &[17]\\
HCOCH$_2$OH:H$_2$O (1:18) & 10 & 1.0 & NASA &[18]\\
\hline
\hline
\end{tabular}
}
\tablefoot{[1] \citet{Oberg2007}, \citet{Gerakines1996}; [2] \citet{Hudgins1993}; [3] \citet{Bisschop2007}; [4] \citet{Boogert1997}; [5] \citet{Rocha2017}; [6] \citet{Hudson2014c2h2}; [7] \citet{Hudson2014eth}; [8] \citet{Novozamsky2001}; [9] \citet{Galvez2010}; [10] \citet{Scheltinga2018}; [11] \citet{Rachid2022}; [12] \citet{Scheltinga2021}; [13] No reference found - taken from the NASA Ice Database (Pure: \url{https://science.gsfc.nasa.gov/691/cosmicice/spectra/refspec/Acids/CH3COOH/ACETIC-W.txt}, Mixture: \url{https://science.gsfc.nasa.gov/691/cosmicice/spectra/8_compounds/Combined_spectra_2018-12-20.xlsx}); [14] \citet{Rachid2021}; [15] \citet{Hudson2019}; [16] \citet{Yarnall2020}; [17] \citet{slav2023}; [18] \citet{Hudson2005}. $^a$LIDA: The Leiden Ice Database for Astrochemistry (\url{https://icedb.strw.leidenuniv.nl/}); OCdb: The Optical Constant Database (\url{https://ocdb.smce.nasa.gov/}); UNIVAP: \url{https://www1.univap.br/gaa/nkabs-database/data.htm}; NASA Cosmic Ice Laboratory: \url{https://science.gsfc.nasa.gov/691/cosmicice/spectra.html}}
\end{table*}

\section{Acetic acid and OCN$^-$ band strengths}
\label{Ap_bs}
Most of the band strengths for the molecules detected in the range addressed in this paper are available in the literature. However, to the best of our knowledge, the band strengths of acetic acid and OCN$^-$ between 6.8 and 8.6~$\mu$m were not calculated before. In particular, the band strength of acetic acid at 5.8~$\mu$m is often assumed to be the same as, or corrected from, the gas-phase acetic acid \citep[e.g.,][]{Oberg2009, Chuang2020} based on \citet{Marechal1987}.

\begin{figure}
   \centering
   \includegraphics[width=\hsize]{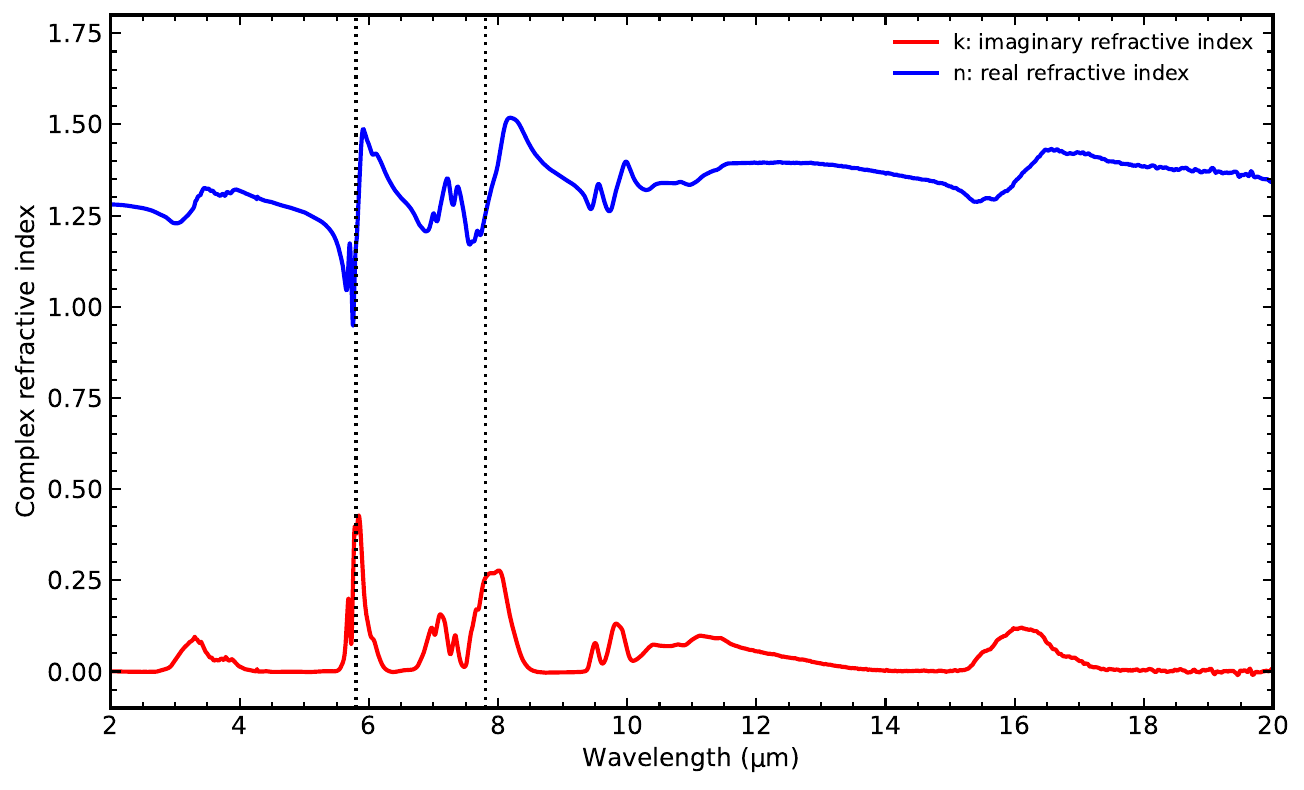}
      \caption{Optical contacts of acetic acid. Vertical dotted lines indicate the features with band strength calculated in this paper.}
         \label{actacid_nk}
   \end{figure}

We derive the absolute ($A$) and apparent ($A'$) band strengths of acetic acid. The absolute band strengths are derived from the imaginary refractive index, by the following equation:
\begin{equation}
    A = \frac{m}{\rho N_A} \int_{\nu_1}^{\nu_2} 4\pi\nu k(\nu) d\nu
\end{equation}
where $m$ is the molar mass of acetic acid in g mol$^{-1}$ (60.052), $\rho$ is the density of acetic acid \citep[0.892 g cm$^{-3}$;][]{Hudson2020bs}, $N_A$ is the Avogrado's number, $\nu$ is the wavenumber and $k$ is the imaginary refractive index. We derive $k$ using the recent version of the \texttt{NKABS} code \citep{Rocha2014}. This code calculates the real ($n$) and imaginary ($k$) refractive index from the absorbance spectrum ($Abs_{\nu}$). The input spectrum is taken from \citet{Hudson2019}, who also estimated the thickness of the ice ($d = 2.1~\mu$m) and the refractive index around 700~nm ($n_0 = 1.29$). The $n$ and $k$ values are shown in Figure~\ref{actacid_nk}. Finally, the band strengths derived for two acetic acid bands are listed in Table~\ref{act_bs}.

\begin{table}[h!]
\caption{\label{act_bs} Absolute and apparent band strengths of acetic acid.}
\renewcommand{\arraystretch}{1.4}
\centering 
\begin{tabular}{lcc}
\hline\hline
$\lambda_{\rm{central}}$ & A$_{abs}$ (cm molec$^{-1}$) & A$_{apr}$ (cm molec$^{-1}$)\\
\hline
5.8 & 7.5 $\times$ 10$^{-17}$ & 7.3 $\times$ 10$^{-17}$\\
7.82 & 4.6 $\times$ 10$^{-17}$ & 4.6 $\times$ 10$^{-17}$\\
\hline
\hline
\end{tabular}
\end{table}

The apparent band strength is calculated by:
\begin{equation}
    A' = 2.303 \frac{m}{\rho N_A d} \int Abs_{\nu} d\nu
\end{equation}

As one can see, both $A$ and $A'$ values for acetic acid are similar.

In the case of OCN$^-$, we derive the apparent band strength at 7.62~$\mu$m (1312~cm$^{-1}$). The OCN$^-$ band at 4.59~$\mu$m (2175~cm$^{-1}$) is used as a reference since the band strength is known \citep[1.3$\times$10$^{-16}$ cm molecule$^{-1}$;][]{vanBroekhuizen2005}. We use the following equation to derive the OCN$^-$ apparent band strength at 7.62~$\mu$m:
\begin{equation}
    A'_{7.62\mu m} = 1.3 \times 10^{-16} \frac{\int Abs_{1312 \rm{cm}^{-1}} d\nu}{\int Abs_{2175 \rm{cm}^{-1}} d\nu},
\end{equation}
where $Abs_{2175 \rm{cm}^{-1}}$ and $Abs_{1312 \rm{cm}^{-1}}$ are the OCN$^-$ bands at the given wavelengths.

\section{Laboratory spectra: removing H$_2$O and CH$_3$OH ice features}
\label{lab_rem}
\begin{figure}
   \centering
   \includegraphics[width=\hsize]{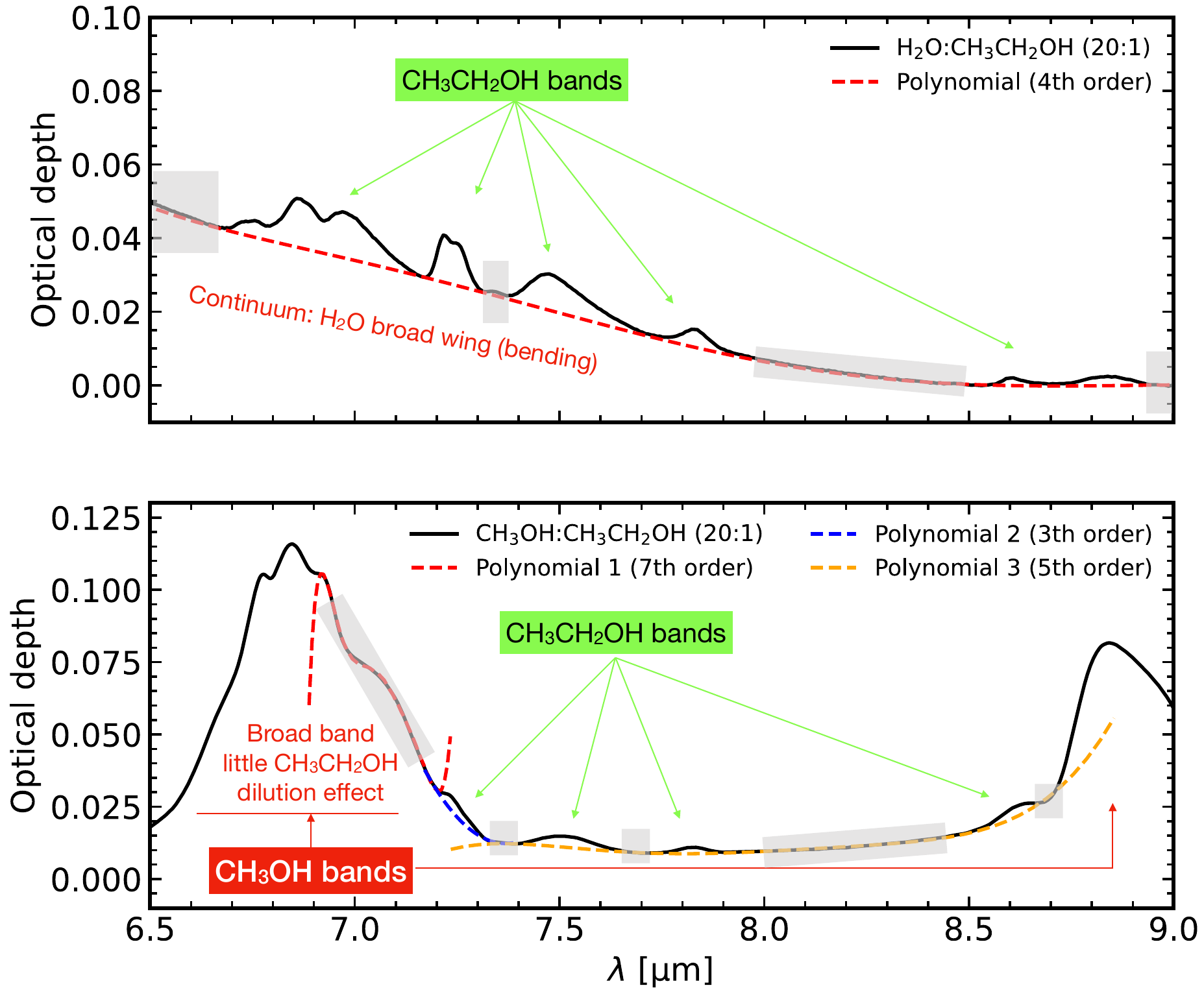}
      \caption{Isolating the CH$_3$CH$_2$OH ice bands from H$_2$O and CH$_3$OH ices. Polynomial functions anchored at the points indicated by the grey areas are used. The top panel shows the ethanol mixed with H$_2$O ice, and the bottom panel displays the mixture with CH$_3$OH.}
         \label{contcoms_fig}
   \end{figure}

In this section, we demonstrate how to remove the H$_2$O and CH$_3$OH features of COMs spectra containing these two molecules in the range between 6.5 and 9.0~$\mu$m. This process is necessary when analysing the observational spectrum between 6.8 and 8.6~$\mu$m using local continuum subtraction. Figure~\ref{contcoms_fig} (top) shows the H$_2$O:CH$_3$CH$_2$OH ice spectrum \citep{Scheltinga2018} and the polynomial (4th-order) function used to trace a baseline under the CH$_3$CH$_2$OH bands. In this spectrum, the ethanol features overlap with the broad shoulder of the H$_2$O ice bending mode, which is represented by the polynomial fit. In Figure~\ref{contcoms_fig} (bottom), we show the CH$_3$OH:CH$_3$CH$_2$OH IR spectrum \citep{Scheltinga2018}. Since both molecules are alcohols they share functional groups, which makes it harder to isolate the ethanol features of methanol. Because of the high dilution factor (20:1), it is not feasible to disentangle the features of the two molecules below 6.8~$\mu$m, and therefore, we separate the CH$_3$CH$_2$OH bands between 6.9 and 8.6~$\mu$m. The step is performed in three stages: (i) high-order polynomial (7th) between 6.9$-$7.15~$\mu$m to remove the CH$_3$OH shoulder, (ii) 3rd-order polynomial between 7.15$-$7.45 to isolate the ethanol band at 7.2~$\mu$m and account for the strong curvature in the data, and (iii) 5th-order polynomial between 7.45-8.7~$\mu$m to extract the other ethanol bands.

\section{Laboratory baseline correction: avoiding spurious features}
\label{spurious_sec}

\begin{figure}
   \centering
   \includegraphics[width=\hsize]{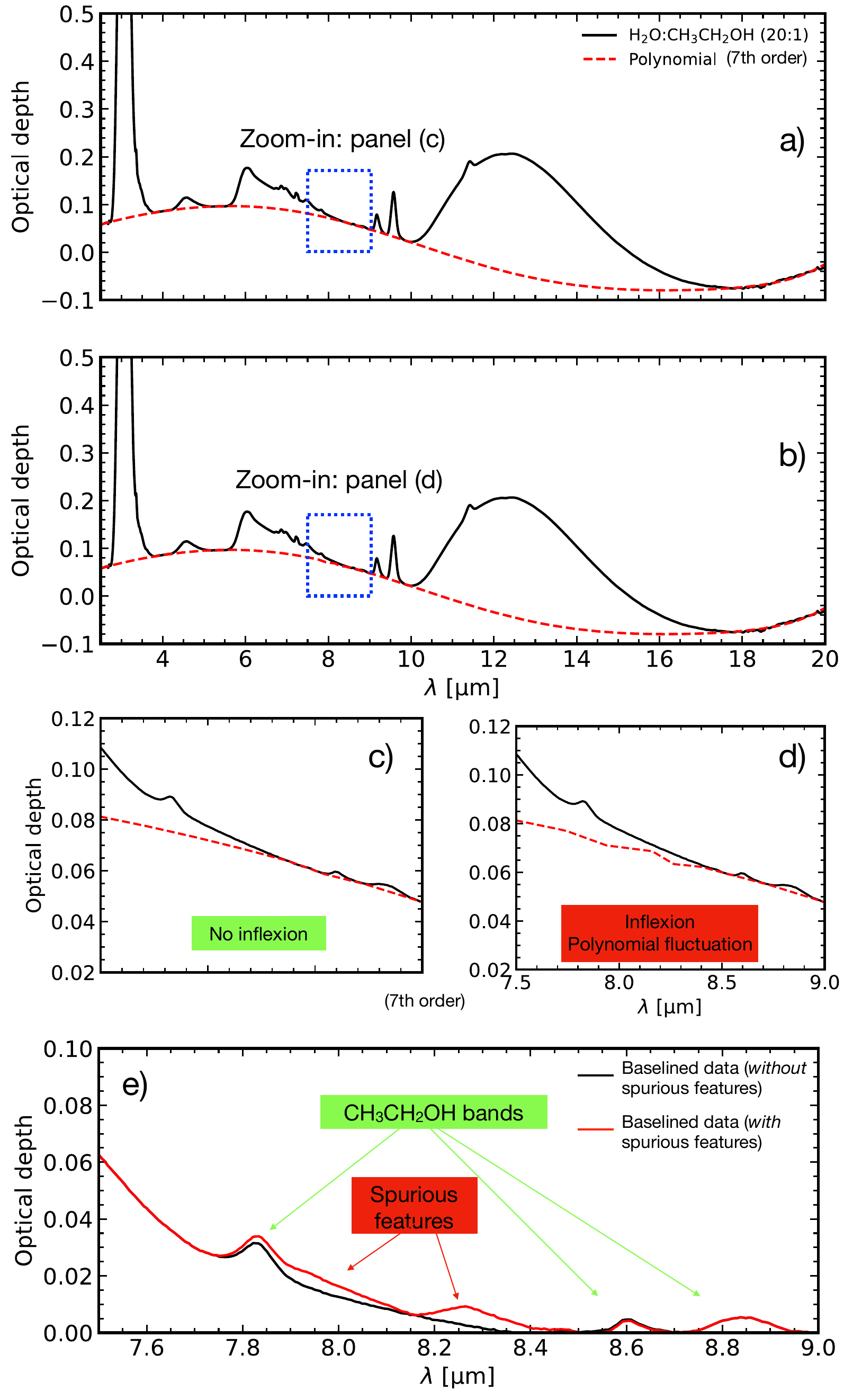}
      \caption{Baseline correction of the IR ice spectrum of H$_2$O:CH$_3$CH$_2$OH data. Panels a and b show the IR ice spectra and the baselines. The blue dotted squares highlight the zoom-ins shown in panels c and d displaying polynomial baselines without and with little inflexion, respectively. Panel e demonstrates the effect of non-accurate baseline subtraction in the ice spectrum.}
         \label{spu_fig}
   \end{figure}

IR spectra of ices are recorded in the laboratory using Fourier Transform Infrared Spectroscopy (FTIR) and interference effects are corrected using a spline or polynomial function, the so-called baseline correction. In Figure~\ref{spu_fig} we show the baseline correction of the H$_2$O:CH$_3$CH$_2$OH IR spectrum at 15~K. Panels a and b show the laboratory spectrum and a 7th-order polynomial used to correct the interference effect. The difference between these two panels is that in panel a, we use as many points as possible to trace the baseline, whereas in panel b, we use only half of the points available in the range of 7.8$-$8.5~$\mu$m. This creates a small fluctuation in the polynomial function. Zoom-ins of both cases are shown in panels c and d, respectively. In panel e, we show the baseline data. The spectrum with no polynomial inflexion contains the real CH$_3$CH$_2$OH IR features, whereas the other data has spurious features at 8.0 and 8.3~$\mu$m. This demonstrates that care must be taken when correcting IR spectrum baselines to avoid creating spurious features.

\section{Incremental version of the fits for IRAS~2A and IRAS~23385}
\label{fits_increment}
\begin{figure*}
   \centering
   \includegraphics[width=15.5cm]{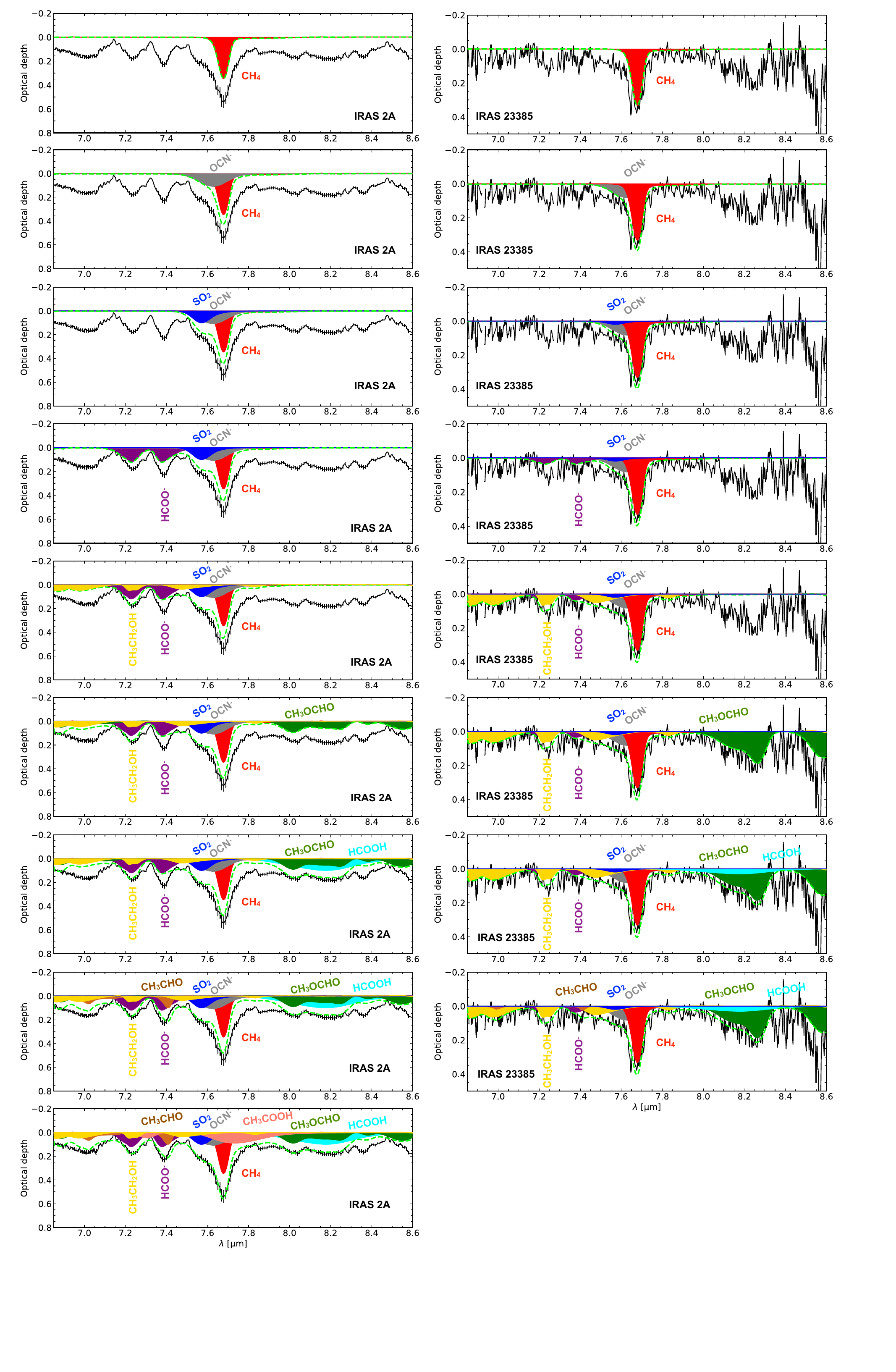}
      \caption{Incremental version from top to bottom of the best fits with the \texttt{ENIIGMA} fitting tool for IRAS~2A (left) and IRAS~23385 (right). The molecule label follows the colour code of the laboratory components in the fit.}
         \label{incremental}
   \end{figure*}

Figure~\ref{incremental} displays the best fits for IRAS~2A (left) and IRAS~23385 (right) by adding one component at a time in each panel. This allows understanding better how each component contributes to the fit.

\section{HCOO$^-$ at 14, 150 and 210~K}
\label{formateionapd}
\begin{figure}
   \centering
   \includegraphics[width=8cm]{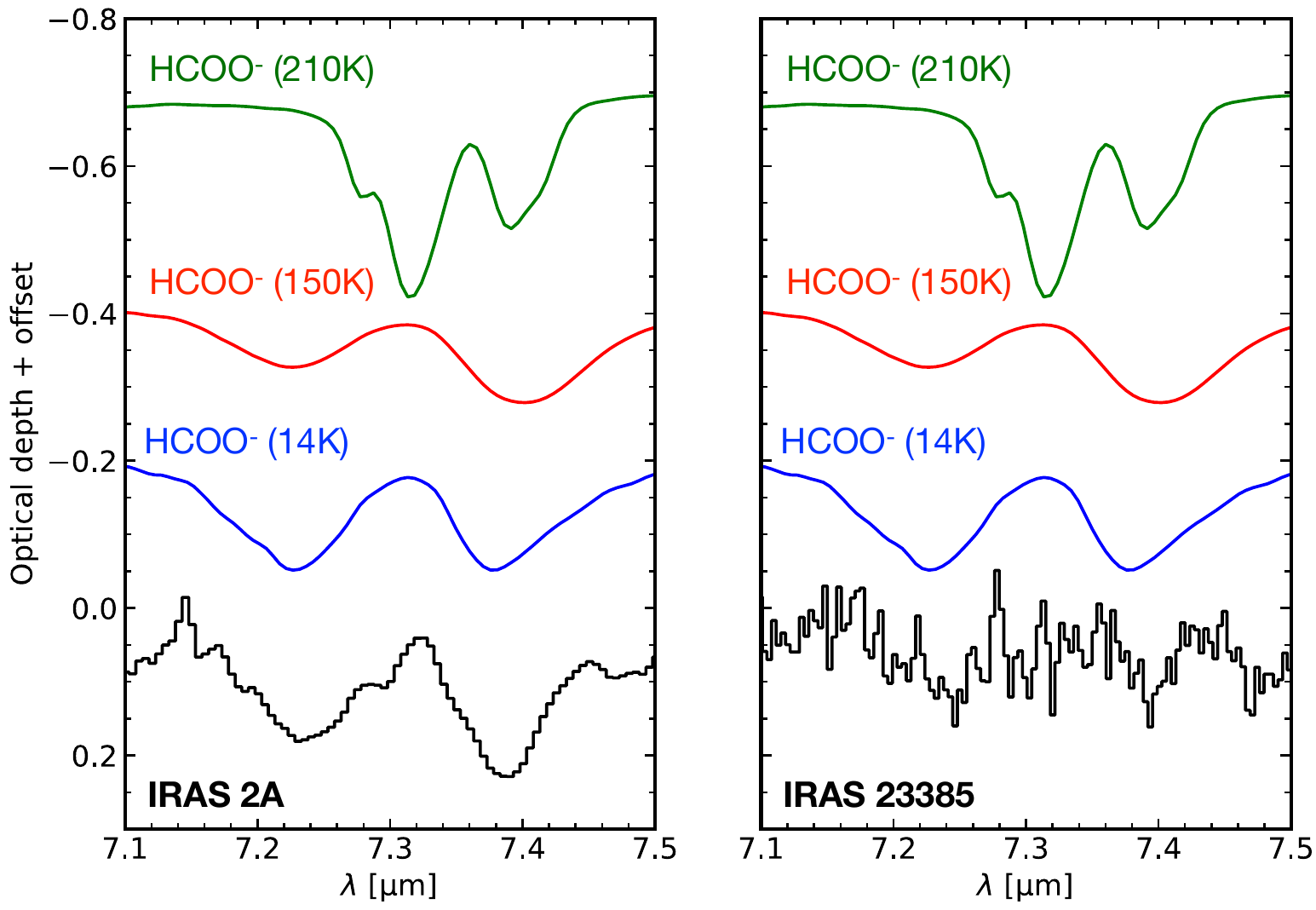}
      \caption{Comparison between the 7.2 and 7.4~$\mu$m band of IRAS~2A (left) and IRAS~23385 (right) with the HCOO$^-$ absorption profiles at 14, 150 and 210~K. These IR ice spectra are taken from \citet{Galvez2010}.}
         \label{formateionfig}
   \end{figure}

We compare the absorption profiles of HCOO$^-$ at three temperatures (14~K, 150~K, 210~K) with the 7.2 and 7.4~$\mu$m bands in IRAS~2A and IRAS~23385 (Figure~\ref{formateionfig}). Our analysis shows that the spectrum at 14~K provides the best fit. The spectrum at 150~K is excluded because the second peak ($\sim$ 7.4~$\mu$m) is broader and shifted compared to the observations. Likewise, the band shape of the highest temperature data (150~K) does not match with both protostars.  

\section{Confidence intervals}
\label{confidence_ap}

\begin{figure}
   \centering
   \includegraphics[width=\hsize]{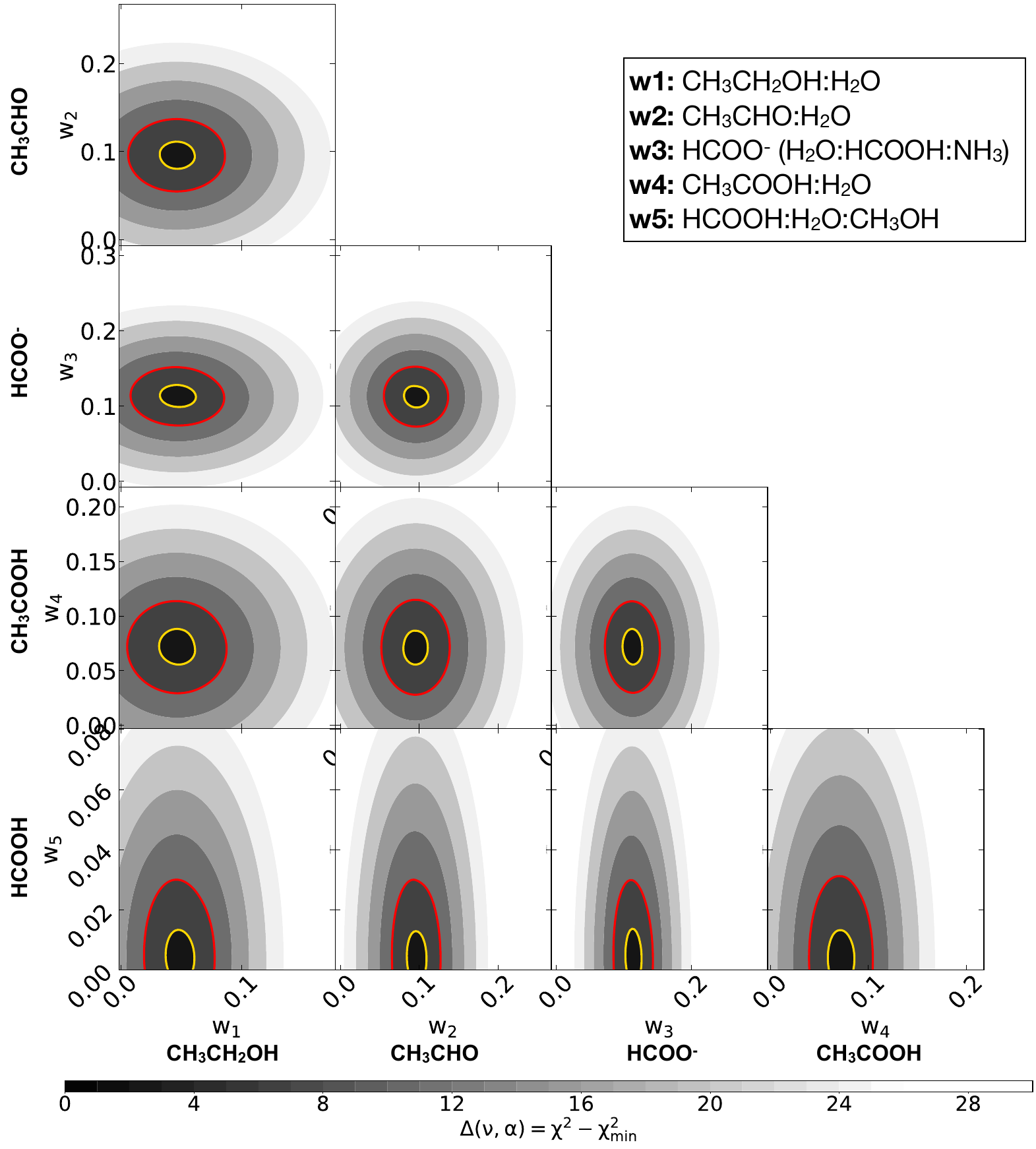}
   \includegraphics[width=\hsize]{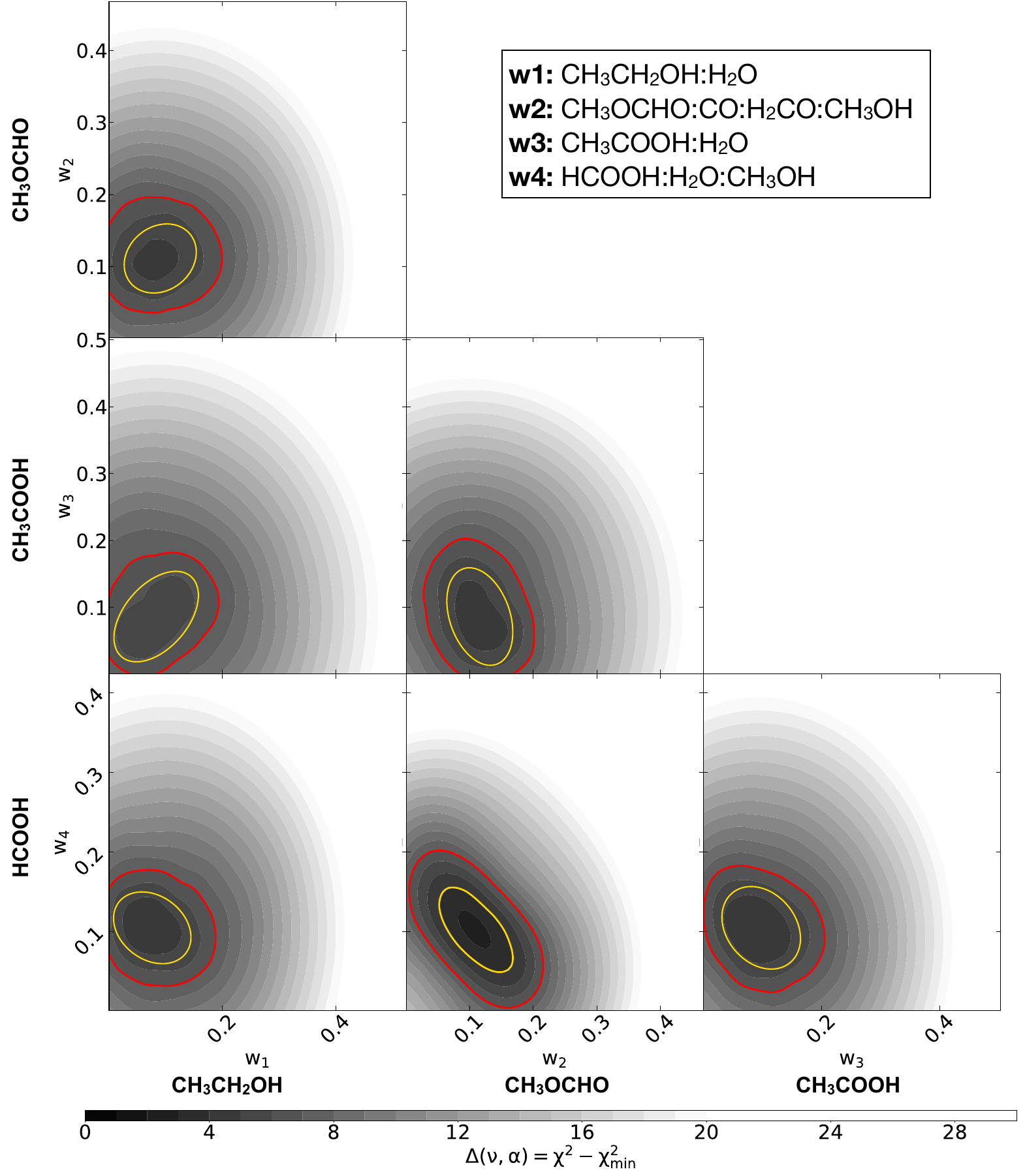}
      \caption{The top and bottom corner plots show the IRAS~2A coefficient confidence intervals for the range between 6.86-7.5 (top) and 7.8$-$8.6~$\mu$m (bottom), respectively.}
         \label{Conf_coeff_ir2a}
   \end{figure}

\begin{figure}
   \centering
   \includegraphics[width=\hsize]{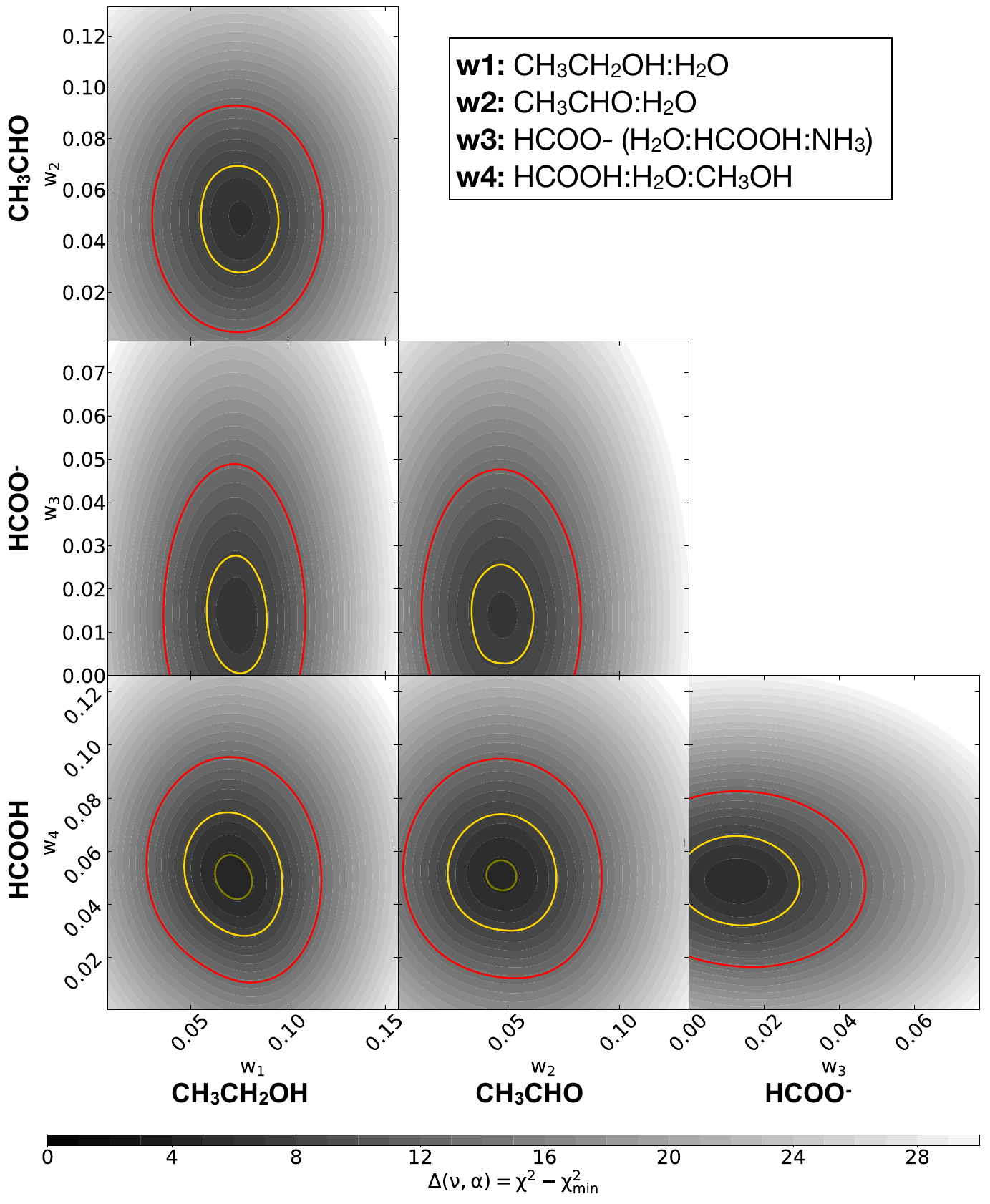} 
   \includegraphics[width=\hsize]{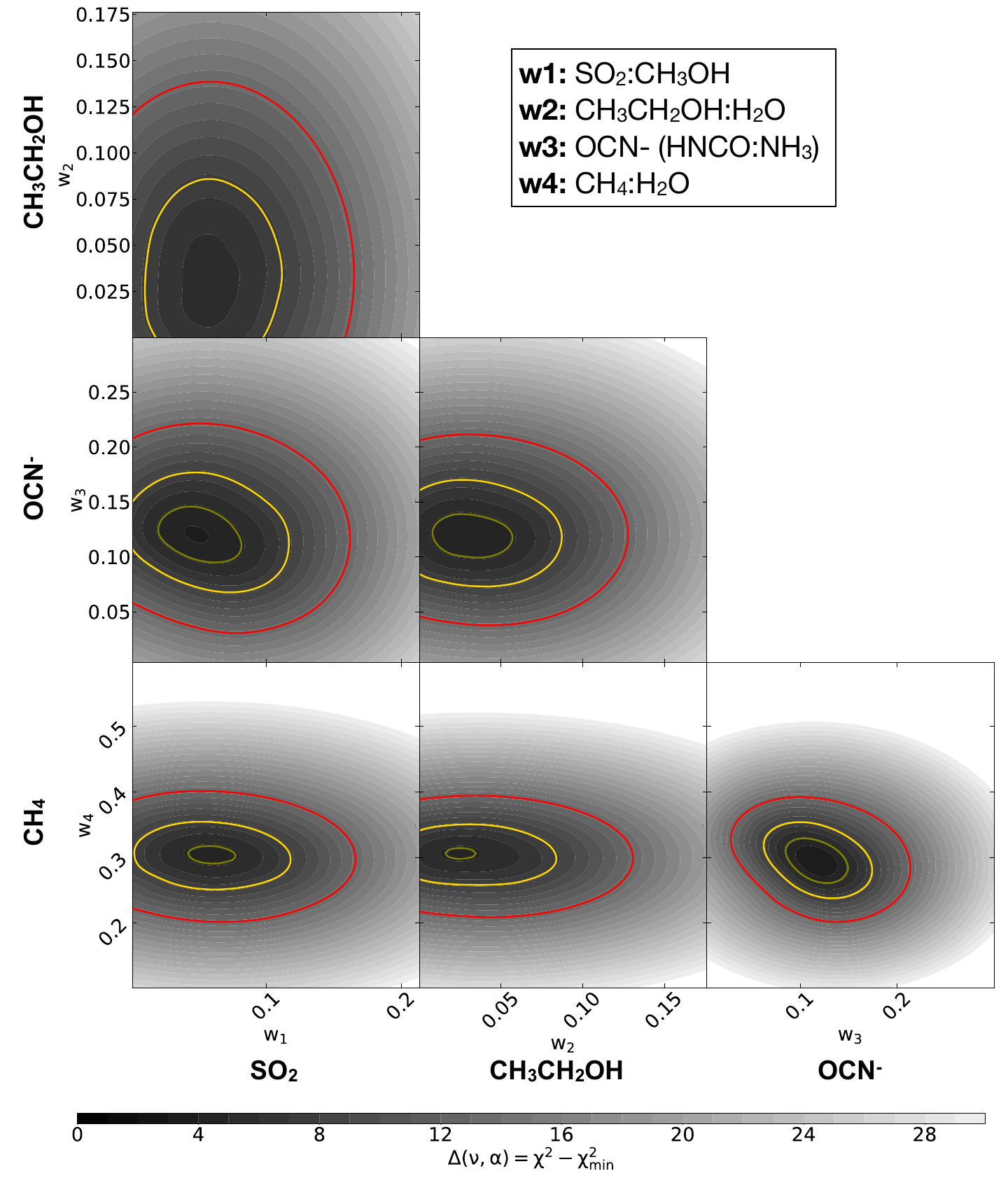}
      \caption{The top and bottom corner plots show the IRAS~23385 coefficient confidence intervals for the range between 6.86$-$7.5 (top) and 7.5$-$7.8~$\mu$m (bottom), respectively.}
        \label{Conf_coeff_ir23385_p1}
   \end{figure}

\begin{figure}
   \centering
   \includegraphics[width=\hsize]{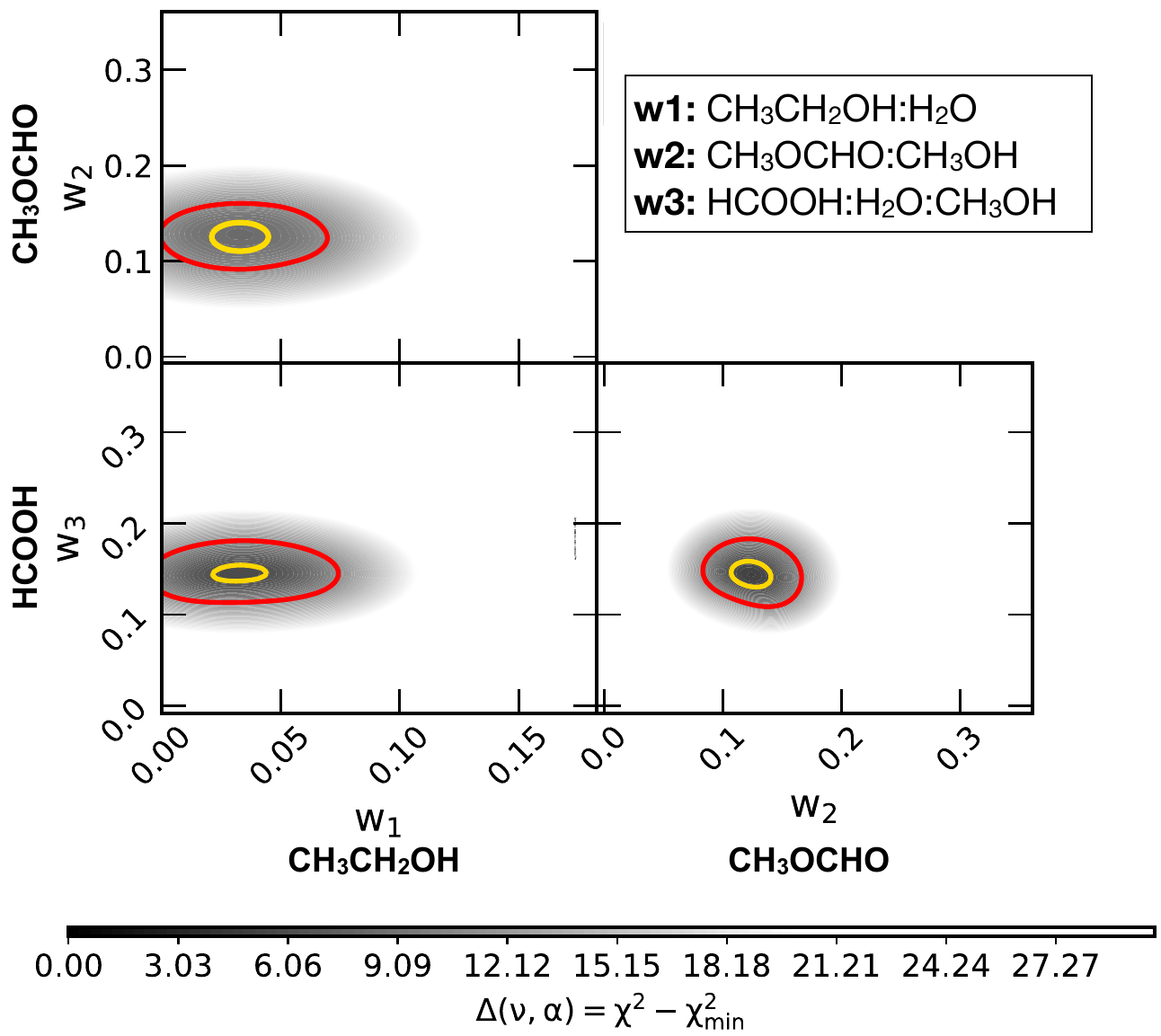}
      \caption{Same as in Figure~\ref{Conf_coeff_ir23385_p1}, but for the range 7.8$-$8.6~$\mu$m.}
       \label{Conf_coeff_ir23385_p2}
   \end{figure}

The top and bottom panels in Figure~\ref{Conf_coeff_ir2a} show the confidence intervals for IRAS~2A in the ranges of 6.8$-$7.5 and 7.8$-$8.6, respectively. The confidence intervals for IRAS~23385 are shown in Figures~\ref{Conf_coeff_ir23385_p1} and \ref{Conf_coeff_ir23385_p2}. For IRAS~2A, it can be noted that all components are essential to the fit, and cannot be excluded. On the other hand, for IRAS~23385, the SO$_2$ band can be excluded as a solution if the OCN$^-$ band is slightly intense.

\section{Different local continuum profiles between 6.8 and 8.6~$\mu$m - IRAS~2A}
\label{cont_effect}

\begin{figure}
   \centering
   \includegraphics[width=\hsize]{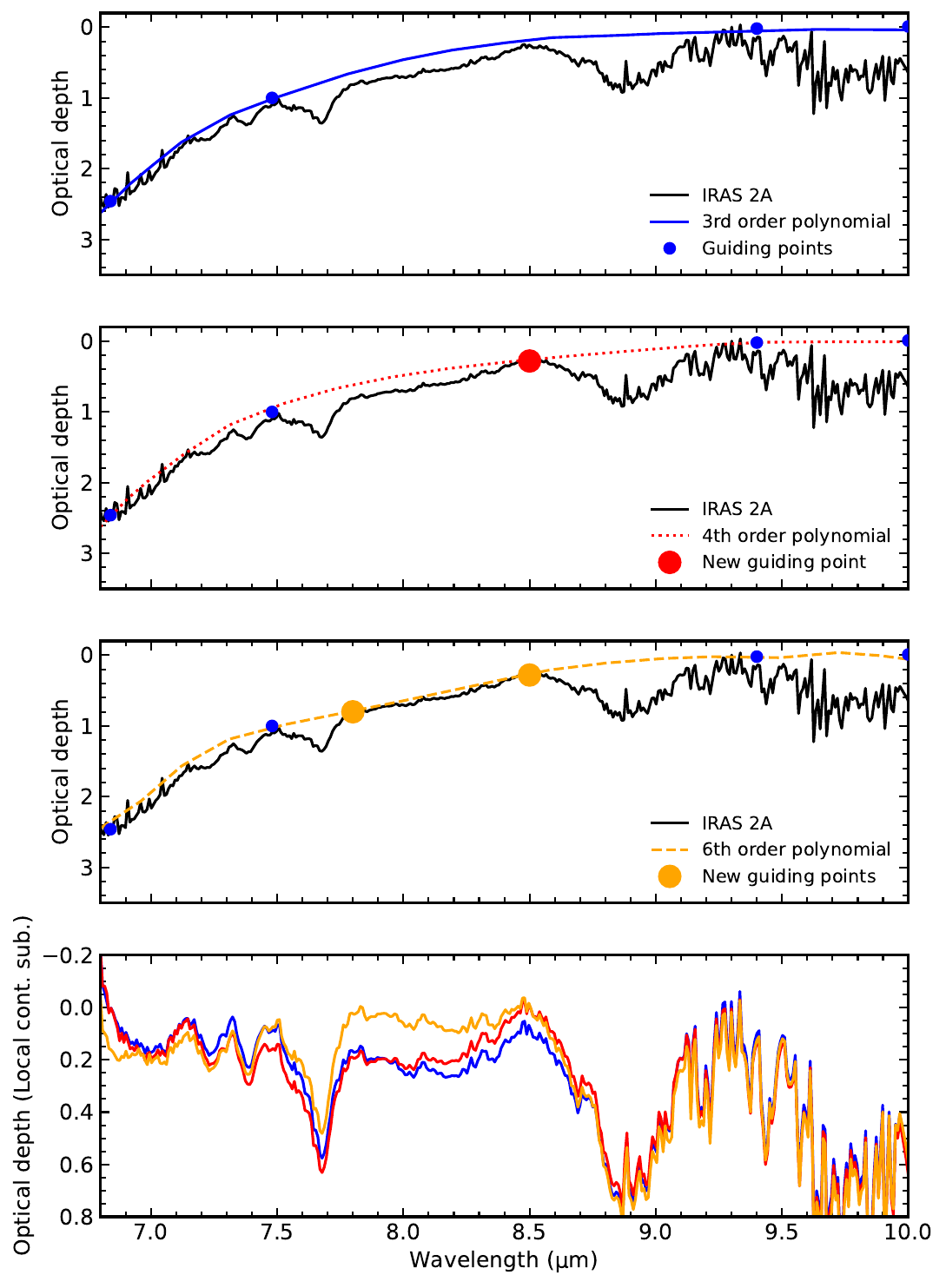}
      \caption{Effect of the continuum choice on the local continuum subtracted spectrum of IRAS~2A. The first panel shows the continuum adopted as the best model in this paper. The second and third panels show two other continuum options by adding the red and orange points, respectively. The optical depth subtracted spectra are shown in the fourth panel.}
         \label{diff_cont}
   \end{figure}

The first three panels of Figure~\ref{diff_cont} show different continuum profiles between 6.8-8.6~$\mu$m in the IRAS~2A spectrum. The top panel is the version adopted for the analysis in this paper that traced a third-order polynomial to the guiding points. The second panel displays the fourth-order polynomial where the red dot is added to the guiding points. In this case, the continuum is slightly elevated at shorter wavelengths to accommodate the fit to the extra point at 8.5~$\mu$m. The third panel presents the continuum when two extra points are added ($\lambda = 7.8$~$\mu$m and 8.5~$\mu$m), and a sixth-order polynomial is used. All subtracted spectra using these three approaches are shown in the bottom panel of Figure~\ref{diff_cont}.

\begin{figure}
   \centering
   \includegraphics[width=\hsize]{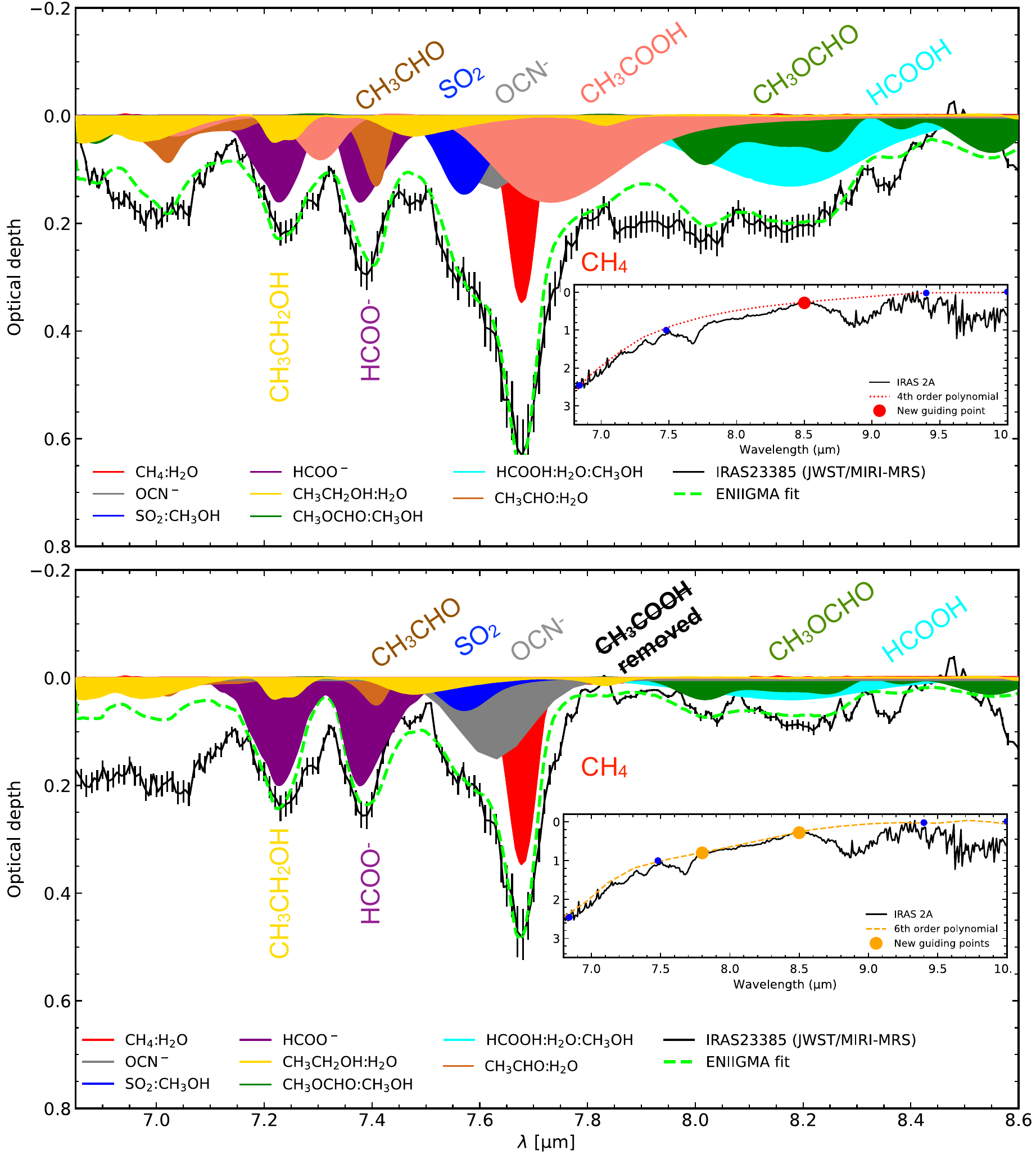}
      \caption{Alternative fits of IRAS~2A optical depth spectrum with different local continuum choices (see Figure~\ref{diff_cont}). The top panel shows the fits with all the components after subtracting the red local continuum with an extra point at 8.5~$\mu$m. The bottom panel shows the same as in the top panel, but considering two extra points for the local continuum (7.8 and 8.5~$\mu$m). Only CH$_3$COOH is excluded in this fit.}
         \label{diff_fits}
   \end{figure}

New fits for IRAS~2A are performed on the other two optical depth spectra obtained from different local continuum choices, which are shown in Figure~\ref{diff_fits}. The panel at the top shows that all components remain needed to reach the best fit. Only the 8.5~$\mu$m band of CH$_3$OCHO exceeds the observations, and this is caused by the guiding point added at 8.5~$\mu$m. The bottom panel of Figure~\ref{diff_fits} shows another fit, where two extra points are added at 7.8 and 8.5~$\mu$m. This version of the fit keeps all components, except CH$_3$COOH, which is excluded because of the anchor point added at 7.8~$\mu$m, where CH$_3$COOH has a strong feature. Other issues are seen in this fit, for example, around 7~$\mu$m, 7.3~$\mu$m, and a poor fit between 7.8 and 8.6~$\mu$m. Despite all these differences, this analysis shows that CH$_3$CHO, CH$_3$CH$_2$OH and CH$_3$OCHO are still robust detections, and cannot be excluded from the fits of the IRAS~2A spectrum.

\section{Water and methanol ice column densities}
\label{water_cd}
\begin{figure}
   \centering
   \includegraphics[width=\hsize]{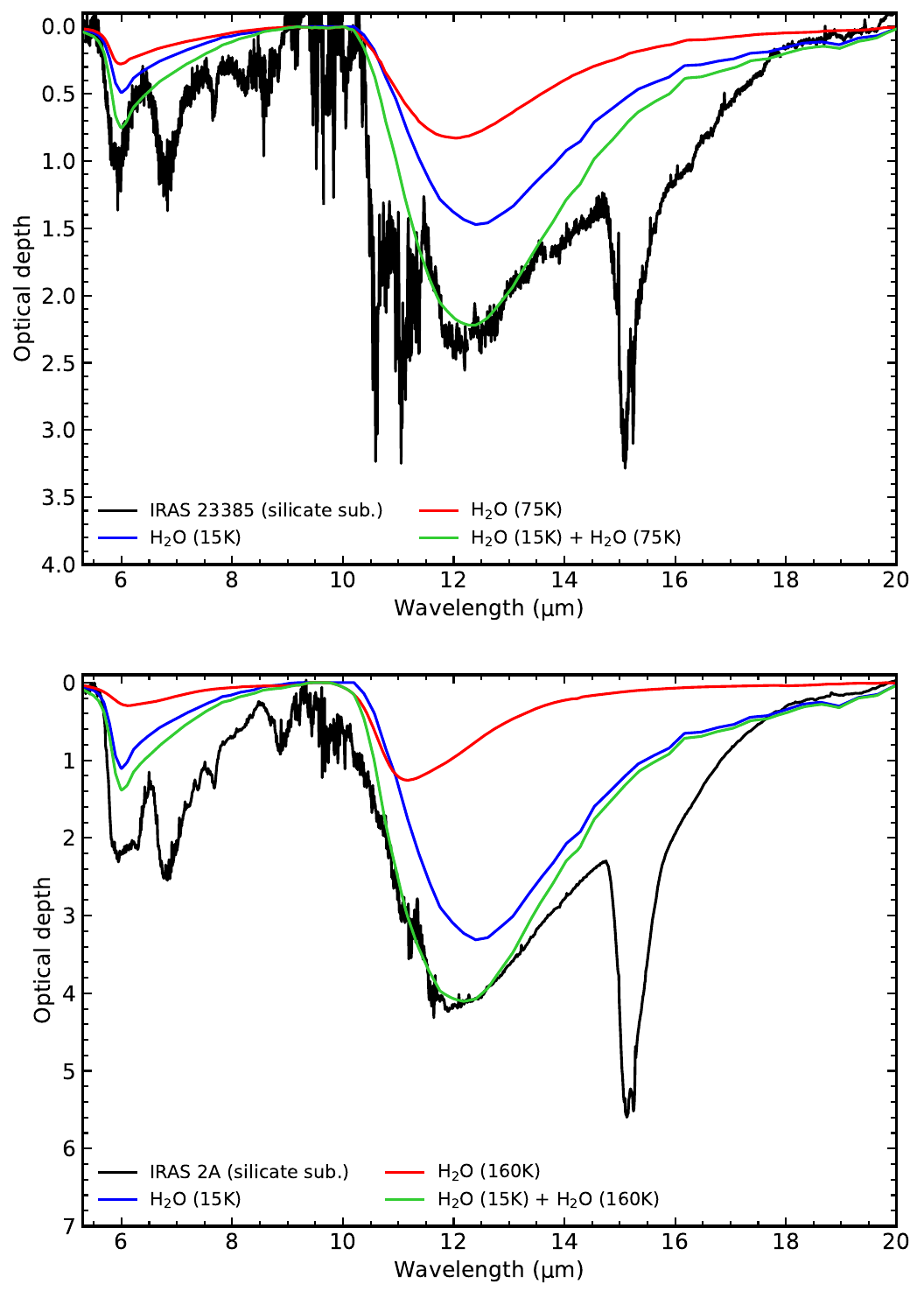}
      \caption{Fits for the H$_2$O ice libration band for IRAS~23385 (top) and IRAS~2A (bottom). The best fit is found by combining two H$_2$O ice grain-shaped corrected spectra: 15K and 75~K for IRAS~23385 and 15~K and 160~K for IRAS~2A.}
         \label{H2Olib}
   \end{figure}

The water ice column densities in both protostars are calculated from the libration mode around 12.5~$\mu$m. To determine the profile of the band, we combine water ice spectra at different temperatures (Figure~\ref{H2Olib}). As pointed out by \citet{Boogert2008}, this band is sensitive to the grain geometry. For this reason, we assume small spherical water ice grains which is consistent with \citet{Boogert2008} to fit the libration mode best. The optical constants for ices at 15, 75, and 160~K are taken from \citet{Rocha2022}. The water libration band is fitted with two components representing different temperatures. It is likely that the libration band is sensitive to a range of temperatures between 15 and 160~K, but addressing this is beyond the scope of this work. Despite this simplification, one can note that only cold water ice is not enough to fit the libration band of IRAS~23385 and IRAS~2A. In particular, IRAS~2A has a strong blue wing excess that requires H$_2$O ice at 160~K. The water ice column density is shown in Table~\ref{ice_cd}.

\begin{figure}
   \centering
   \includegraphics[width=\hsize]{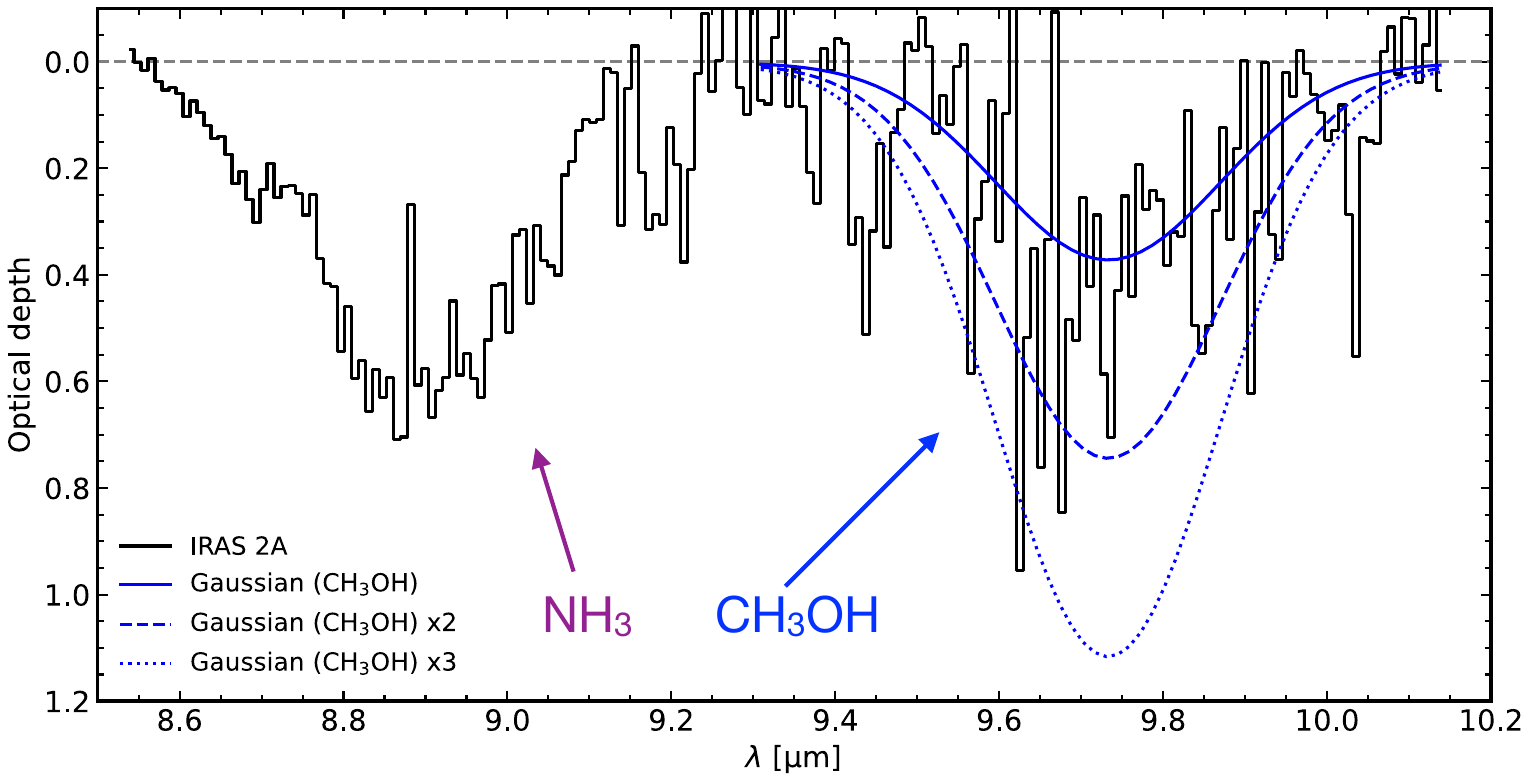}
      \caption{NH$_3$ and CH$_3$OH features in the bottom of the silicate band and H$_2$O ice subtracted spectra of IRAS~2A. Three Gaussian profiles are scaled to the CH$_3$OH band to indicate different ice column densities.}
         \label{methanol_cd}
   \end{figure}

In the case of CH$_3$OH ice, we use the band at 9.8~$\mu$m to derive a column density. We fit a Gaussian profile to the feature at 9.8~$\mu$m (solid curve in Figure~\ref{methanol_cd}) and multiply it by a factor of 2 (dashed curve) and 3 (dotted curve). It is unlikely that the solid curve accounts for all CH$_3$OH absorption, and therefore the CH$_3$OH ice column densities used in this paper correspond to the other two Gaussian profiles.

\section{Comparison with hydrocarbons}
\label{hydro}

\begin{figure}
   \centering
   \includegraphics[width=\hsize]{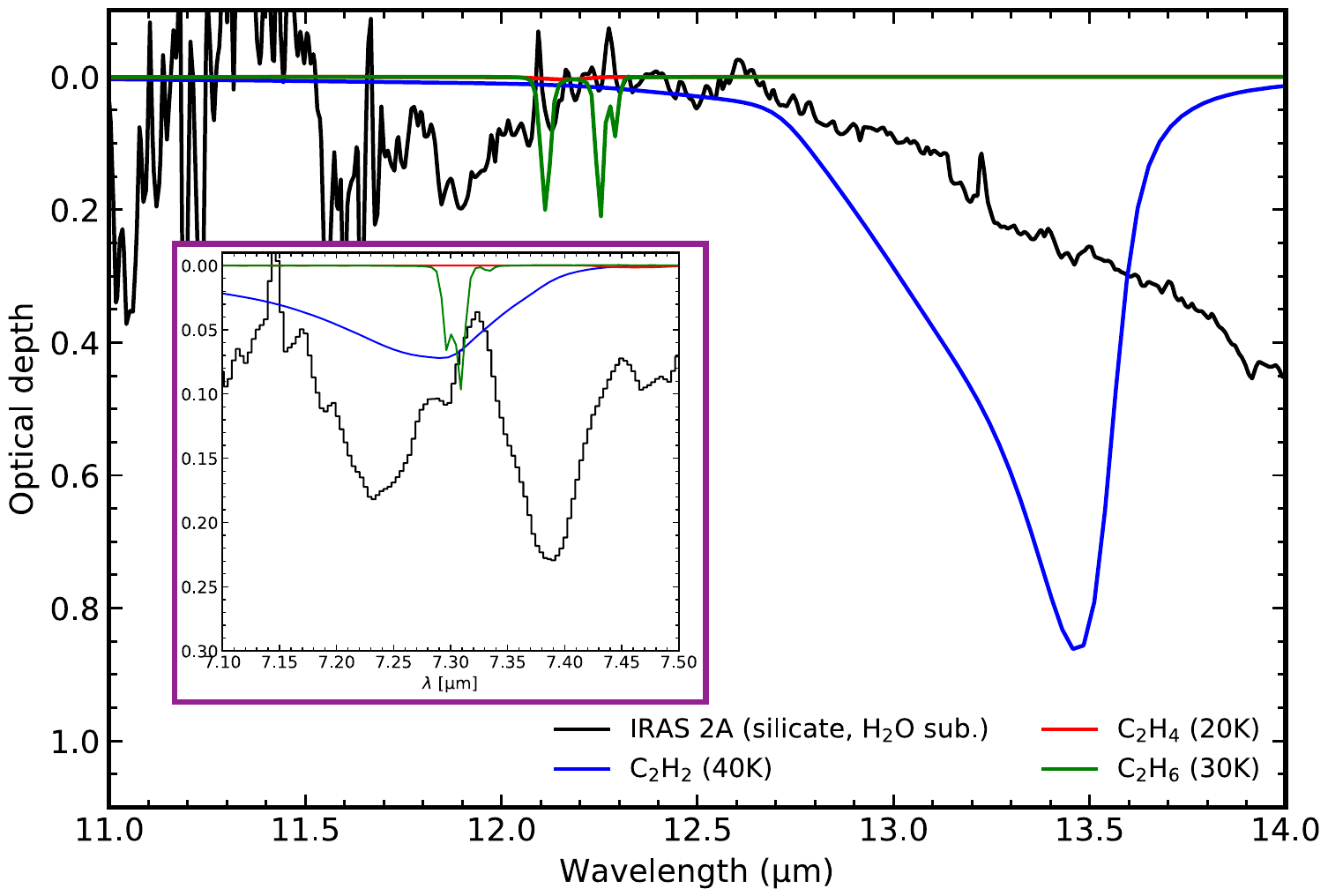}
      \caption{Comparison between the 7.2 and 7.4~$\mu$m band of IRAS~2A (silicate and H$_2$O ice subtracted) and hydrocarbons (C$_2$H$_2$, C$_2$H$_4$ and C$_2$H$_6$).}
         \label{hydrocarb}
   \end{figure}

Because of the degeneracy intrinsic to the ice fittings, particularly regarding COMs that share the same functional groups, we show in this section a comparison of hydrocarbons (C$_x$H$_x$) with the IRAS~2A spectrum. These molecules participated in the global fits between 6.8 and 8.6~$\mu$m, but are discarded as solutions. Therefore, these comparisons serve as an additional check that these components are not part of the global minimum solution. Figure~\ref{hydrocarb} shows scaled IR spectra of pure C$_2$H$_2$, C$_2$H$_4$, C$_2$H$_6$ compared to IRAS~2A. The scaling factors are arbitrarily chosen to match the IRAS~2A absorption profile at 7.3~$\mu$m. It can be seen that C$_2$H$_4$ does not have any contribution to the 7.2 and 7.4~$\mu$m. On the other hand, the absorption bands of C$_2$H$_2$ and C$_2$H$_6$ that could be hidden at 7.3~$\mu$m, exceed the IRAS~2A absorption profiles around 12 and 13.5~$\mu$m.

\end{document}